\documentclass[review]{elsarticle}
\usepackage[latin1]{inputenc} 
\usepackage{listings, tcolorbox}

\usepackage{lineno,hyperref}
\usepackage{amssymb}
\usepackage{color}
\usepackage{amsmath}
\usepackage{tikz}
\usepackage{amsfonts}
\usepackage{mathrsfs}
\usepackage{amsthm}

\usepackage{bbding}
\usepackage{mathrsfs}
\usepackage{amsmath, amsfonts, amssymb, mathrsfs, txfonts}
\usepackage{graphicx, subfigure}
\usepackage{wasysym}
\usepackage{color}
\usepackage{bm}
\usepackage{enumerate}

\usepackage{pifont}
\usepackage{txfonts}
\usepackage{amsmath}
\usepackage{graphicx}
\usepackage{amsfonts}
\usepackage{amssymb}
\usepackage{mathrsfs,psfrag,eepic,epsfig}
\usepackage{epstopdf}

\biboptions{numbers,sort&compress}

\theoremstyle{definition}

\modulolinenumbers[5]

\journal{Journal of \LaTeX\ Templates}

\makeatletter \@addtoreset{equation}{section}

\begin{document}

\begin{frontmatter}
\title{Inverse scattering transform for the integrable nonlocal Lakshmanan-Porsezian-Daniel equation}
\tnotetext[mytitlenote]{Project supported by the Fundamental Research Fund for the Central Universities under the grant No. 2019ZDPY07.\\
\hspace*{3ex}$^{*}$Corresponding author.\\
\hspace*{3ex}\emph{E-mail addresses}:  sftian@cumt.edu.cn and
shoufu2006@126.com (S. F. Tian)}

\author{Wei-Kang Xun and Shou-Fu Tian$^{*}$}
\address{
School of Mathematics and Institute of Mathematical Physics, China University of Mining and Technology, Xuzhou 221116, People's Republic of China
 }

\begin{abstract}
  In this work, a generalized nonlocal  Lakshmanan-Porsezian-Daniel (LPD) equation   is introduced, and its integrability  as an infinite dimensional Hamilton dynamic system is established.  Motivated by the ideas of  Ablowitz and Musslimani (2016 Nonlinearity \textbf{29} 915),  we successfully derive the inverse  scattering transform (IST) of  the nonlocal LPD equation.    The direct scattering problem  of the   equation is first constructed, and some important symmetries of the eigenfunctions and  the scattering data are discussed. By using a  novel Left-Right Riemann-Hilbert (RH) problem,  the inverse scattering problem is analyzed, and the potential function is recovered.  By introducing  the special conditions  of    reflectionless  case,  the time-periodic soliton solutions formula of   the   equation is derived successfully. Take $J=\overline{J}=1,2,3$ and $4$ for example,  we obtain some  interesting  phenomenon such as  breather-type solitons, arc solitons,  three soliton and four soliton. Furthermore, the influence of parameter $\delta$ on these solutions is further considered  via the graphical analysis.  Finally, the eigenvalues and conserved quantities are investigated  under  a few special initial conditions.
\end{abstract}


\end{frontmatter}

\clearpage
\tableofcontents
\clearpage

\section{Introduction}
 Nonlinear integrable evolution equations  exist in all aspects of scientific research  and play a essential role in  modern physical branches.  There are numerous  nonlinear integrable  evolution equations  which are applied into  fluid mechanics, elasticity, lattice dynamics, electromagnetics, etc.  For example, the Korteweg-de Vries (KdV)  and modified Korteweg-de Vries (mKdV)  equations describe the evolution of weakly dispersive and small amplitude waves in quadratic and cubic nonlinear media, respectively \cite{Ablowitz1981Solitons}.    The KdV equation is more famous for its application in shallow water waves.  Besides,  the integrable cubic nonlinear Schr\"{o}dinger (NLS) equation which is well-known for its application to the evolution of weakly nonlinear and quasi-monochromatic wave trains in media with cubic nonlinearities \cite{dianov1988nonlinear,Kivshar2003Optical}. Besides,  the Kadomtsev-Petviashvili (KP) equation which aries in plasma physics and internal waves \cite{Ablowitz1981Solitons,Ablowitz1991Solitons} is applied to describe the evolution of weakly dispersive and small amplitude waves with additional weak transverse variation \cite{ablowitz2012nonlinear,kadomtsev1970stability}. Based on the importance of  nonlinear integrable evolution equations,  these are   the focus of scholars' research from the beginning to the end.  In order to solve these equations,  many novel and  effective methods have been produced,   such as   Hirota bilinear method \cite{hirota1980direct},  Darboux  and B\"{a}cklund transformation \cite{matveev1979darboux} and inverse  scattering transform(IST) \cite{Ablowitz1981Solitons,Ablowitz1991Solitons,fokas2012unified,novikov1984theory}.

 However, there is a special kind of equation called nonlocal equation among many nonlinear integrable equations.  As the name suggests, nonlinear  integrable  nonlocal  equation refers to the nonlinear integrable evolution equation with nonlocal nonlinear term, for example, $q(x,t)$  is replacled by $q^{\ast}(-x,t)$, $q(-x,-t)$ or $q^{\ast}(-x,-t)$.  In Ref. \cite{ablowitz2013integrable}, Ablowitz and Musslimani had found  a new class of
nonlocal integrable
NLS hierarchy with the infinite number of
conservation laws by introducing a new symmetry
reduction $r(x,t)=q^{\ast}(-x, t)$.
   According to the different inversion relation,   integrable nonlocal nonlinear equations roughly include the following categories:  real (complex) reverse time nonlocal equation,  real (complex)  reverse space nonlocal equation ans  real(complex) reverse space-time nonlocal equation \cite{ablowitz2017integrable}.
    Recently  there are  several new nonlocal  system have
 been  analyzed, including multidimensional versions of the NLS equation \cite{Fokas-nonlocal,Tian-mmas}, nonlocal reverse-time NLS equations \cite{Ma}, nonlocal mKdV equation \cite{13,Yanzy,Faneg}, nonlocal sine-Gordon equation\cite{14}, nonlocal Davey-Stewartson  equation \cite{15,Zhouzx}, nonlocal Alice-Bob systems \cite{Lou}, nonlocal (2+1)-D breaking solitons hierarchy \cite{Zhangdj} and nonlocal  integrable equations \cite{Yangjk}.

   As for the method of solving  the nonlocal equation, the above mentioned methods are not all effective to it,  and  the most classical and effective method  is  IST. IST associates a compatible pair of linear equations  with the integrable nonlinear equation. One of the equations  is used to determine suitably analytic eigenfunctions and    transform the initial data to appropriate scattering data. The other linear equation is used to complete the evolution  of the scattering data. Based on the linear equations (Lax pair),  one can find the exact solutions of origin objective equations  successfully \cite{Ablowitz1981Solitons}.

    In this work, we consider   IST for the  integrable  nonlocal  Lakshmanan-Porsezian-Daniel (LPD) equation
\begin{equation}\label{LPD}
   q_t(x,t) + \frac{1}{2} i q_{xx}(x,t) -i \gamma  q^2(x,t) q^\ast(-x,t)  -\delta H[q(x,t)] =0,
\end{equation}
where
\begin{equation}
\begin{aligned}
    H[q(x,t)]=& -i q_{xxxx}(x,t) +6i \gamma q^{\ast}(-x,t) q^2(x,t) + 4i  \gamma q(x,t)   q_x(x,t) q^{\ast}_{x}(-x,t)  \\
                & +8i \gamma q^{\ast}(-x,t) q(x,t) q_{xx}(x,t)+2i \gamma  q^2(x,t) q^{\ast}_{xx}(-x,t)   -6 i  q^{\ast2}(-x,t)  q^3(x,t),  \\
\end{aligned}
\end{equation}
which is an NLS
type equation with higher order nonlinear terms, such as   
fourth-order dispersion,  second-order dispersion,  cubic and quintic nonlinearities.
The LPD equation describes the nonlinear effect more clearly in
Refs. \cite{LPD1,LPD2,Hejs-2}. 
      The integrable nonlocal LPD equation whose the  potential functions satisfy $r(x,t)=q^{\ast}(-x,t)$  is studied via Darboux transformation in \cite{liu2016dynamical}.  The authors demonstrated  the integrability of the nonlocal LPD equation, provided its Lax pair, and presented its rational soliton solutions and self-potential function. However, in this work, by using  a  ingenious  method,  we analyze infinite number of conserved quantities and conservation laws for nonlocal LPD equation whose  potential functions satisfy $r(x,t)=\gamma q^{\ast}(-x,t)$, where $\gamma=\pm 1$. Furthermore,  by using  IST  method \cite{ablowitz2016inverse}, we obtain the time-periodic  pure soliton solutions of the integrable  nonlocal LPD equation whose    potential functions satisfy $r(x,t)=-q^{\ast}(-x,t)$.  Generally speaking,  the significance of this work is to improve the previous research on the soliton solutions and some important properties of integrable nonlocal LPD equation reported in \cite{liu2016dynamical}, which is   very helpful for us to better understand and master  this class equations.

This work is organized as follows. In Section 2, the Lax pair and the compatibility condition of nonlocal LPD equation are given. Besides,  some other properties of nonlocal LPD equation are listed in the end  of this section. In  Section 3,  by introducing  a novel method, we derive the global conservation laws and the local conservation laws which establish the integrability of  the objective equation. In Section 4,  the direct scattering problem  of the nonlocal LPD equation is constructed and some other important symmetries of the eigenfunctions and  the scattering data are discussed.  Afterwards,  by using the Left-Right RH method, the inverse scattering problem is established  and the potential function is recovered successfully.  Next, in Section 8, we  discuss the time-periodic pure soliton solutions under the reflectionless case. Moreover, in order to understand the soliton solutions,  we select $J=\overline{J}=1,2,3$ and $4$ as example to show the graphics of the soliton solutions vividly. In this process, we find some interesting phenomenon such as   breather-type solitons, arc solitons,  three soliton and four soliton.    What's more, the influence of parameter $\delta$ on the soliton solutions is considered.  In Section 9,   we consider some special cases of initial conditions and derive the eigenvalues and conserved quantities. Finally, the conclusions and the acknowledgement are given in the last two sections.

\section{Lax representation and compatibility condition}
   We  begin  our  discussion   by considering the following scattering problem
\begin{equation}\label{Lax}
\left\{
       \begin{aligned}
           \phi_x= L \phi, & \quad L= -\zeta J + U,            \\
           \phi_t= M \phi, & \quad  M = \zeta^2 J - \zeta U + \frac{1}{2} V + \delta V_p,
       \end{aligned}
       \right.
\end{equation}
with
\begin{equation}
   \begin{aligned}
     & J=\begin{pmatrix} i & 0 \\ 0 & -i  \end{pmatrix},    &  V=\begin{pmatrix} i q(x,t)r(x,t)  & -i  q_x(x,t)  \\    -i r_x(x,t) &  -i q(x,t) r(x,t)    \end{pmatrix},  \\
     & U=\begin{pmatrix} 0  &  q(x,t) \\  r(x,t) & 0  \end{pmatrix},      &   V_p(x,t)=\begin{pmatrix}    i A_p(x,t)   & B_p(x,t)   \\ -C_p(x,t)    &  -i A_p(x,t)     \end{pmatrix}, \\
    \end{aligned}
\end{equation}
\begin{equation}
\begin{aligned}
   A_p = & -8 \zeta^4 - 4 r(x,t)q(x,t)\zeta^2 - 2 i r(x,t)q_x(x,t)\zeta - 2 i q(x,t) r(x,t) \zeta  \\
        &  - 3 q^2(x,t) r^2(x,t) +q_x(x,t) r_x(x,t) + q(x,t) r(x,t) + r(x,t) q_{xx}(x,t),                       \\
   B_p= &  8 q(x,t) \zeta^3 + 4 i q_x(x,t) \zeta^2 -2 q_{xx}(x,t) \zeta +4r(x,t) q^2(x,t)  -i q_{xxx}(x,t) \\
           &      +6i q(x,t) r(x,t) q_x(x,t),          \\
   C_p= &  -8 r(x,t) \zeta^3 -4i r_x(x,t) \zeta^2 +2 r_{xx}(x,t)\zeta -4 r^2(x,t) q(x,t) \zeta + i r_{xxx}(x,t)  \\
         &  -6i r(x,t) r_x(x,t) q(x,t),
\end{aligned}
\end{equation}
where $\phi(x,t)=(\phi_1(x,t),\phi_2(x,t))^{T}$ is a two-component vector and the potential functions $q(x,t)$ and $r(x,t)$  are complex functions.
Under the symmetry reduction $r(x,t)=\gamma q^\ast(-x,t)(\gamma=\pm 1)$, the zero curve equation $L_t - M_x +[L,M]=0$  leads to the objective equation   \eqref{LPD}.

In what follows, we list  a few important properties of the above equation:
\begin{itemize}
  \item Time-reverse symmetry: if $q(x,t)$ is a solution, then $q^{\ast}(x,-t)$ is also a solution.
  \item Space-reverse symmetry: if $q(x,t)$ is a solution, then $q(-x, t)$ is also a solution.
  \item Gauge invariance: if $q(x,t)$ is a solution, then $e^{ i \rho_0}q(x, t)$ is also a solution with  real and constant $\rho_0$.
  \item Spatial translation invariance: if $q(x,t)$ is a solution, then $q(x+ix_0, t)$ is also a solution for any real and constant $x_0$.
  \item PT-symmetry: if $q(x,t)$ is a solution, then $q^{\ast}(-x,-t)$ is also a solution.  It is noted that $V(x,t)=\gamma q(x,t) q^{\ast}(-x,t)$ satisfy the special symmetry $V(x,t)=V^{\ast}(-x,t)$, which is referred to as a self-induced potential in the classical optics.
\end{itemize}

\section{Infinite number of conserved quantities and conservation laws }
As we all know,  a finite number of  local conservation laws  and global conservation laws of the nonlinear  integrable equation is helpful to establish  its integrability as an infinite dimensional Hamilton dynamic system.  In this section,  we will explain  how to obtain the local and global conservation laws of Eq. \eqref{LPD}.

\subsection{Global  conservation laws}
   The infinite number of conserved quantities of Eq. \eqref{LPD}   is  derived  as follows. Before our operation,  we suppose that the potential function $q(x,t)$  decays rapidly at infinity.  At the same time, the solutions of the scattering transform can be obtained by defining the four  functions which  satisfy the following boundary condition
\begin{equation} \label{boun}
   \begin{aligned}
     & \lim_{x \to  -\infty }  \phi(x,t)= \begin{pmatrix} 1 \\ 0 \end{pmatrix} e^{-i\zeta x}, \qquad  & & \lim_{x \to  -\infty }  \overline{\phi}(x,t)= \begin{pmatrix} 0 \\ 1 \end{pmatrix} e^{i\zeta x},\\
     &  \lim_{x \to  +\infty }  \psi(x,t)= \begin{pmatrix} 0 \\ 1 \end{pmatrix} e^{i\zeta x}, \qquad  & & \lim_{x \to  +\infty }  \overline{\psi}(x,t)= \begin{pmatrix} 1 \\ 0 \end{pmatrix} e^{-i\zeta x},
   \end{aligned}
\end{equation}
where $ \overline{\phi}(x,t)$  is not the  complex conjugate of  $\phi (x,t)$, and  $\phi^\ast(x,t)$ denotes the complex conjugate of  $\phi (x,t)$.
 If  $\phi(x,t)= \left( \phi_1(x,t), \phi_2(x,t)    \right)^{T}$ is the solution  to  Eq. \eqref{Lax} which satisfies  the above boundary conditions, we can obtain that $\phi_1(x,t)e^{i \zeta x}$ is analytic for $ \mathrm{Im} (\zeta) \geq 0$ and approaches to $1$ as $x \to  \pm \infty$.
 Substituting $\phi_1(x,t)=\exp \left(-i\zeta x +\varphi(x,t)  \right)$ into Eq. \eqref{Lax}, we can find that the function $\mu(x,t)=\varphi_x(x,t)$ satisfies the Riccati equation
 \begin{equation}\label{Riccati}
    q \frac{\partial}{\partial x}  \left( \frac{\mu}{q} \right) + \mu^2 - qr - 2 i \zeta \mu = 0.
 \end{equation}
For $  \mathrm{Im} (\zeta)  > 0$,  we have  $\lim_{|\zeta| \to \infty} \varphi(x,\zeta)=0$.  Substituting the expansion $\mu(x,\zeta)= \sum_{n=0}^{\infty} \frac{\mu_n(x,t)}{(2i \zeta)^{n+1}}$  into Eq. \eqref{Riccati} and equating the powers of $\zeta$,   we find
\begin{equation}\label{conser1}
   \begin{aligned}
       \mu_0(x,t)= & -  q(x,t) r(x,t) =- \gamma q(x,t) q^{\ast}(-x,t), \\
       \mu_1(x,t)= & -  q(x,t) r_x(x,t) =  \gamma q(x,t) q_x^{\ast}(-x,t),
   \end{aligned}
\end{equation}
and
\begin{equation}\label{conser2}
 \mu_{n+1} =  q(x,t) \frac{\partial}{\partial x}  \left( \frac{\mu_n}{q} \right) + \sum_{m=0}^{n-1} \mu_{m} \mu_{n-m-1}, \qquad  n \geq 1.
\end{equation}
From the boundary conditions   it follows
\begin{equation}
   \lim_{x \to -\infty} \phi_1(x,\zeta)e^{i\zeta x}=1,  \qquad  \lim_{x \to -\infty} \varphi(x,\zeta)=0,
\end{equation}
then we can obtain
\begin{equation}\label{conser3}
  \ln a(\zeta)= \ln(\phi_1(x,t)e^{i\zeta x}) = \sum_{n=0}^{\infty} \frac{\mathcal{C}_{n}}{(2i \zeta)^{n+1}},   \qquad \mathcal{C}_{n} = \int_{-\infty}^{+\infty} \mu_{n}(x,t) dx.
\end{equation}
Since $\phi_1(x,t)e^{i\zeta x}$ is time independent for $k$ with $ \mathrm{Im} \zeta>0$,   then the above $\mathcal{C}_n$ is also time independent for $\zeta$ with $ \mathrm{Im} \zeta>0$. According to Eqs. \eqref{conser1}, \eqref{conser2} and \eqref{conser3},   we can obtain  all conserved quantities. More explicitly, the first few  conserved quantities are listed as follows:
\begin{equation}
   \begin{aligned}
       \mathcal{C}_0 &= - \gamma \int_{-\infty}^{+\infty} q(x,t) q^{\ast}(-x,t) dx,  \qquad  \mathcal{C}_1 =   \gamma  \int_{-\infty}^{+\infty}     q(x,t) q_x^{\ast}(-x,t)    dx, \\
       \mathcal{C}_2 &= - \gamma \int_{-\infty}^{+\infty} \left(   q(x,t) q_{xx}^{\ast}(-x,t)- \gamma q^2(x,t) q^{\ast2}(-x,t) \right) dx,  \\
       \mathcal{C}_3 & = \gamma \int_{-\infty}^{+\infty}  \left( q(x,t)q^{\ast}(-x,t) + \gamma  q(x,t) q_{x}(x,t) q^{\ast2}(-x,t) -4 q^{2}(x,t) q^{\ast}(-x,t) q^{\ast}_{x}(-x,t) \right) dx, \\
        \mathcal{C}_4 & = \gamma  \int_{-\infty}^{+\infty}  \left( -q(x,t)q^{\ast}_{xxxx}(-x,t) +5 \gamma q^{2}(x,t) q^{\ast2}_{x}(-x,t) + 6 \gamma q^{2}(x,t)q^{\ast}(-x,t) q^{\ast}_{xx}(-x,t) \right. \\
         & \left. + \gamma  q(x,t) q_{xx}(x,t) q^{\ast2}(-x,t) -6 \gamma q(x,t) q^{\ast}(-x,t) q_{x}(x,t) q^{\ast}_{x}(-x,t) -2 q^{3}(x,t) q^{\ast3}(-x,t)     \right) dx.
   \end{aligned}
\end{equation}

\subsection{Local  conservation  laws}
In order to obtain the local conservation laws, we  consider the time-dependent problem
\begin{equation}
    \phi_{1t}= A \phi_1 + B \phi_2,
\end{equation}
where $A,B$ denote the $(1,1)-$ and $(1,2)-$entry of $M$ in Eq. \eqref{Lax}, respectively. According to the expression of $\mu$ and $\phi$, we find
\begin{equation}\label{conser4}
   \partial_t \mu(x,t) = \partial_x  \left( A_{nonloc} + \frac{B_{nonloc} \mu(x,t)}{q(x,t)} \right),
\end{equation}
where
\begin{equation}\label{para}
   \begin{aligned}
      A_{nonloc} = & -8i \delta  \zeta^4 +i \zeta^2 \left(  1- 4 \delta \gamma q q^\ast(-x,t)  \right)   +2 \delta \gamma \zeta \left(  q_x q^\ast(-x,t) -q q_x^\ast(-x,t)   \right)           \\
        &    +i \left( \frac{\gamma}{2} q q^\ast(-x,t) -3 \delta  q^2 q^{\ast2}(-x,t) -\delta \gamma q^\ast_x(-x,t) q_x(x,t) +\delta \gamma q q^\ast_{xx}(-x,t)   \right),                     \\
      B_{nonloc} =& 8 \delta q \zeta^3 + 4i \delta  q_x  \zeta^2 + \left( 4\delta \gamma q^2 q^\ast(-x,t) - q(x,t) - 2 q_{xx}   \right) \zeta     \\
                  &  + i \left(  6\gamma q q_x q^\ast(-x,t) + \frac{1}{2} q_x  -q_{xxx}   \right).           \\
   \end{aligned}
\end{equation}
Substituting  Eq.  \eqref{para} and the expansion of $\mu$    into Eq. \eqref{conser4},   we  have
\begin{equation}
  \partial_t \left( \sum_{n=0}^{\infty} \frac{\mu_n(x,t)}{(2i \zeta)^{n+1}}   \right) = \partial_x  \left( A_{nonloc} + \frac{B_{nonloc}}{q(x,t)} \left( \sum_{n=0}^{\infty} \frac{\mu_n(x,t)}{(2i \zeta)^{n+1}} \right) \right),
\end{equation}
from which we  obtain
\begin{equation}\label{local}
  \partial_{t} (\mu_n)  = i \partial_x \left( \mu_n S_1 + \frac{1}{2} \mu_{n+1} S_2 -  \delta \frac{q_x}{q} \mu_{n+2} +  \delta \mu_{n+3}  \right), \quad n=0,1,2,3,\dots,
\end{equation}
where
\begin{equation}
      S_1=    6  \gamma   q_x q^\ast(-x,t) + \frac{  q_x }{2 q} -  \frac{q_{xxx}}{q},   \quad S_2 =     1 + 2   \frac{q_{xx}}{q} - 4\delta \gamma q q^\ast(-x,t).
\end{equation}
We can write the conservation laws   \eqref{local} as the form
\begin{equation}
  \frac{\partial{\mathcal{T}}}{\partial{t}} = - i \frac{\partial{\mathcal{X}}}{\partial{x}},
\end{equation}
where $\mathcal{T}=\mu_n$ and $\mathcal{X}=-\mu_n S_1 - \frac{1}{2} \mu_{n+1} S_2 + \delta \frac{q_x}{q} \mu_{n+2} - \delta \mu_{n+3}$   $(n=0,1,2,3,\dots)$ are the so-called  densities and fluxes,   respectively.  The first three local conservation laws are
\begin{equation}
\begin{aligned}
     \mathcal{T} =\mu_0, \quad   \mathcal{X} = - S_1 \mu_0 - \frac{1}{2} S_2 \mu_1 + \delta \frac{q_x}{q} \mu_2  -\delta \mu_3,  \\
   \mathcal{T} = \mu_1, \quad  \mathcal{X} = - S_1 \mu_1 - \frac{1}{2} S_2 \mu_2 + \delta \frac{q_x}{q} \mu_3  -\delta \mu_4,  \\
\mathcal{T} = \mu_2, \quad  \mathcal{X} = - S_1 \mu_2 - \frac{1}{2} S_2 \mu_3 + \delta \frac{q_x}{q} \mu_4  -\delta \mu_5,
  \end{aligned}
\end{equation}
where
\begin{equation}
     \begin{aligned}
         \mu_0 = &- \gamma q(x,t) q^{\ast}(-x,t),  \qquad  \mu_1 =  \gamma q(x,t) q_x^{\ast}(-x,t),  \\
         \mu_2= & -\gamma q(x,t) q^{\ast}_{xx}(-x,t) +q^{2}(x,t) q^{\ast2}(-x,t),  \\
         \mu_3= & \gamma q(x,t) q^{\ast}_{xxx}(-x,t) +q(x,t)q_{x}(x,t)q^{\ast2}(-x,t)-4 q^2(x,t) q^{\ast}(-x,t) q^{\ast}_{x}(-x,t), \\
         \mu_4 =&  -\gamma q(x,t) q^{\ast}_{xxxx}(-x,t) + q(x,t) q^{\ast2}(-x,t) q_{xx}(x,t) -6 q(x,t) q_{x}(x,t) q^{\ast}(-x,t) q^{\ast}_{x}(-x,t)\\
            & +5 q^2(x,t) q^{\ast2}_{x}(-x,t) +6q^2(x,t) q^{\ast}(-x,t) q^{\ast}_{xx}(-x,t) -2\gamma q^3(x,t) q^{\ast3}(-x,t), \\
     \mu_5 =&  \gamma q^2 q^{\ast}_{xxxxx}(-x,t)  +q^2 q_{xxx} q^{\ast2}(-x,t) - 6 q^3 q^{\ast}(-x,t) q^{\ast}_{xxx}(-x,t)-2 \gamma q^2 q_x  q^{\ast3}(-x,t)    \\
     & - 2q^2 q^{\ast}(-x,t) q^{\ast}_{xxx}(-x,t) - 8 q^2 q_{xx} q^{\ast}(-x,t) q^{\ast}_{x}(-x,t) + 12 q^2 q_x q^{\ast}(-x,t) q^{\ast}_{xx}(-x,t) \\
       & - 2 q^2 q^{\ast}_{x}(-x,t) q^{\ast}_{xx}(-x,t) -16 q^3 q^{\ast}_{x}(-x,t) q^{\ast}_{xx}(-x,t)  + 11 q^2 q_x q^{\ast2}_{x}(-x,t)  \\
       & - 4 \gamma q^3 q_x q^{\ast3}(-x,t) + 6 \gamma q^4 q^{\ast}_{x}(-x,t) q^{\ast2}(-x,t)  + 10 \gamma q^3 q^{\ast2}(-x,t) q^{\ast}_{x}(-x,t).
     \end{aligned}
\end{equation}

\section{Direct scattering problem}
In the following sections, we  will consider the scattering problem of the system of Eq.  \eqref{Lax}. For the convenience of  the  discussion, we define the following Jost functions
\begin{equation}\label{eigen}
   \begin{aligned}
  & M(x,\zeta)= e^{i \zeta x} \phi(x,\zeta),  &  \quad   \overline{M}(x,\zeta)= e^{- i \zeta x} \overline{\phi}(x,\zeta),              \\
  & N(x,\zeta)= e^{- i \zeta x}  \psi(x,\zeta),&  \quad   \overline{N}(x,\zeta)= e^{  i \zeta x}  \overline{\psi}(x,\zeta),
   \end{aligned}
\end{equation}
which satisfy the constant boundary condition induced from Eq. \eqref{boun}.  Furthermore, we can see that the above functions satisfy a linear integral equations and show that  $M(x,\zeta)$, $N(x,\zeta)$ are analytic in the upper half complex $\zeta$ plane  whereas  $\overline{M}(x,\zeta)$, $\overline{N}(x,\zeta)$ are analytic in the lower half complex $\zeta$ plane \cite{ablowitz2004discrete}.  Moreover, the large $\zeta$ behavior of the Jost functions are  given by \cite{ablowitz2004discrete}
\begin{equation}\label{large}
  \begin{aligned}
     M(x,\zeta)= & \begin{pmatrix} 1-\frac{1}{2i\zeta} \int_{-\infty}^{x} r(z)q(z)dz  \\ -\frac{r(x)}{2i \zeta}  \end{pmatrix} + O\left(  \zeta^{-2} \right),\\
   N(x,\zeta)= & \begin{pmatrix}  \frac{q(x)}{2i\zeta}  \\  1-\frac{1}{2i\zeta} \int_{x}^{+\infty} r(z)q(z)dz   \end{pmatrix} + O\left(  \zeta^{-2} \right),\\
 \overline{M}(x,\zeta)= & \begin{pmatrix}  \frac{q(x)}{2i\zeta}  \\  1-\frac{1}{2i\zeta} \int_{-\infty}^{x} r(z)q(z)dz   \end{pmatrix} + O\left(  \zeta^{-2} \right),\\
 \overline{N}(x,\zeta)= & \begin{pmatrix} 1 + \frac{1}{2i\zeta} \int_{x}^{+ \infty} r(z)q(z)dz  \\ -\frac{r(x)}{2i \zeta}  \end{pmatrix} + O\left(  \zeta^{-2} \right).
  \end{aligned}
\end{equation}

From the boundary condition  \eqref{boun},  it is seen that the solutions  $\phi(x,\zeta)$ and $\overline{\phi}(x,\zeta)$ of the scattering problem   \eqref{Lax} are linearly dependent. Similarly, the solutions  $\psi(x,\zeta)$ and $\overline{\psi}(x,\zeta)$ of the scattering problem  \eqref{Lax}  are linearly dependent.  Since the scattering problem  \eqref{Lax} is  a second order linearly ordinary differential equation,  $\{ \phi,  \overline{\phi} \}$ and  $\{ \psi,  \overline{\psi}\}$ are linearly dependent. Moreover, we can express the relation of  them  as follows
  \begin{equation}\label{left}
    \begin{aligned}
    \phi(x,\zeta)=a(\zeta) \overline{\psi}(x,\zeta) + b(\zeta) \psi(x,\zeta),  \\
     \overline{\phi}(x,\zeta)=\overline{a}(\zeta)  \psi(x,\zeta) + \overline{b}(\zeta) \overline{\psi}(x,\zeta),
    \end{aligned}
  \end{equation}
where $a(\zeta)$, $\overline{a}(\zeta)$, $b(\zeta)$ and  $\overline{b}(\zeta)$ are the scattering data,  then we can have
\begin{equation}\label{wro}
    \begin{aligned}
       a(\zeta) = W  \left( \phi(x,\zeta), \psi(x,\zeta)  \right),  \qquad & \overline{a}(\zeta)= W \left(  \overline{\psi}(x,\zeta),  \overline{\phi}(x,\zeta)     \right), \\
       b(\zeta) =  W  \left(        \overline{\psi}(x,\zeta),   \phi(x,\zeta)        \right),  \qquad  & \overline{b}(\zeta)= W \left(      \overline{\phi}(x,\zeta), \psi(x,\zeta)     \right),
    \end{aligned}
\end{equation}
where $W(u,v)=u_1 v_2 -u_2 v_1$ represents  the  Wronskian determinant.  Furthermore, it can be seen that $a(\zeta)$ and $\overline{a}(\zeta)$ are analytic in the upper half complex plane and the lower half complex plane, respectively,  while  $b(\zeta)$  and $\overline{b}(\zeta)$ cannot be  extend off the real $\zeta$ axis. In addition, the scattering data satisfy  the relation $a(\zeta)\overline{a}(\zeta)-b(\zeta)\overline{b}(\zeta)=1$  for $\mathrm{Im} \zeta =0$.

\section{Symmetry reduction  $r(x,t)=\gamma q^{\ast}(-x,t)$: eigenfunctions and the scattering data}
\subsection{Symmetry of the eigenfunctions}
Next, we make efforts to establish the symmetry properties of the eigenfunctions under the symmetry reduction $r(x,t)=\gamma q^{\ast}(-x,t)$, $\gamma=\pm 1$.  We suppose that $v(x,\zeta)=\left(    v_{1}(x,t),  v_2(x,t)         \right)^{T}$ is the solution of Eq. \eqref{Lax}, then
$\left( v_2^{\ast}(-x,-k^{\ast}),   -\gamma v_{1}^{\ast}(-x,-k^{\ast})    \right)^{T}$ is also the solution of Eq. \eqref{Lax}.  Since  the solutions of the scattering problem are   uniquely determined by  the boundary condition \eqref{boun}, we have the following important symmetry
\begin{equation}\label{eigenf}
 \begin{aligned}
  \psi(x,\zeta)= \begin{pmatrix} 0 & -\gamma \\ 1 &  0      \end{pmatrix} \phi^{\ast}(-x,-k^\ast), \\   \overline{\psi}(x,\zeta)= \begin{pmatrix} 0 &
  1 \\  -\gamma  &  0      \end{pmatrix} \overline{\phi}^{\ast}(-x,-k^\ast).
  \end{aligned}
 \end{equation}
From Eq. \eqref{eigen}, we obtain the symmetry of the Jost functions
\begin{equation}\label{symeig}
\begin{aligned}
      N(x,\zeta)= \begin{pmatrix} 0 & -\gamma \\ 1 &  0      \end{pmatrix} M^{\ast}(-x,-k^\ast), \\
         \overline{N}(x,\zeta)= \begin{pmatrix} 0 & 1 \\  -\gamma  &  0      \end{pmatrix} \overline{M}^{\ast}(-x,-k^\ast).
\end{aligned}
\end{equation}

\subsection{Symmetry of the  scattering data}
  According to  the symmetry of the eigenfunctions \eqref{eigenf} and  the Wronskian representations of the scattering data \eqref{wro}, we get
\begin{equation}
  a(\zeta)=a^{\ast}(-\zeta^{\ast}), \quad   \overline{a}(\zeta)=\overline{a}^{\ast}(-\zeta^{\ast}), \quad   \overline{b}(\zeta)=\gamma b^{\ast}(-\zeta^{\ast}),
\end{equation}
which means that if $\zeta_{k}=\xi_{k}+i \eta_{k}$ is a zero of $a(\zeta)$ in the upper half complex plane, then $-\zeta_{k}^{\ast}=-\xi_{k}+i \eta_{k}$ is also a zero of $a(\zeta)$ in the upper half complex plane. Similarly, $\overline{\zeta}_{k}$ is a zero of $\overline{a}(\zeta)$ in the lower half complex plane, then $-\overline{\zeta}_{k}^{\ast}$ is also a zero of $\overline{a}(\zeta)$ in the lower half complex plane.

\section{Inverse scattering problem: Left-Right RH approach}
\subsection{Left scattering problem}
 At first, we  recall   Eq. \eqref{left}
  \begin{equation}
    \begin{aligned}
    \phi(x,\zeta)=a(\zeta) \overline{\psi}(x,\zeta) + b(\zeta) \psi(x,\zeta),  \\
     \overline{\phi}(x,\zeta)=\overline{a}(\zeta)  \psi(x,\zeta) + \overline{b}(\zeta) \overline{\psi}(x,\zeta),
    \end{aligned}
  \end{equation}
the above system  can be  rewritten as the following  matrix  form
\begin{equation}
   \Phi(x,\zeta)= S_{L} \Psi(x,\zeta),
\end{equation}
where $\Phi(x,\zeta)=\left(  \phi(x,\zeta), \overline{\phi}(x,\zeta)   \right)^{T}$, $\Psi(x,\zeta)=\left(\overline{\psi}(x,\zeta), \psi(x,\zeta)    \right)^{T}$ and $S_{L}(\zeta)$ is the left scattering matrix
\begin{equation}
  S_{L}(\zeta)= \begin{pmatrix}   a(\zeta)  &    b(\zeta)   \\   \overline{b}(\zeta)  &  \overline{a}(\zeta)     \end{pmatrix}.
\end{equation}
Following the results reported in \cite{ablowitz2016inverse},  we can formulate the corresponding RH problem on the left and  obtain  the following linear integral equations  which
represent the functions $N(x,\zeta)$ and $\overline{N}(x,\zeta)$:
\begin{equation}\label{so1}
   \begin{aligned}
     N(x,\zeta)= & \begin{pmatrix}  0 \\  1 \end{pmatrix} +\sum_{l=1}^{\overline{J}} \frac{\overline{C}_{l} \overline{N}(x,\overline{\zeta}_{l}) e^{-2i \overline{\zeta}_{l}x}}{\zeta-\overline{\zeta}_{l}} - \frac{1}{2\pi i} \int_{-\infty}^{\infty} \frac{\overline{\rho}(\xi) e^{ - 2i \xi x} \overline{N}(x,\xi)}{\xi -(\zeta + i 0)} d\xi,  \\
     \overline{N}(x,\zeta)=& \begin{pmatrix}  1 \\  0  \end{pmatrix} +\sum_{l=1}^{J} \frac{C_{l} N(x,\zeta_{l}) e^{2i \zeta_{l}x}}{\zeta-\zeta_{l}} - \frac{1}{2\pi i} \int_{-\infty}^{\infty} \frac{\rho(\xi) e^{  2i \xi x}  N(x,\xi)}{\xi -(\zeta - i 0)} d\xi,
   \end{aligned}
\end{equation}
where $\rho(\zeta)$ and $\overline{\rho}(\zeta)$ are the left reflection  coefficients defined by
\begin{equation}\label{leftreflec}
   \rho(\zeta)=\frac{b(\zeta)}{a(\zeta)}, \qquad   \overline{\rho}(\zeta)=\frac{\overline{b}(\zeta)}{\overline{a}(\zeta)},
\end{equation}
and $C_{l}$ and $\overline{C}_{l}$ are the left norming constants defined by
\begin{equation}
  C_{l}= \frac{b(\zeta_{l})}{ a'(\zeta_l)}, \qquad  \overline{C}_{l}= \frac{\overline{b}(\overline{\zeta}_{l})}{ \overline{a}'(\overline{\zeta}_l)},
\end{equation}
where $a'(\zeta_l)$ and $\overline{a}'(\overline{\zeta}_l)$ denote the derivative  at $\zeta_l$  and $\overline{\zeta}_l$, respectively.

\subsection{Time evolution of the scattering data: Left scattering problem}
According to   Eq. \eqref{Lax}, we  derive the time evolution of the scattering data
\begin{equation}\label{timeevolu1}
 \begin{aligned}
  a(\zeta,t)=a(\zeta,0), & \quad   b(\zeta,t)= e^{(16i\delta \zeta^4 - 2i\zeta^2)t} b(\zeta,0),\\
    \overline{a}(\zeta,t)=\overline{a}(\zeta,0), & \quad  \overline{b}(\zeta,t)= e^{(-16i\delta \zeta^4 + 2i\zeta^2)t} \overline{b}(\zeta,0).
  \end{aligned}
\end{equation}
In what follows, we obtain the time evolution of the left reflection coefficients $\rho(\zeta)$  and $\overline{\rho}(\zeta) $ and the left norming constants  $C_{l}$ and $\overline{C}_{l}$ according to Eqs. \eqref{leftreflec} and  \eqref{timeevolu1}
\begin{equation}
 \begin{aligned}
   C_{l} =& C_{l}(0) e^{(16i\delta \zeta^4 - 2i\zeta^2)t}, & \quad \rho(\zeta,t)=&e^{(16i\delta \zeta^4 - 2i\zeta^2)t} b(\zeta,0)/a(\zeta,0), \\
   \overline{C}_{l} = & \overline{C}_{l}(0) e^{(-16i\delta \overline{\zeta}^4 + 2i\overline{\zeta}^2)t}, & \quad \overline{\rho}(\zeta,t)=& e^{(-16i\delta \zeta^4 + 2i\zeta^2)t}\overline{b}(\zeta,0)/\overline{a}(\zeta,0).
  \end{aligned}
\end{equation}

\subsection{Right scattering problem}
 Next, we consider the following  system
\begin{equation}
 \begin{aligned}
    \psi(x,\zeta) = & \alpha(\zeta) \overline{\phi}(x,\zeta)+ \beta(\zeta) \phi(x,\zeta), \\
    \overline{\psi}(x,\zeta) =  & \overline{\alpha}(\zeta) \phi(x,\zeta) + \overline{\beta}(\zeta) \overline{\phi}(x,\zeta),
 \end{aligned}
\end{equation}
where $\alpha(\zeta)$, $\overline{\alpha}(\zeta)$, $\beta(\zeta)$ and $\overline{\beta}(\zeta)$ are the right scattering data. Similarly, we can rewrite the above system as the matrix form
\begin{equation}
  \Psi(x,\zeta)= S_{R} \Phi(x,\zeta),
\end{equation}
where $\Psi(x,\zeta)=\left(  \psi(x,\zeta),   \overline{\psi}(x,\zeta)  \right)^{T}$,  $\Phi(x,\zeta)=\left(     \overline{\phi}(x,\zeta),  \phi(x,\zeta) \right)^{T}$  and $S_R$ is the right scattering matrix
\begin{equation}
   S_{R}= \begin{pmatrix}    \alpha(\zeta)  &  \beta(\zeta)   \\ \overline{\beta}(\zeta) &  \overline{\alpha}(\zeta)    \end{pmatrix}.
\end{equation}
We can formulate the corresponding RH problem on the  right and  obtain  the following linear integral equations  which
govern the functions $M(x,\zeta)$ and $\overline{M}(x,\zeta)$:
\begin{equation}
  \begin{aligned}
     M(x,\zeta)= & \begin{pmatrix} 1 \\ 0  \end{pmatrix} + \sum_{l=1}^{\overline{J}} \frac{\overline{B}_{l}\overline{M}(x,\overline{\zeta}_{l})e^{2i\overline{\zeta}_{l}x}}{\zeta-\overline{\zeta}_{l}} -\frac{1}{2\pi i} \int_{-\infty}^{\infty} \frac{\overline{R}(\xi)e^{2i\xi x} \overline{M}(x,\xi) }{\xi-(\zeta +i0)} d\xi,   \\
       \overline{M}(x,\zeta)=& \begin{pmatrix} 0 \\ 1  \end{pmatrix} + \sum_{l=1}^{J} \frac{B_{l} M(x,\overline{\zeta}_{l})e^{-2i \zeta_{l}x}}{\zeta - \zeta_{l}} + \frac{1}{2\pi i} \int_{-\infty}^{\infty} \frac{R(\xi)e^{-2i\xi x} M(x,\xi) }{\xi-(\zeta - i0)} d\xi,
      \end{aligned}
\end{equation}
where $R(\zeta)$ and $\overline{R}(\zeta)$ are the right reflection coefficients given by
\begin{equation}\label{s1}
  R(\zeta)= \frac{\beta(\zeta)}{\alpha(\zeta)}, \quad  \overline{R}(\zeta)=\frac{\overline{\beta}(\zeta)}{\overline{\alpha}(\zeta)},
\end{equation}
and $B_{l}$  and $\overline{B}_{l}$ are the right norming constants defined by
\begin{equation}\label{s2}
  B_{l}=\frac{\beta(\zeta_{l})}{\alpha^{'}(\zeta_{l})},  \quad  \overline{B}_{l}=\frac{\overline{\beta}(\overline{\zeta}_{l})}{\overline{\alpha}^{'}(\overline{\zeta}_{l})}.
\end{equation}

\subsection{Time evolution of the scattering data: Right scattering problem}
Similar to the left case, we obtain the time evolution of the right scattering data
\begin{equation}\label{timeevolu2}
 \begin{aligned}
  \alpha(\zeta,t)=\alpha(\zeta,0), & \quad   \beta(\zeta,t)= e^{(16i\delta \zeta^4 - 2i\zeta^2)t} \beta(\zeta,0),\\
  \overline{\alpha}(\zeta,t)=\overline{\alpha}(\zeta,0), & \quad  \overline{\beta}(\zeta,t)= e^{(-16i\delta \zeta^4 + 2i\zeta^2)t} \overline{\beta}(\zeta,0).
  \end{aligned}
\end{equation}
According to Eqs. \eqref{timeevolu2}, \eqref{s1} and \eqref{s2}, the time evolution of the right  reflection coefficients and  norming constants can be obtained by
\begin{equation}
 \begin{aligned}
   B_{l} =& B_{l}(0) e^{(16i\delta \zeta^4 - 2i\zeta^2)t}, & \quad R(\zeta,t)=&e^{(16i\delta \zeta^4 - 2i\zeta^2)t} \beta(\zeta,0)/\alpha(\zeta,0), \\
   \overline{B}_{l} = & \overline{B}_{l}(0) e^{(-16i\delta \overline{\zeta}^4 - 2i\overline{\zeta}^2)t}, & \quad \overline{R}(\zeta,t)=& e^{(-16i\delta \zeta^4 + 2i\zeta^2)t}\overline{\beta}(\zeta,0)/\overline{\alpha}(\zeta,0).
  \end{aligned}
\end{equation}

\subsection{Relationship between the reflection coefficients}
According to the matrix forms of the left and the right scattering problem, we have the relationship between the left and the right scattering matrix $S_{R}=S_{L}^{-1}$, more explicitly,
\begin{equation}
   \begin{aligned}
       a(\zeta)=\alpha(\zeta), \quad &  \overline{a}(\zeta)=\overline{\alpha}(\zeta),   \\
       \overline{\beta}(\zeta)= -b(\zeta), \quad &    \beta(\zeta)= -\overline{b}(\zeta).
   \end{aligned}
\end{equation}
Furthermore,  we have
\begin{equation}
   \begin{aligned}
    R(\zeta)= \frac{\beta(\zeta)}{\alpha(\zeta)}=  -\frac{\overline{b}(\zeta)}{a(\zeta)} = -\gamma \frac{ b^{\ast}(-\zeta^{\ast})}{a^{\ast}(-\zeta^{\ast})} = - \gamma \rho^{\ast}(-\zeta^{\ast}),   \\
    \overline{R}(\zeta) = \frac{\overline{\beta}(\zeta)}{\overline{\alpha}(\zeta)} = - \frac{b(\zeta)}{ \overline{a}(\zeta)} = - \gamma \frac{\overline{b}^{\ast}(-\zeta^\ast)}{\overline{a}^{\ast}(-\zeta^{\ast})}= -\gamma \overline{\rho}^{\ast}(-\zeta^{\ast}).
   \end{aligned}
\end{equation}

\subsection{Additional symmetry  between the eigenfunctions}
Suppose that $\zeta_l$ is the eigenvalue of $a(\zeta)$ in the upper complex plane, i.e., $a(k_l)=0$, the eigenfunction $\phi(x,\zeta)$ and $\psi(x,zeta)$ are linear dependent, $\phi(x,\zeta_l)=b(\zeta_l) \psi(x,\zeta_l)$. Moreover,
\begin{equation}
\begin{aligned}
   M_1(x,\zeta_{l})= b(\zeta_l) N_1(x,\zeta_l) e^{2ik_l x}, \\
    M_2(x,\zeta_{l})= b(\zeta_l) N_2(x,\zeta_l) e^{2ik_l x},
\end{aligned}
\end{equation}
then
\begin{equation}
  M_1(x,\zeta_{l})  N_2(x,\zeta_l) =  M_2(x,\zeta_{l})  N_1(x,\zeta_l).
\end{equation}
With the aid of Eq. \eqref{symeig}, we obtain
\begin{equation}
  N_2^{\ast}(-x,\zeta_{l})  N_2(x,\zeta_l) =  N_1^{\ast}(-x,\zeta_{l})  N_1(x,\zeta_l).
\end{equation}
Similarly, the  other important  conclusion is given as follows
\begin{equation}
  \overline{M}_2^{\ast}(-x,\overline{\zeta}_{l})  \overline{M}_2(x,\overline{\zeta}_l) =  \overline{M}_1^{\ast}(-x,\overline{\zeta}_{l})  \overline{M}_1(x,\overline{\zeta}_l).
\end{equation}

\section{Recovery of the potentials}
Based on the above results, we can recover the potential functions $q(x,t)$ and $r(x,t)$ successfully.  At first,  recall from Eq. \eqref{so1} that
\begin{equation}
    \overline{N}(x,\zeta)=  \begin{pmatrix}  1 \\  0  \end{pmatrix} +\sum_{l=1}^{J} \frac{C_{l} N(x,\zeta_{l}) e^{2i \zeta_{l}x}}{\zeta-\zeta_{l}} - \frac{1}{2\pi i} \int_{-\infty}^{\infty} \frac{\rho(\xi) e^{  2i \xi x}  N(x,\xi)}{\xi -(\zeta - i 0)} d\xi.
\end{equation}
The large $k$ behavior of  $\overline{N}_2(x,\zeta)$ is determined by
\begin{equation}\label{large2}
  \overline{N}_2(x,\zeta)  \sim \frac{1}{\zeta} \sum_{l=1}^{J} C_{l} N_2(x,\zeta_l) e^{2i \zeta_l x} -   \frac{1}{2\pi i \zeta} \int_{-\infty}^{\infty} \rho(\xi) e^{2i \xi x} N_2(x,\xi) d\xi.
\end{equation}
According to Eq. \eqref{large}, we obtain
\begin{equation}
   \overline{N}_2(x,\zeta) \sim -\frac{r(x)}{2i \zeta},
\end{equation}
thus,  we can recover the potential  $r(x)$  by
\begin{equation}\label{rp}
  r(x) \sim - 2i \left(     \sum_{l=1}^{J} C_{l} N_2(x,\zeta_l) e^{2i \zeta_l x} -   \frac{1}{2\pi i} \int_{-\infty}^{\infty} \rho(\xi) e^{2i \xi x} N_2(x,\xi) d\xi   \right).
\end{equation}
With the asymptotic relation   \eqref{large}
\begin{equation}
  \overline{M}_1(x,\zeta) \sim \frac{q(x)}{2i\zeta},
\end{equation}
and the symmetry relation $\overline{M}_1(x,\zeta)= -\gamma  \overline{N}_2^{\ast}(-x,-\zeta^{\ast})$,    we obtain the following asymptotic relation of $q(x)$:
\begin{equation}
 q(x) \sim  - 2i \gamma \zeta \overline{N}_2^{\ast}(-x,-\zeta^{\ast}), \quad \gamma = \pm 1.
\end{equation}
From Eq. \eqref{large2},  we obtain
\begin{equation}\label{potential}
  q(x) =  2i \gamma \sum_{l=1}^{J}C_{l}^{\ast}N_2^{\ast}(-x,\zeta_l) e^{2i \zeta_l^{\ast} x} + \frac{\gamma}{\pi} \int_{-\infty}^{\infty} \rho^{\ast}(\xi) e^{2i\xi x}N_2^{\ast}(-x,\xi) d\xi.
\end{equation}
According to Eqs. \eqref{rp} and \eqref{potential}, it can be seen that the symmetry relation $r(x)=\gamma q^{\ast}(-x)$ still  holds.

\section{Soliton solutions}
 In this section,  we mainly discuss the pure soliton solutions of  nonlocal integrable LPD equation. It is noted that pure soliton solutions  arise when the reflection coefficients $\rho(\zeta)$ and $\overline{\rho}(\zeta)$ vanish.  Besides, it can be proved that  these  types of soliton solutions are only be obtained when $\gamma = - 1$(c.f. \cite{ablowitz2016inverse}). According to  the special conditions mentioned  above, the formula of   pure soliton solutions is  obtained by
\begin{equation}\label{soliton}
  q(x) =  -  2i   \sum_{l=1}^{J}C_{l}^{\ast}N_2^{\ast}(-x,\zeta_l) e^{2i \zeta_l^{\ast} x}.
\end{equation}

In order to facilitate the discussion of the properties of soliton solutions,  it is necessary to obtain the explicit expression of some critical parameters
\begin{equation}
   C_{j}= \frac{b_{j}}{a^{'}_{j}}e^{(16i\delta \zeta^4 - 2i\zeta^2)t},  \qquad  \overline{C}_{j}=\frac{\overline{b}_{j}}{\overline{a}^{'}_{j}}e^{(-16i\delta \overline{\zeta}^4 + 2i\overline{\zeta}^2)t},
\end{equation}
where
\begin{equation}
  \begin{aligned}
      b_{j}=e^{\theta_j},  & \quad a^{'}(\zeta)=  \frac{\prod_{j=1}^{N}(\zeta-\zeta_{j})}{\prod_{j=1}^{N}(\zeta- \overline{\zeta}_{j})} \sum_{l=1}^{N} \frac{(\zeta_{l} - \overline{\zeta}_{l})}{(\zeta-\zeta_{l})(\zeta - \overline{\zeta}_{l})},  \\
            \overline{b}_{j}=e^{\overline{\theta}_j},  & \quad \overline{a}^{'}(\zeta)= \frac{\prod_{j=1}^{N}(\zeta-\overline{\zeta}_{j})}{\prod_{j=1}^{N}(\zeta - \zeta_{j})} \sum_{l=1}^{N} \frac{(\overline{\zeta}_{l} - \zeta_{l})}{ (\zeta-\overline{\zeta}_{l})(\zeta-\zeta_{l})},
  \end{aligned}
\end{equation}
with $\theta_{j}$ and $\overline{\theta}_{j}$ are the amplitude of $b_{j}$ and $\overline{b_j}$, respectively \cite{ablowitz2016inverse}.

Next, we will take some special parameters to give the explicit expression of soliton solutions and  present them graphically with the aid of  mathematic software, which is helpful for studying the properties of soliton solutions.

\subsection{One soliton solutions}

In this subsection, we discuss  the one-soliton solutions of the nonlocal LPD equations  by taking $J=\overline{J}=1$ in Eqs. \eqref{soliton}.
Such solution corresponds to soliton eigenvalues
\begin{equation}
  \begin{aligned}
       \zeta_1 = \xi_1 + i \eta_1,    \quad \eta_1 >0,  \qquad     \overline{\zeta}_1  = \overline{\xi}_1 + i \overline{\eta}_1,  \quad   \overline{\eta}_1 < 0.
  \end{aligned}
\end{equation}
Taking $J=1$ for Eq. \eqref{soliton}, we obtain
\begin{equation}
   q(x) =  -  2i    C_{1}^{\ast}N_2^{\ast}(-x,\zeta_l) e^{2i \zeta_1^{\ast} x},
\end{equation}
where $C_{1}$ and $\overline{C}_{1}$ are the norming constants (in $x$) whose time evolution is determined by
\begin{equation}
\begin{aligned}
   C_{1}(t) & =C_{1}(0) e^{(16i\delta \zeta_1^4 - 2i\zeta_1^2)t}= e^{\theta_1} (\zeta_1-\overline{\zeta}_{1})e^{(16i\delta \zeta_1^4 - 2i\zeta_1^2)t}, \\  \overline{C}_1(t) & =\overline{C}_1(0) e^{(-16i\delta \overline{\zeta}_1^4 +2i\overline{\zeta}_1^2 )t} =i e^{\overline{\theta}_1}(\overline{\zeta}_1-\zeta_1)e^{(-16i\delta \overline{\zeta}_1^4 +2i\overline{\zeta}_1^2 )t},
\end{aligned}
\end{equation}
and the expression of $N_2^{\ast}(-x,\zeta_1)$ can be obtained by setting $J=1$ into Eq. \eqref{soliton}
\begin{equation}
   N_{2}^{\ast}(-x,\zeta_1)= \frac{|\zeta_1 - \overline{\zeta}_1|^2}{|\zeta_1 - \overline{\zeta}_1|^2-C_1^{\ast}\overline{C}_1^{\ast}e^{2i(\overline{\zeta}_1+ \zeta_1^{\ast})x}},
\end{equation}
then we find the one soliton solution
\begin{equation}
  q(x,t)=-2 i  \frac{C_{1}^{\ast}(t) e^{2i k_{1}^{\ast} x} |\zeta_1 - \overline{\zeta}_1|^2}{|\zeta_1 - \overline{\zeta}_1|^2- C_{1}^{\ast}(t)\overline{C}_1^{\ast}(t) e^{2i(\overline{\zeta}_1+ \zeta_1^{\ast})x}}.
\end{equation}

The localized structure, the density and the wave propagation of one soliton solution is shown in Fig. 1. From Fig. 1, we can learn that the single soliton wave propagate almost along the axis of  $x=0$. Moreover, in the process of wave propagate,  the amplitude and the width of the single soliton  are  not changed.

\noindent
{\rotatebox{0}{\includegraphics[width=3.6cm,height=3.5cm,angle=0]{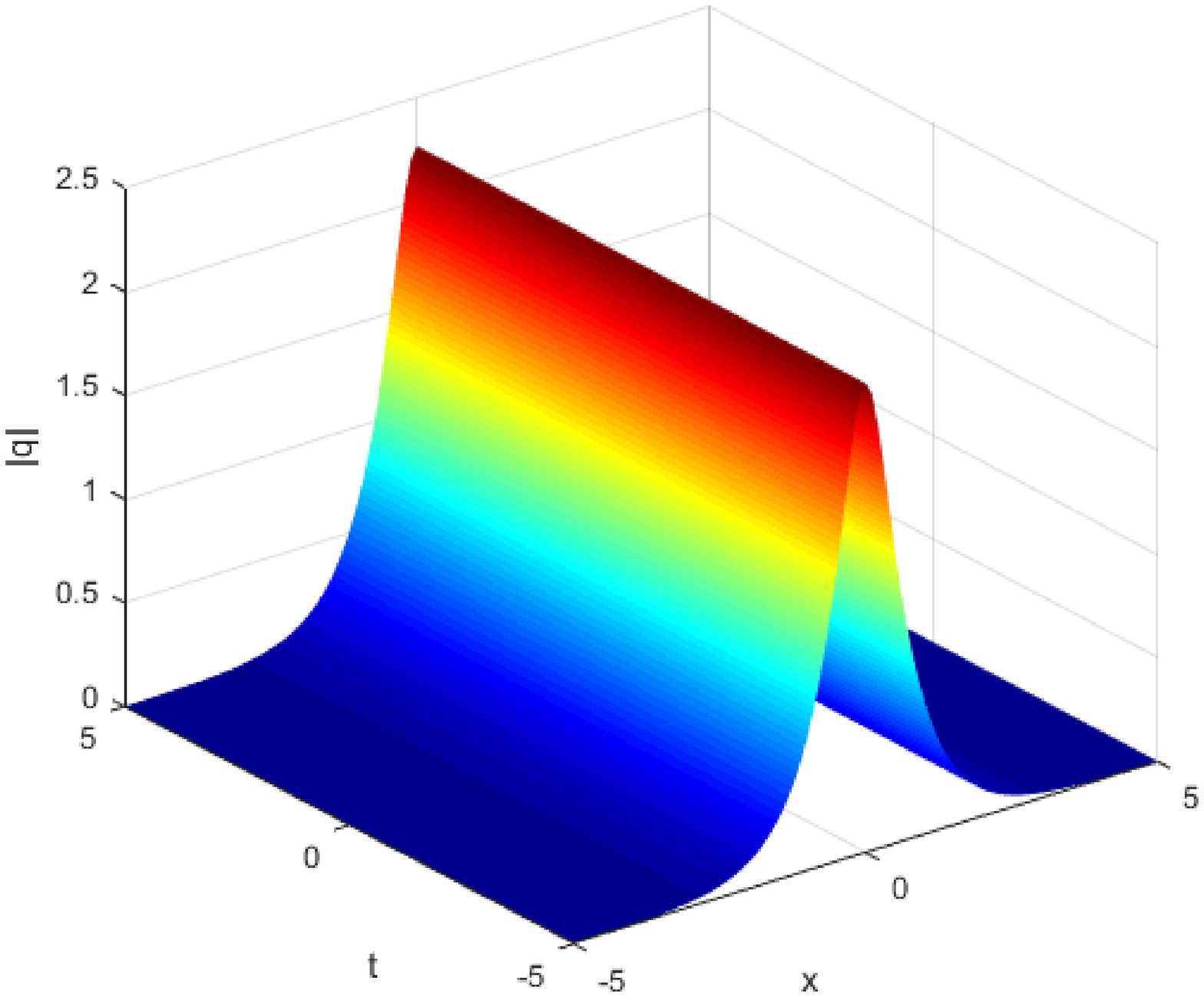}}}
~~~~
{\rotatebox{0}{\includegraphics[width=3.6cm,height=3.5cm,angle=0]{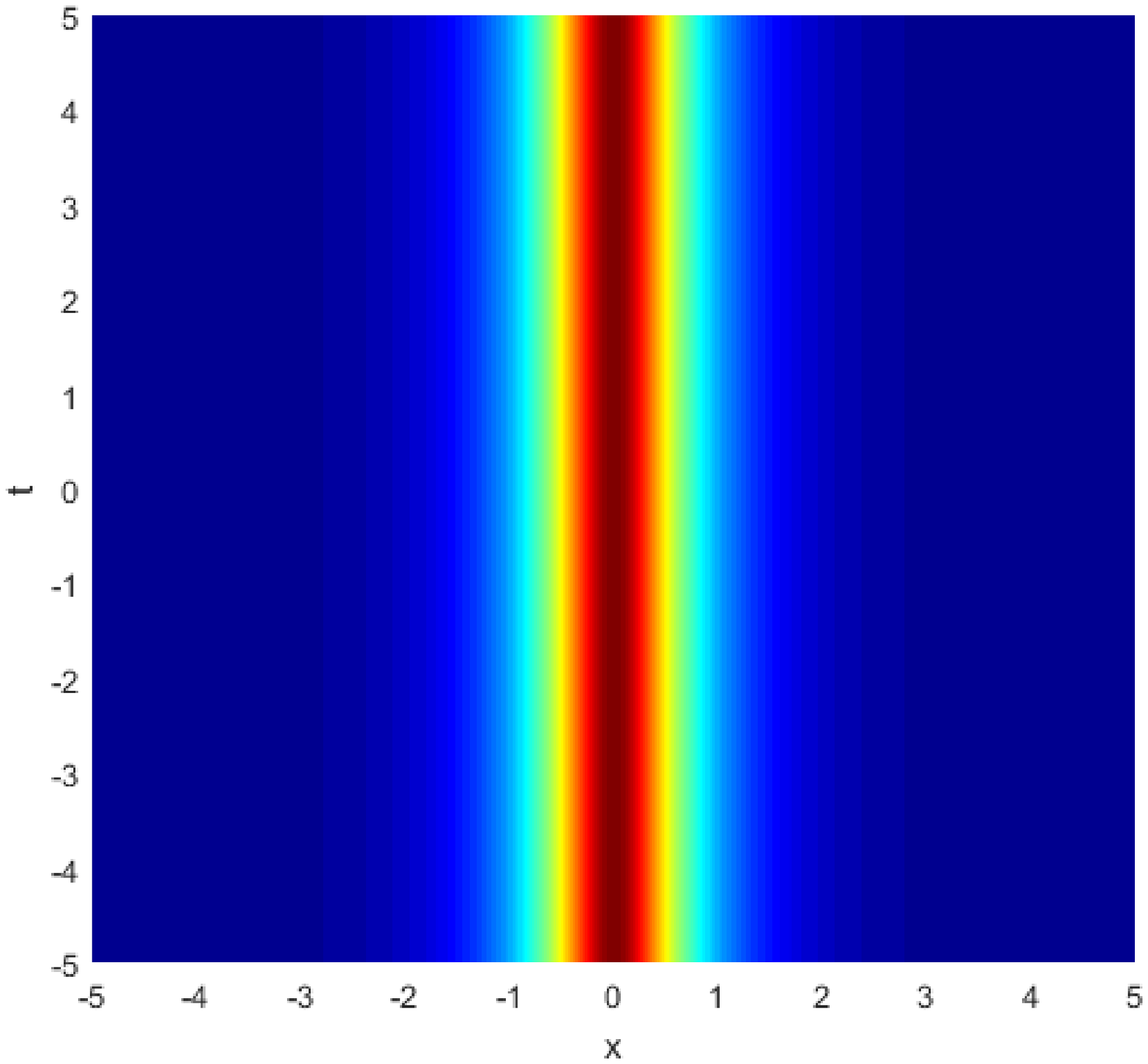}}}
~~~~
{\rotatebox{0}{\includegraphics[width=3.6cm,height=3.5cm,angle=0]{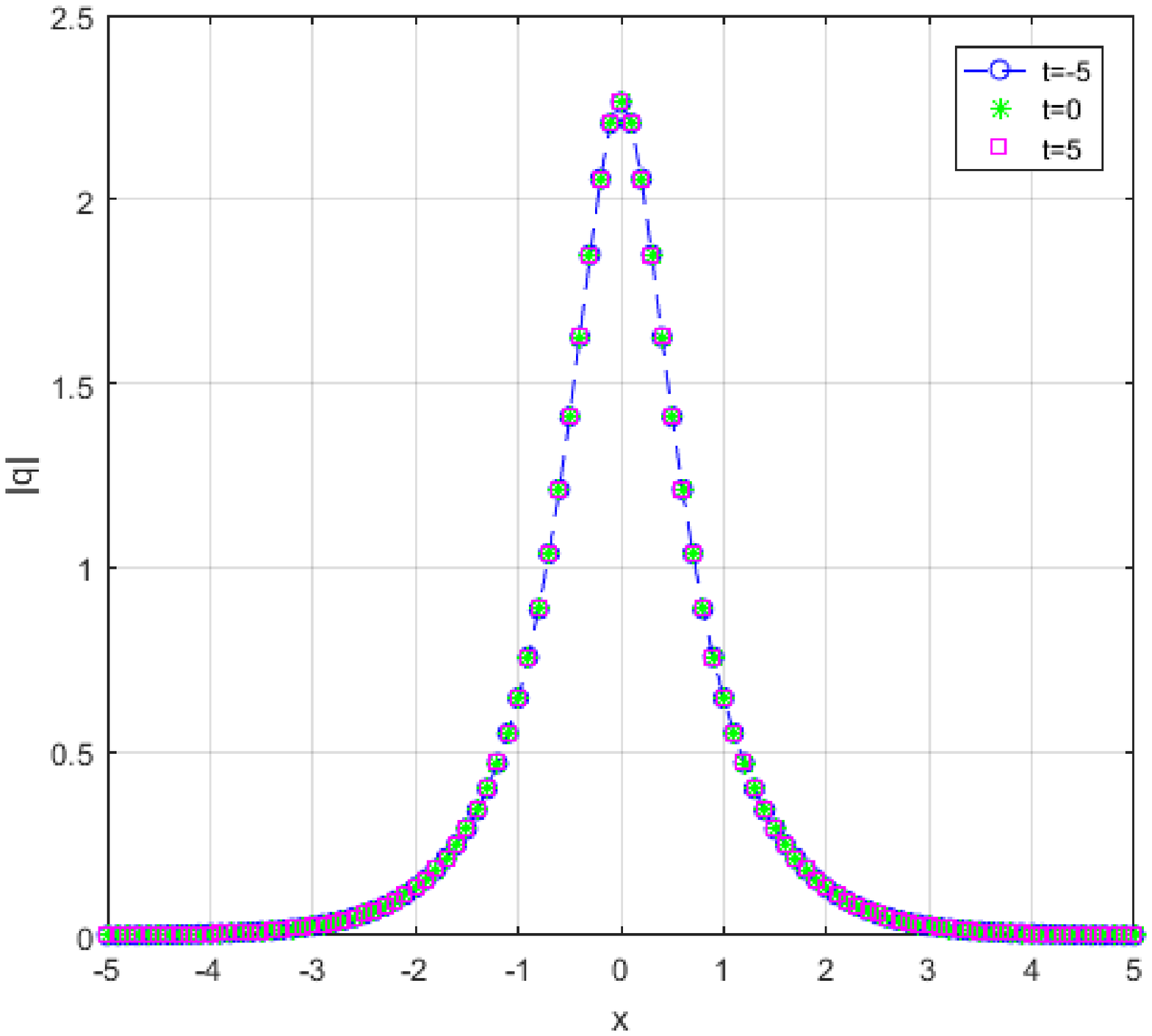}}}

$\ \qquad~~~~~~(\textbf{a})\qquad \ \qquad\qquad\qquad\qquad~(\textbf{b})
\ \qquad\qquad\qquad\qquad\qquad~(\textbf{c})$\\
\noindent { \small \textbf{Figure 1.} One-soliton  solution   with parameters $\delta =1$, $\theta_1=\frac{\pi}{3}$, $\overline{\theta}_1= - \frac{\pi}{3}$,  $\zeta_1=0.8i$  and  $\overline{\zeta}_1=-0.8i$.
$\textbf{(a)}$: the structures of the one-soliton  solution,
$\textbf{(b)}$: the density plot,
$\textbf{(c)}$: the wave propagation of the one-soliton  solution.}

\subsection{Two soliton solutions}
In this subsection,  we consider the soliton solutions of the nonlocal LPD equations \eqref{LPD} with $J=\overline{J}=2$.  Suppose the corresponding eigenvalues as follows
\begin{equation}
   \begin{aligned}
        \zeta_1 =\xi_1 + i \eta_1, & \quad \zeta_2 =\xi_2+ i \eta_2,  & \quad \eta_1, \eta_2>0,  \\
       \overline{ \zeta}_1 =\overline{\xi}_1 + i \overline{\eta}_1, & \quad \overline{\zeta}_2 =\overline{\xi}_2+ i \overline{\eta}_2,  & \quad \overline{\eta}_1, \overline{\eta}_2< 0.  \\
   \end{aligned}
\end{equation}
Setting $J=2$ into Eq. \eqref{soliton}, we find
\begin{equation}\label{two}
   q(x)= -2i C_{1}^{\ast} N_2^{\ast}(-x,\zeta_1)e^{2i\zeta_1^\ast x} -2 i C_{2}^{\ast} N_2^{\ast}(-x,\zeta_2)e^{2i \zeta_2^\ast x},
\end{equation}
where $C_{j}$, $\overline{C}_{j}$, $j=1,2$ are the norming constants whose time evolution is given by
\begin{equation}
   \begin{aligned}
       C_{1}(t)=C_{1}(0)e^{(16i\delta \zeta_1^4 - 2i\zeta_1^2)t}, & \qquad  C_{2}(t)=C_{2}(0)e^{(16i\delta \zeta_2^4 - 2i\zeta_2^2)t},  \\
       \overline{C}_{1}(t)=\overline{C}_{1}(0)e^{(-16i\delta \overline{\zeta}_1^4 + 2i\overline{\zeta}_1^2)t}, & \qquad  \overline{C}_{2}(t)=\overline{C}_{2}(0)e^{(-16i\delta \overline{\zeta}_2^4 + 2i\overline{\zeta}_2^2)t}.
   \end{aligned}
\end{equation}
To obtain the functions $N_2^{\ast}(-x,\zeta_1)$ and $N_2^{\ast}(-x,\zeta_2)$, we need to solve the following system
\begin{equation}
 \left\{
    \begin{aligned}
         \overline{M}_1(x,-\overline{\zeta}_1^{\ast}) =& \alpha_1 N_2^{\ast}(-x,\zeta_1) + \beta_1 N_2^{\ast}(-x,\zeta_2),\\
          \overline{M}_1(x,-\overline{\zeta}_2^{\ast}) =& \alpha_2 N_2^{\ast}(-x,\zeta_1) + \beta_2 N_2^{\ast}(-x,\zeta_2),\\
        N_2^{\ast}(-x,\zeta_1)= &  1+ \overline{\alpha}_1 \overline{M}_1(x,-\overline{\zeta}_1^{\ast}) +  \overline{\beta}_1 \overline{M}_1(x,-\overline{\zeta}_2^{\ast}),  \\
       N_2^{\ast}(-x,\zeta_2)= &  1+ \overline{\alpha}_2 \overline{M}_1(x,-\overline{\zeta}_1^{\ast}) +  \overline{\beta}_2 \overline{M}_1(x,-\overline{\zeta}_2^{\ast}),
    \end{aligned}
 \right.
\end{equation}
where
\begin{equation}
   \begin{aligned}
     \alpha_1=\frac{C_1^{\ast}(t)e^{2i\zeta_1^{\ast}x}}{\overline{\zeta}_1^{\ast}-\zeta_1^{\ast}},  & \qquad \beta_1= \frac{C_2^{\ast}(t)e^{2i\zeta_2^{\ast}x}}{\overline{\zeta}_1^{\ast}-\zeta_2^{\ast}},  \\
      \alpha_2 =\frac{C_1^{\ast}(t)e^{2i\zeta_1^{\ast}x}}{\overline{\zeta}_2^{\ast}-\zeta_1^{\ast}},  & \qquad \beta_2= \frac{C_2^{\ast}(t)e^{2i\zeta_2^{\ast}x}}{\overline{\zeta}_2^{\ast}-\zeta_2^{\ast}},  \\
      \overline{\alpha}_1 = \frac{\overline{C}_1^{\ast}(t)e^{-2i\overline{\zeta}_{1}^{\ast}x}}{\zeta_1^{\ast}-\overline{\zeta}_1^{\ast}}, & \qquad
      \overline{\beta}_1 =  \frac{\overline{C}_2^{\ast}(t)e^{-2i\overline{\zeta}_{2}^{\ast}x}}{\zeta_1^{\ast}-\overline{\zeta}_2^{\ast}}, \\
      \overline{\alpha}_2 = \frac{\overline{C}_1^{\ast}(t)e^{-2i\overline{\zeta}_{1}^{\ast}x}}{\zeta_2^{\ast}-\overline{\zeta}_1^{\ast}}, & \qquad
      \overline{\beta}_2 =  \frac{\overline{C}_2^{\ast}(t)e^{-2i\overline{\zeta}_{2}^{\ast}x}}{\zeta_2^{\ast}-\overline{\zeta}_2^{\ast}}.
   \end{aligned}
\end{equation}
Solving the above system, we  get
\begin{equation}
     N_2^{\ast}(-x,\zeta_1)= \frac{ \lambda_4 - \lambda_2}{\lambda_1 \lambda_4 - \lambda_2 \lambda_3},  \qquad
      N_2^{\ast}(-x,\zeta_2)= \frac{ \lambda_1 - \lambda_3}{\lambda_1 \lambda_4 - \lambda_2 \lambda_3},
\end{equation}
where
\begin{equation}
\left\{
  \begin{aligned}
  \lambda_1 = &1 - \alpha_1 \overline{\alpha}_1 - \alpha_2  \overline{\beta}_1,  \\    \lambda_2 = &- \overline{\alpha}_1 \beta_1 - \beta_2 \overline{\beta}_1,\\
  \lambda_3 = &-\alpha_1 \overline{\alpha}_2 - \alpha_2 \overline{\beta}_2, \\  \lambda_4 = & 1 - \beta_2 \overline{\beta}_2 - \overline{\alpha}_2 \beta_1. \\
  \end{aligned}
\right.
\end{equation}
Substituting the above equations into Eq. \eqref{two}, we can obtain the formula of  two-soliton solutions. \\

\noindent
{\rotatebox{0}{\includegraphics[width=3.6cm,height=3.3cm,angle=0]{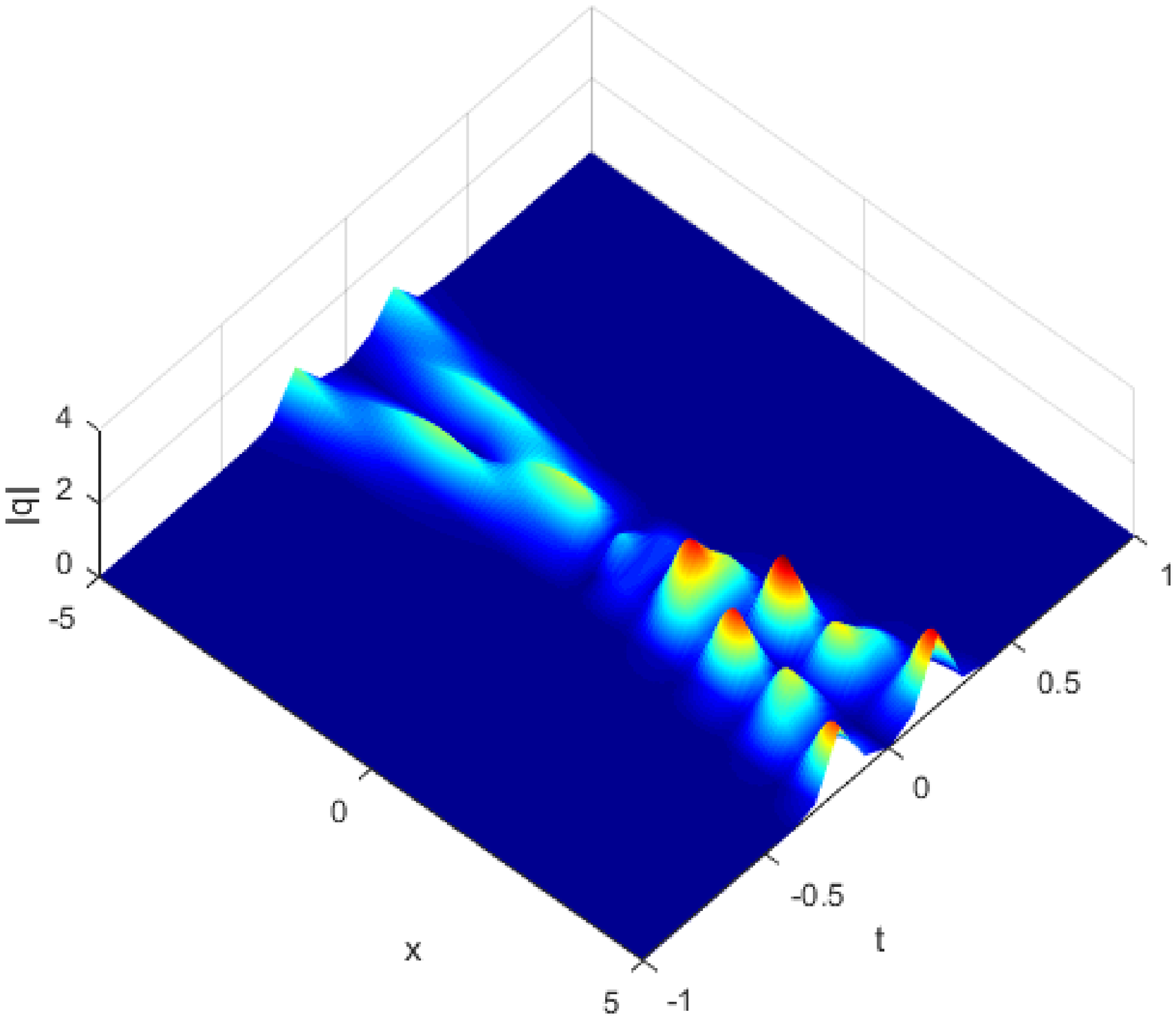}}}
~~~~
{\rotatebox{0}{\includegraphics[width=3.6cm,height=3.3cm,angle=0]{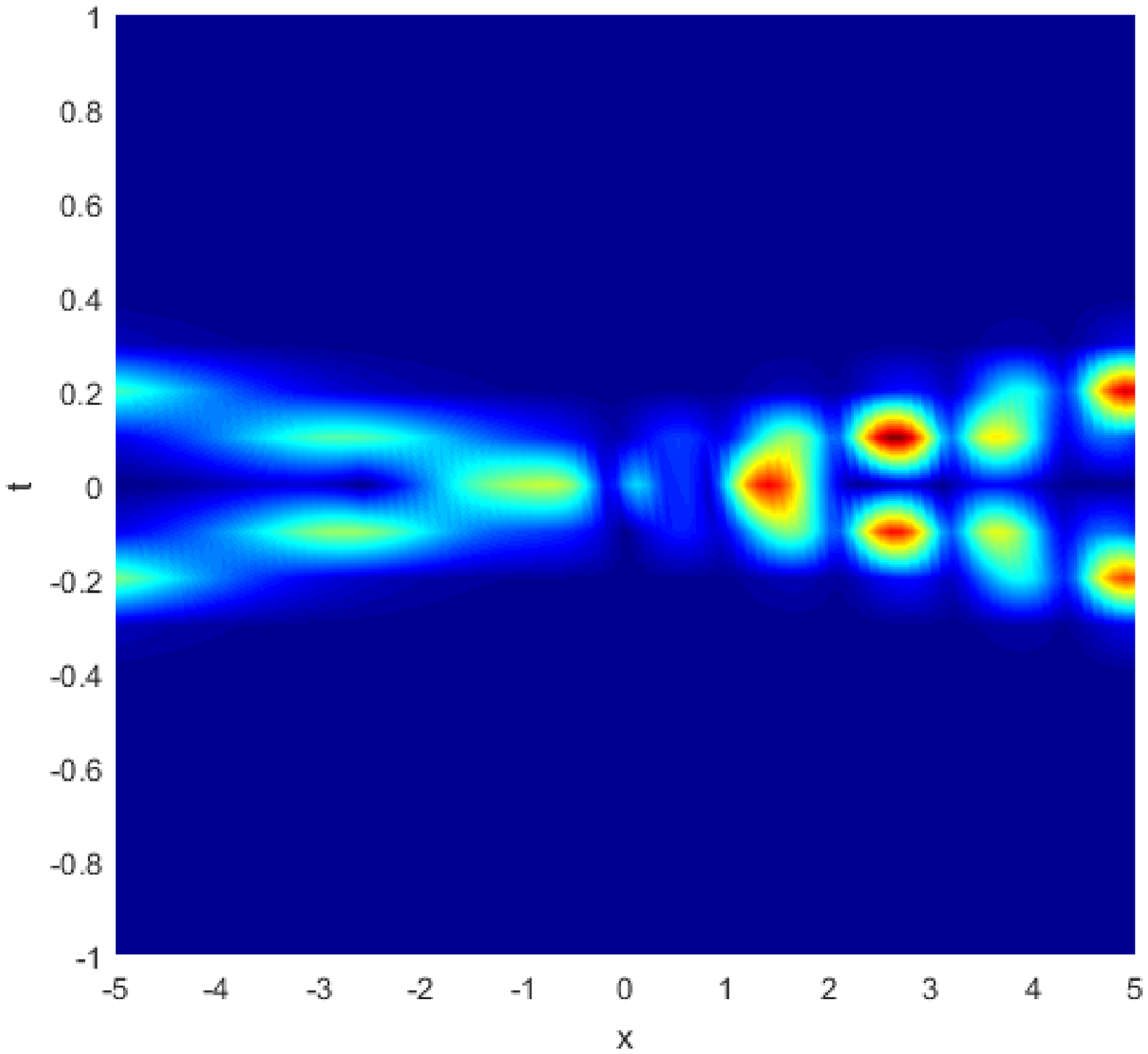}}}
~~~~
{\rotatebox{0}{\includegraphics[width=3.6cm,height=3.3cm,angle=0]{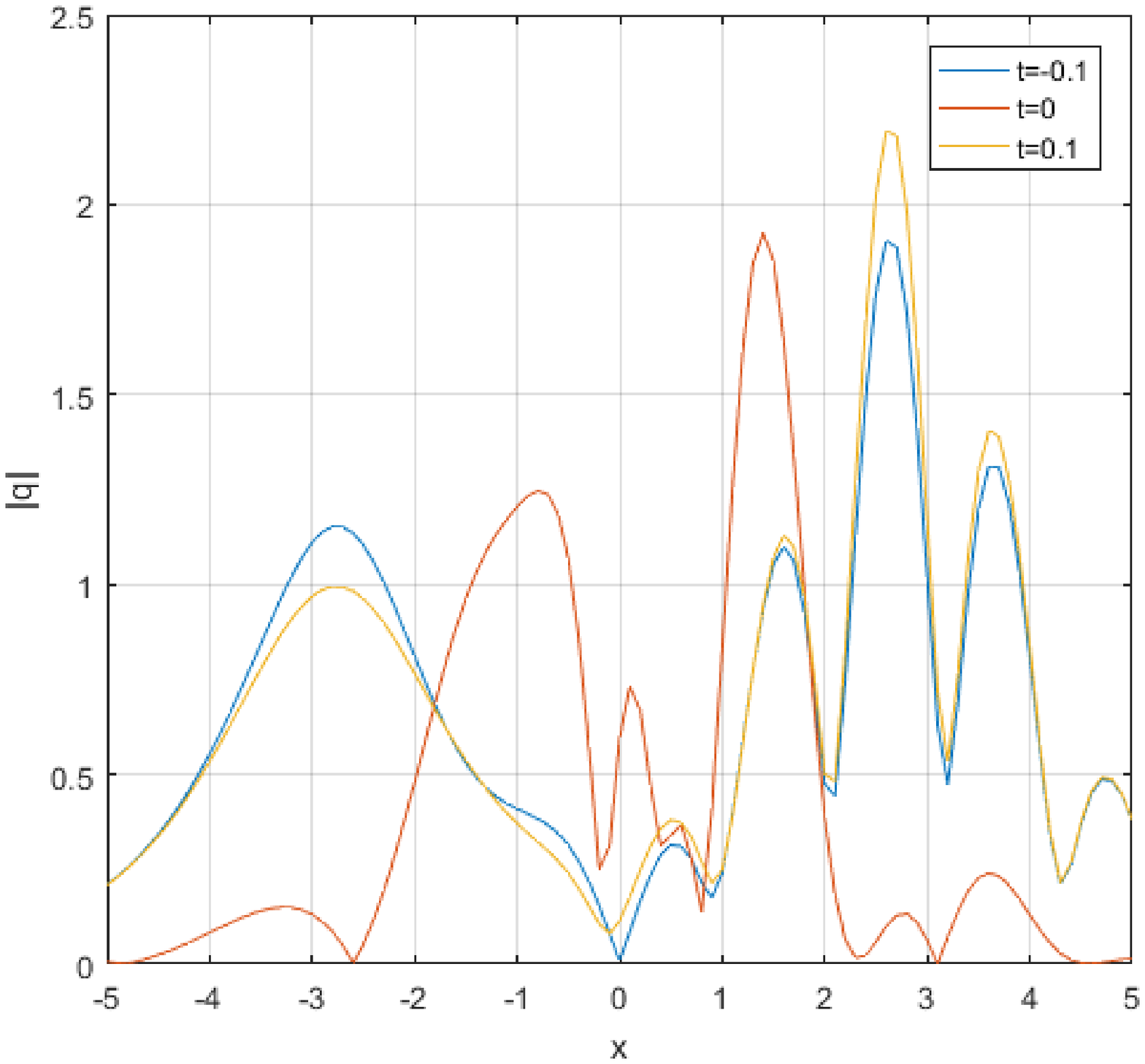}}}

$\ \qquad~~~~~~(\textbf{a})\qquad \ \qquad\qquad\qquad\qquad~(\textbf{b})
\ \qquad\qquad\qquad\qquad\qquad~(\textbf{c})$\\
\noindent
{\rotatebox{0}{\includegraphics[width=3.6cm,height=3.0cm,angle=0]{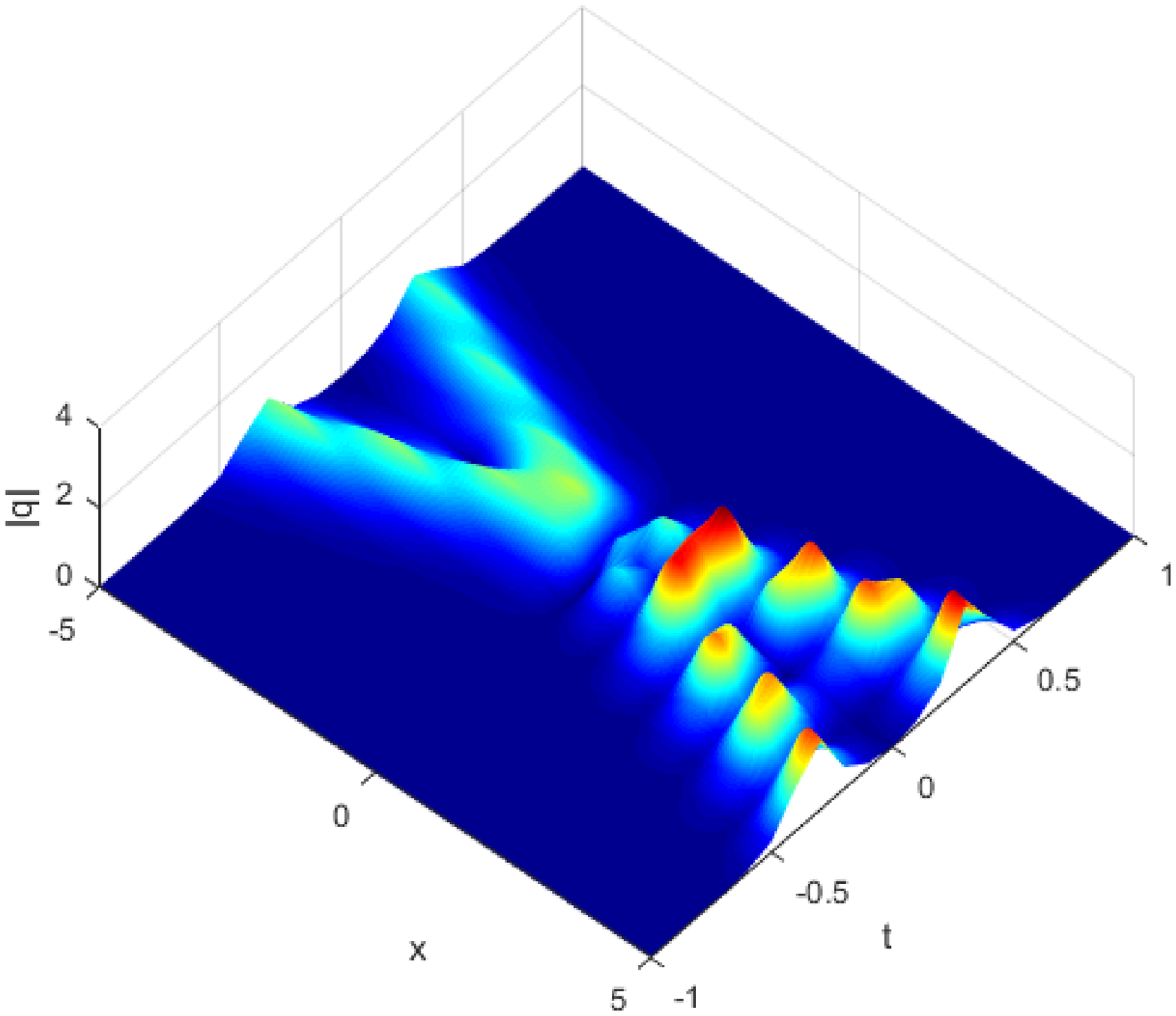}}}
~~~~
{\rotatebox{0}{\includegraphics[width=3.6cm,height=3.0cm,angle=0]{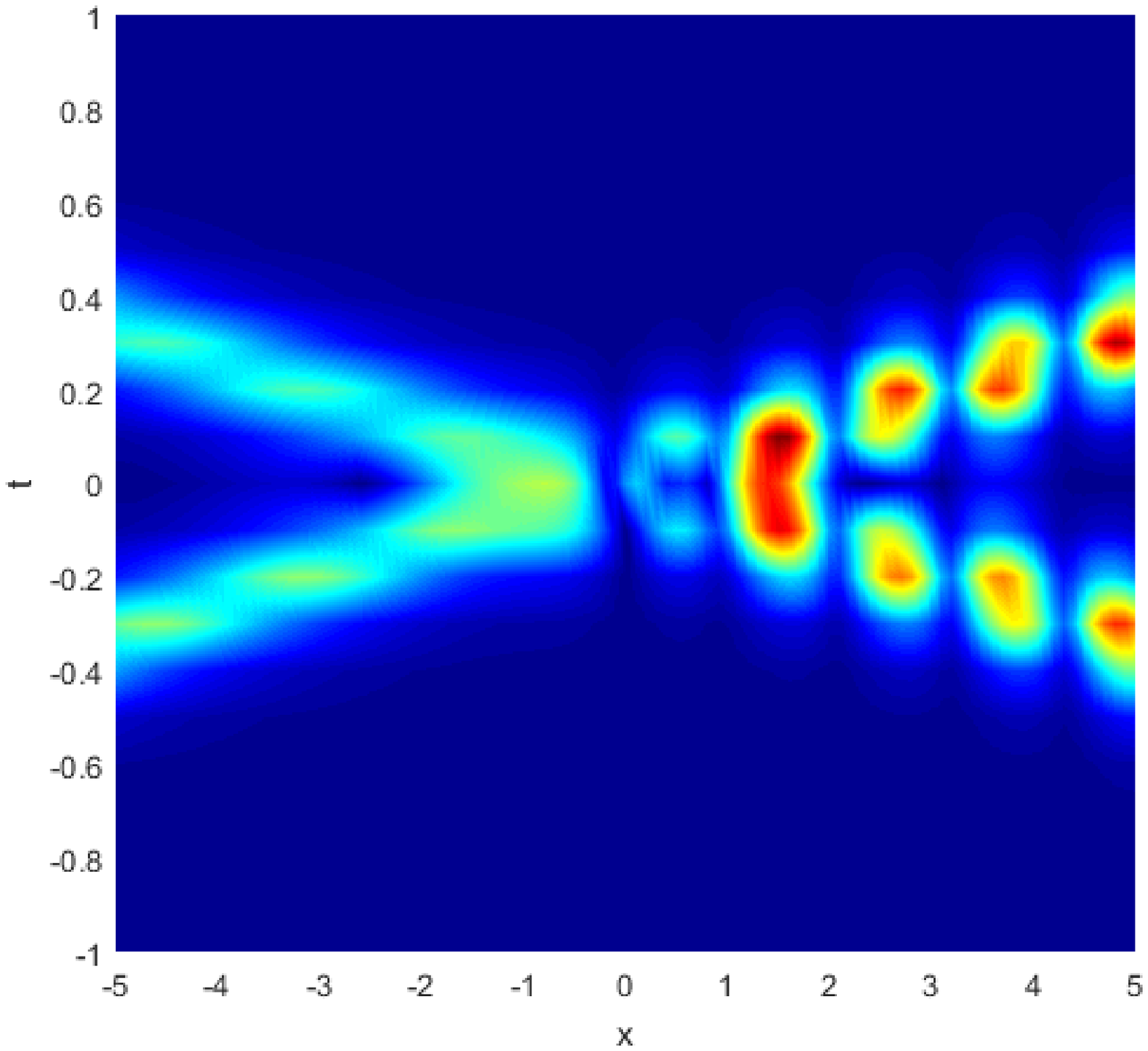}}}
~~~~
{\rotatebox{0}{\includegraphics[width=3.6cm,height=3.0cm,angle=0]{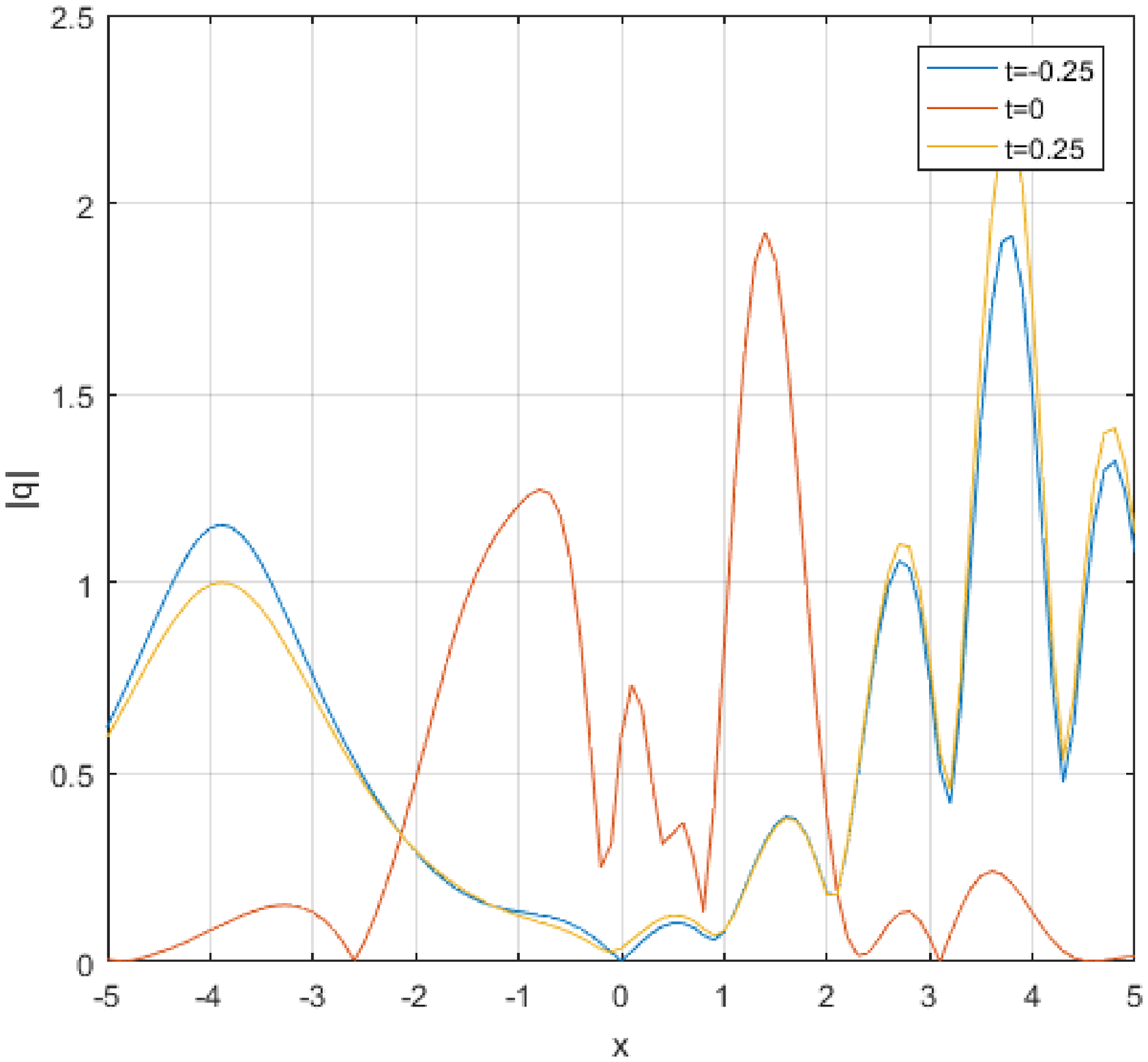}}}

$\ \qquad~~~~~~(\textbf{d})\qquad \ \qquad\qquad\qquad\qquad~(\textbf{e})
\ \qquad\qquad\qquad\qquad\qquad~(\textbf{f})$\\
\noindent
{\rotatebox{0}{\includegraphics[width=3.6cm,height=3.0cm,angle=0]{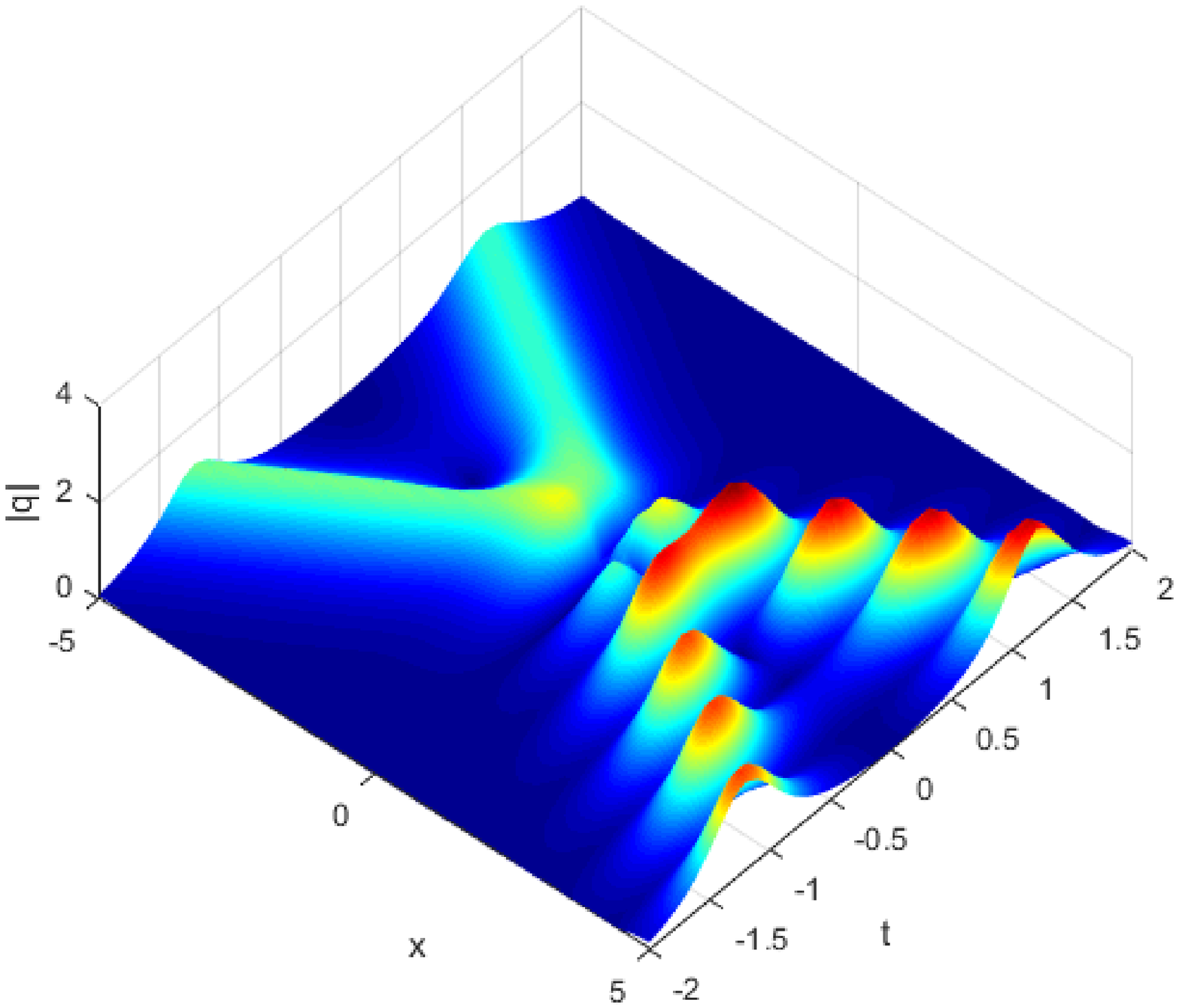}}}
~~~~
{\rotatebox{0}{\includegraphics[width=3.6cm,height=3.0cm,angle=0]{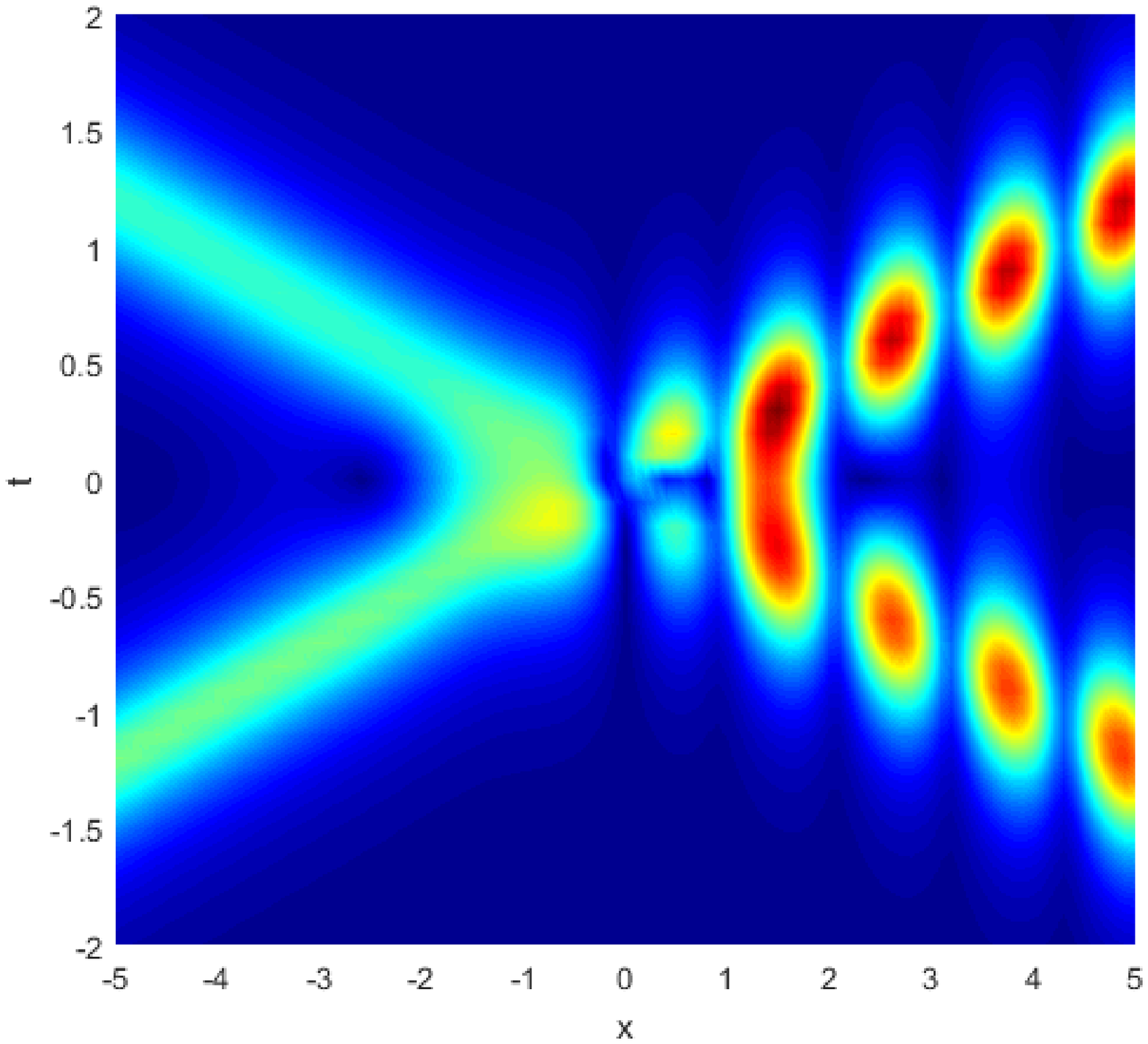}}}
~~~~
{\rotatebox{0}{\includegraphics[width=3.6cm,height=3.0cm,angle=0]{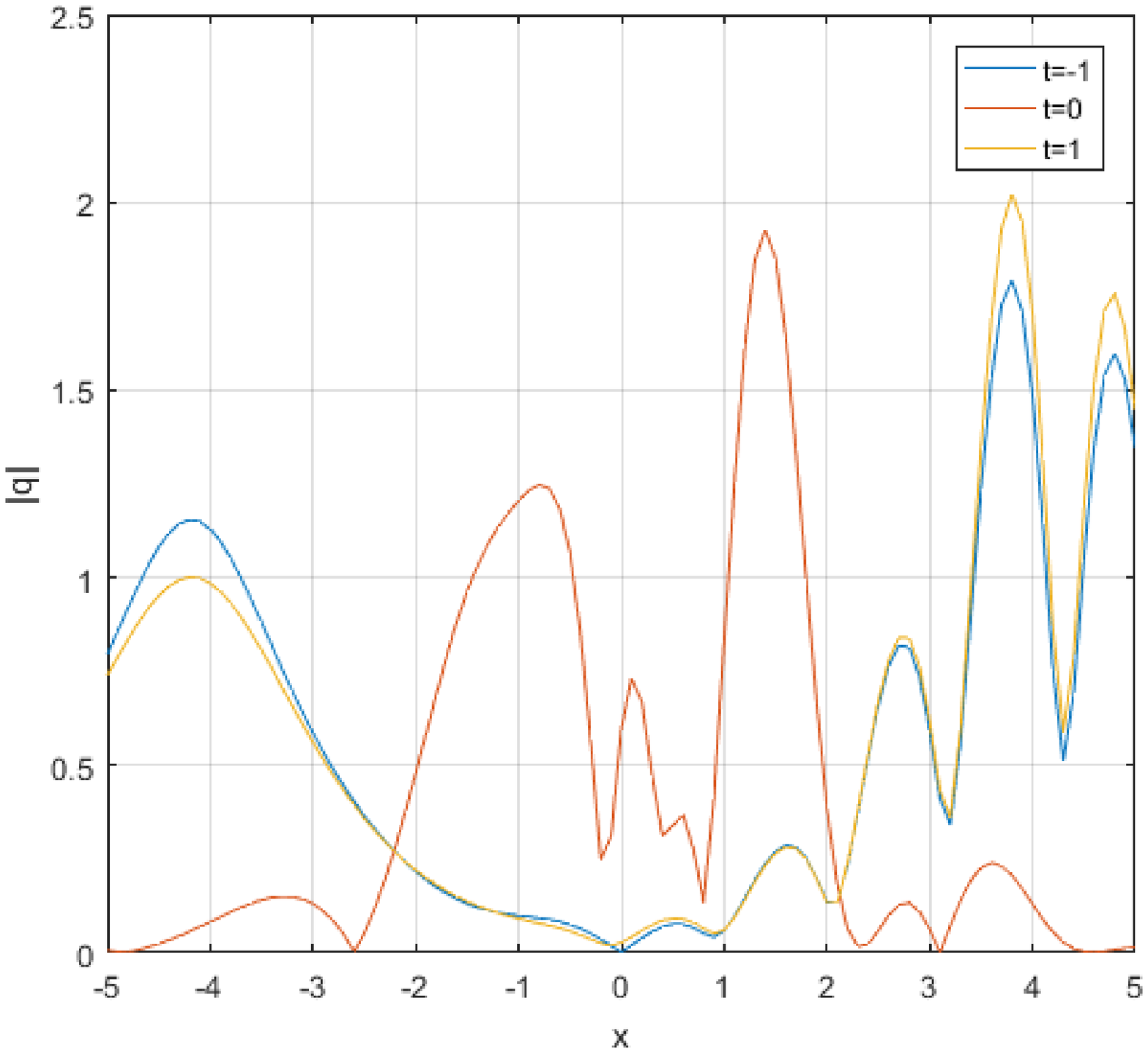}}}

$\ \qquad~~~~~~(\textbf{g})\qquad \ \qquad\qquad\qquad\qquad~(\textbf{h})
\ \qquad\qquad\qquad\qquad\qquad~(\textbf{i})$\\
\noindent { \small \textbf{Figure 2.} Two-soliton  solutions   with parameters $\theta_1=\frac{2}{3}\pi$,   $\theta_2=\frac{3}{8}\pi$,  $\overline{\theta}_1=\frac{2}{3}\pi$,   $\overline{\theta}_2=\frac{3}{5}\pi$,  $\zeta_1=0.7+0.5i$,  $\zeta_2=-0.7+0.5i$,   $\overline{\zeta}_1=0.7-0.5i$ and  $\overline{\zeta}_2=-0.7-0.5i$.
$\textbf{(a)(b)(c)}$: the structures and the wave propagation of the two-soliton  solutions with $\delta=5$,
$\textbf{(d)(e)(f)}$: the structures and the wave propagation of the two-soliton  solutions with $\delta=3$,
$\textbf{(g)(h)(i)}$: the structures and the wave propagation of the two-soliton  solutions with $\delta=1$.}  \\

The local structure, the density and the wave propagation of two soliton solution is shown in Fig. 2.  It is interesting that Fig. 2 shows  the whole process of two solitons meet, collide   elastically and move away. Furthermore, among these two solitons, one is a ordinary soliton and the other is a breather soliton. Besides, we also find a meaningful phenomenon by select different parameter $\delta$. Observe the three density plots carefully, we find that the angle between two solitons will increase as the parameter $\delta$ increases, which reveals the influence of  parameter $\delta$ on  the soliton solution graphically. \\

\noindent
{\rotatebox{0}{\includegraphics[width=3.6cm,height=3.2cm,angle=0]{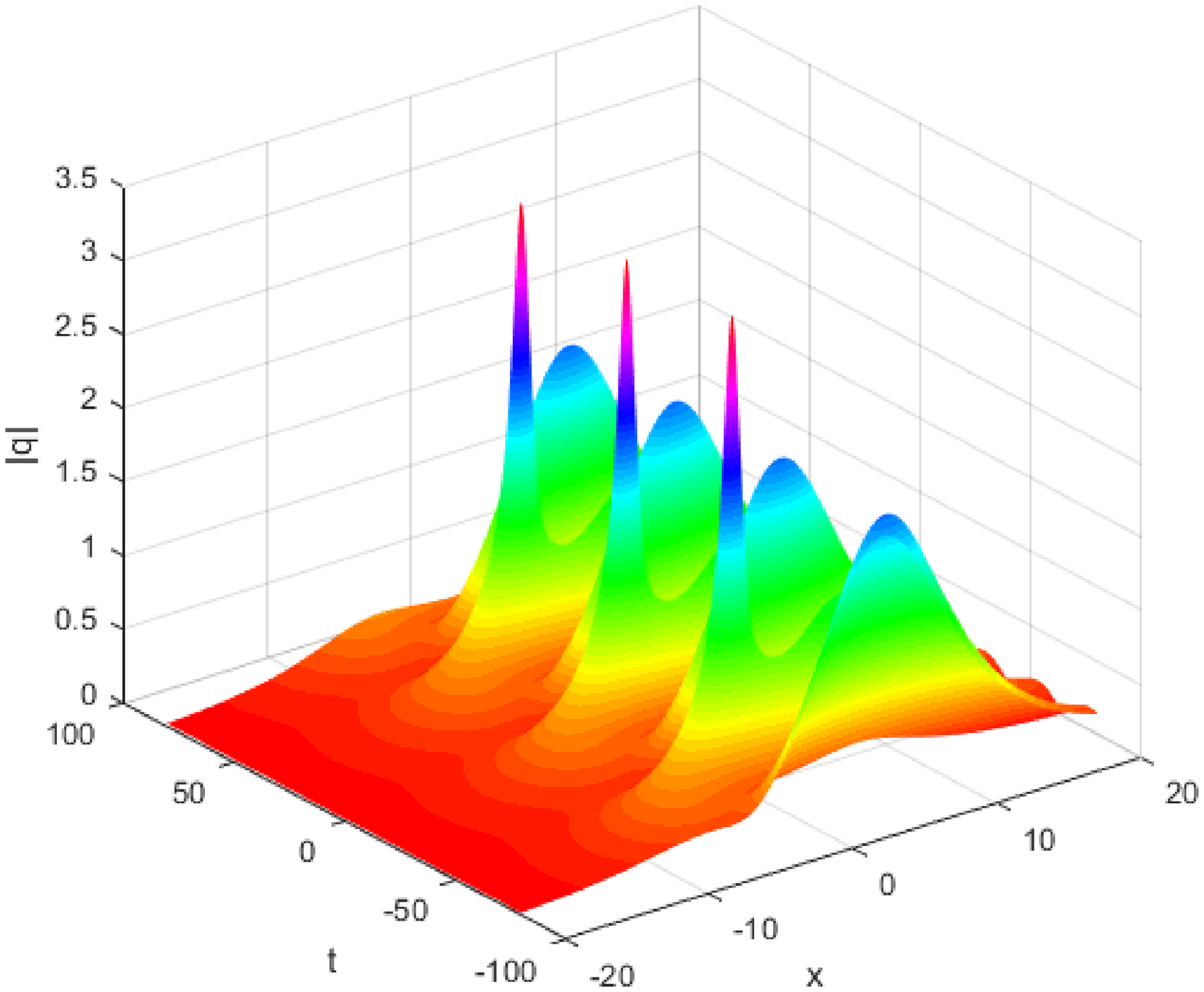}}}
~~~~
{\rotatebox{0}{\includegraphics[width=3.6cm,height=3.2cm,angle=0]{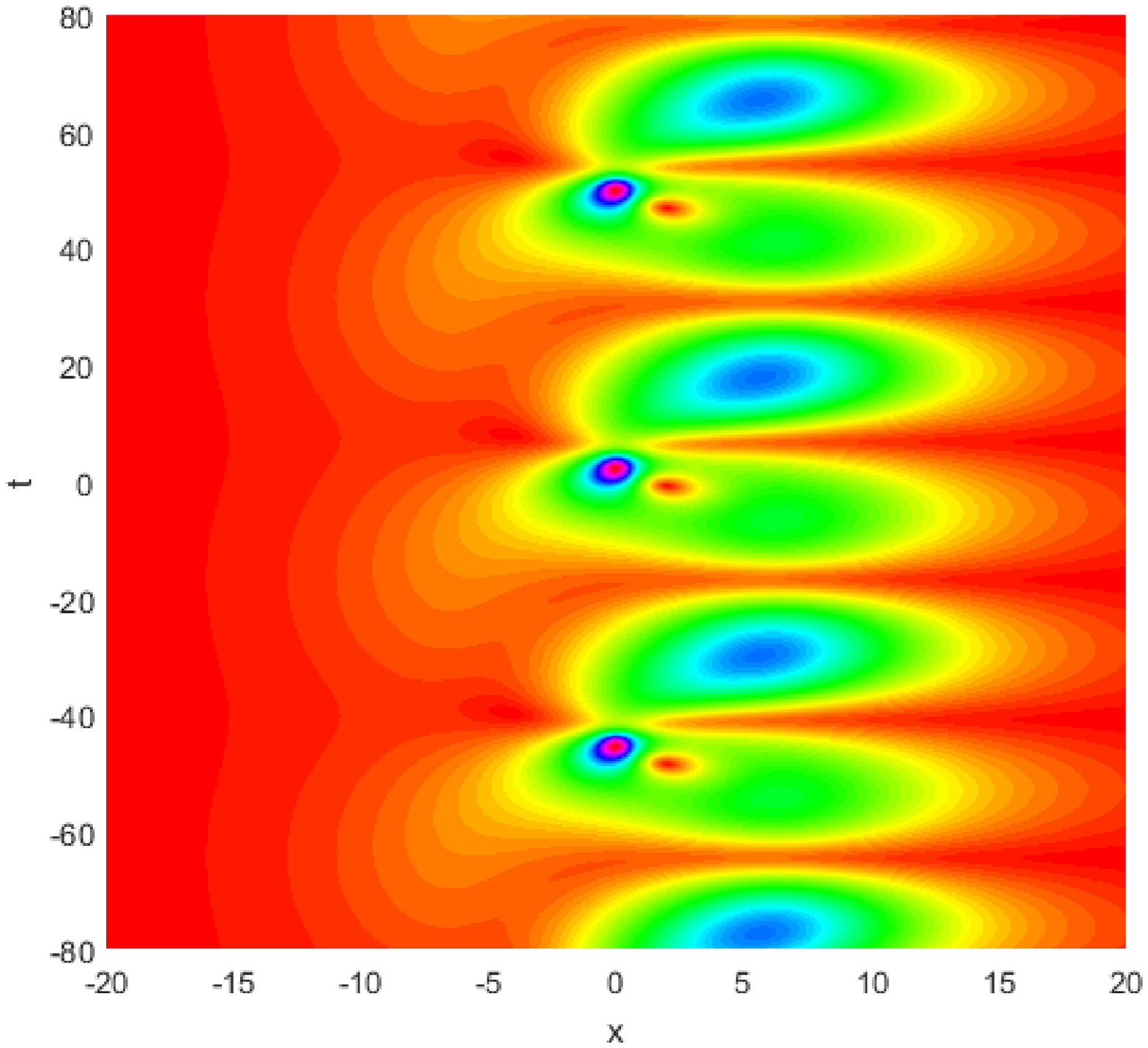}}}
~~~~
{\rotatebox{0}{\includegraphics[width=3.6cm,height=3.2cm,angle=0]{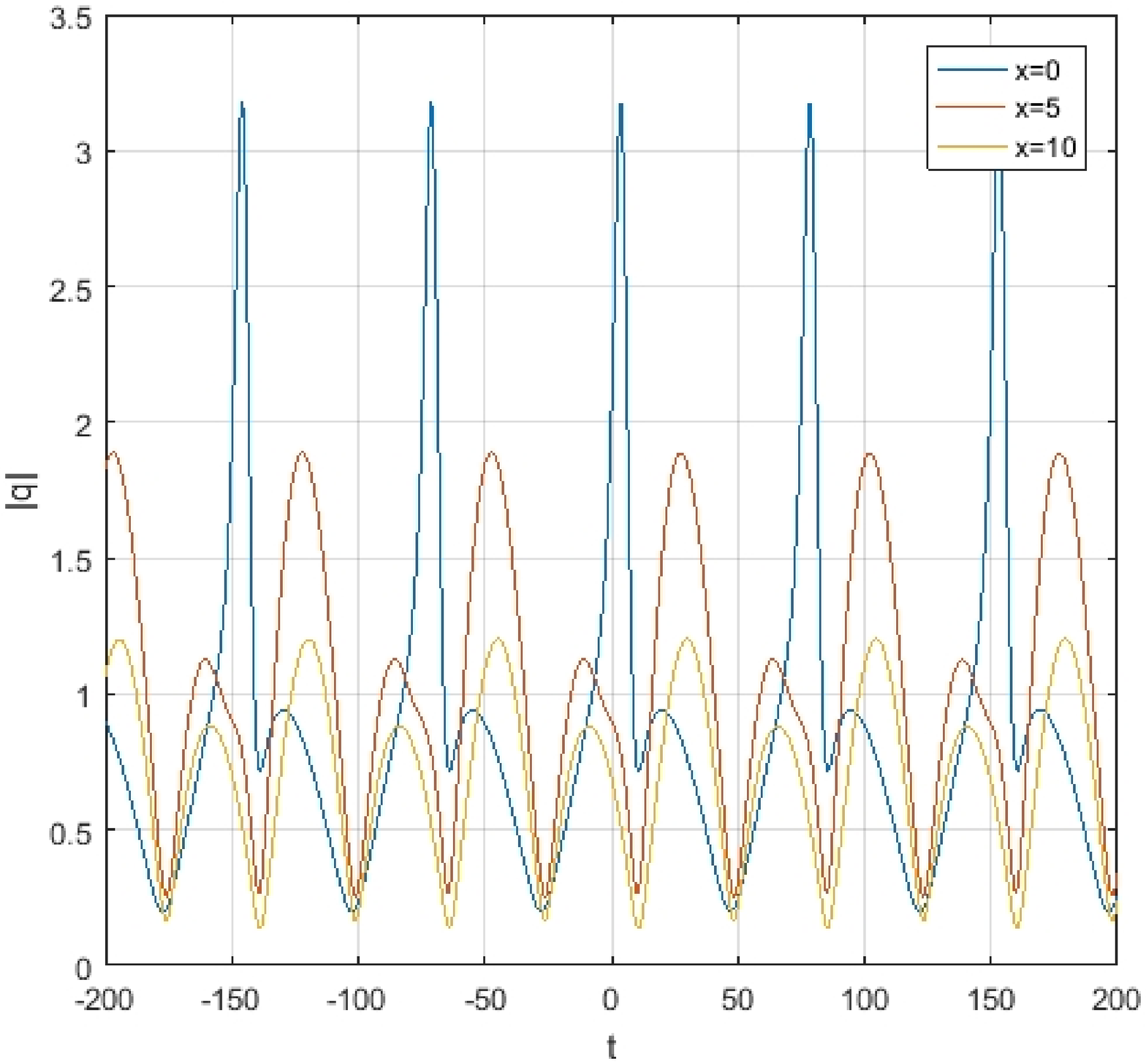}}}

$\ \qquad~~~~~~(\textbf{a})\qquad \ \qquad\qquad\qquad\qquad~(\textbf{b})
\ \qquad\qquad\qquad\qquad\qquad~(\textbf{c})$\\
\noindent { \small \textbf{Figure 3.} Breather-type  solution   with parameters  $\delta=1$, $\theta_1=\frac{2}{3}\pi$,   $\theta_2=\frac{3}{8}\pi$,  $\overline{\theta}_1=\frac{2}{3}\pi$,   $\overline{\theta}_2=\frac{3}{5}\pi$,  $\zeta_1=0.1i$,  $\zeta_2=0.2i$,   $\overline{\zeta}_1=-0.1i$ and  $\overline{\zeta}_2=-0.2i$.
$\textbf{(a)}$: the structures of the breather-type  solution,
$\textbf{(b)}$: the density plot,
$\textbf{(c)}$: the wave propagation of the breather-type  solution.}  \\

By introducing the  appropriate parameters, we get the other interesting discovery which is presented in Fig. 3.  In  Fig. 3,  two different breather-type  solitons spread alternately forward. Furthermore,    the periodicity of the solution is clearly reflected.

\subsection{Three soliton solutions}
In this section,  we consider the three-soliton solutions of the nonlocal LPD equations \eqref{LPD}. Suppose the corresponding eigenvalues as follows
\begin{equation}
   \begin{aligned}
       & \zeta_1 =\xi_1 + i \eta_1, & \quad \zeta_2 =\xi_2+ i \eta_2, & \quad \zeta_3 =\xi_3 + i \eta_3, & \quad \eta_1, \eta_2, \eta_3>0,  \\
    &   \overline{ \zeta}_1 =\overline{\xi}_1 + i \overline{\eta}_1, & \quad \overline{\zeta}_2 =\overline{\xi}_2+ i \overline{\eta}_2,  & \quad \overline{\zeta}_3 =\overline{\xi}_3+ i \overline{\eta}_3,   & \quad \overline{\eta}_1, \overline{\eta}_2, \overline{\eta}_3< 0.  \\
   \end{aligned}
\end{equation}
Setting $J=\overline{J}=3$ into Eq. \eqref{soliton}, we find
\begin{equation}\label{three}
   q(x)= -2i C_{1}^{\ast} N_2^{\ast}(-x,\zeta_1)e^{2i\zeta_1^\ast x} -2 i C_{2}^{\ast} N_2^{\ast}(-x,\zeta_2)e^{2i \zeta_2^\ast x} -2 i C_{3}^{\ast} N_2^{\ast}(-x,\zeta_3)e^{2i \zeta_3^\ast x},
\end{equation}
where $C_{j}$, $\overline{C}_{j}$, $j=1,2,3$ are the norming constants whose time evolution is given by
\begin{equation}
   \begin{aligned}
       C_{1}(t)=C_{1}(0)e^{(16i\delta \zeta_1^4 - 2i\zeta_1^2)t}, &\qquad \overline{C}_{1}(t)=\overline{C}_{1}(0)e^{(-16i\delta \overline{\zeta}_1^4 + 2i\overline{\zeta}_1^2)t}, \\
      \qquad  C_{2}(t)=C_{2}(0)e^{(16i\delta \zeta_2^4 - 2i\zeta_2^2)t},   & \qquad  \overline{C}_{2}(t)=\overline{C}_{2}(0)e^{(-16i\delta \overline{\zeta}_2^4 + 2i\overline{\zeta}_2^2)t}, \\
      C_{3}(t)=C_{3}(0)e^{(16i\delta \zeta_3^4 - 2i\zeta_3^2)t}, & \qquad \overline{C}_{3}(t)=\overline{C}_{3}(0)e^{(-16i\delta \overline{\zeta}_3^4 + 2i\overline{\zeta}_3^2)t}.
   \end{aligned}
\end{equation}
To obtain the functions $N_2^{\ast}(-x,\zeta_1)$,  $N_2^{\ast}(-x,\zeta_2)$  and $N_2^{\ast}(-x,\zeta_3)$, we need to solve the following system
\begin{equation}
\left\{
    \begin{aligned}
        \overline{M}_1(x,-\overline{\zeta}_1^{\ast})  = &\alpha_{11} N_2^{\ast}(-x,\zeta_1) + \alpha_{12} N_2^{\ast}(-x,\zeta_2) + \alpha_{13} N_2^{\ast}(-x,\zeta_3),\\
        \overline{M}_1(x,-\overline{\zeta}_2^{\ast})  =& \alpha_{21} N_2^{\ast}(-x,\zeta_1) + \alpha_{22} N_2^{\ast}(-x,\zeta_2)+ \alpha_{23} N_2^{\ast}(-x,\zeta_3),\\
       \overline{M}_1(x,-\overline{\zeta}_3^{\ast})   =& \alpha_{31} N_2^{\ast}(-x,\zeta_1) + \alpha_{32} N_2^{\ast}(-x,\zeta_2)+ \alpha_{33} N_2^{\ast}(-x,\zeta_3),\\
        N_2^{\ast}(-x,\zeta_1)= &   1 + \beta_{11} \overline{M}_1(x,-\overline{\zeta}_1^{\ast}) +  \beta_{12} \overline{M}_1(x,-\overline{\zeta}_2^{\ast})+  \beta_{13} \overline{M}_1(x,-\overline{\zeta}_3^{\ast}),  \\
        N_2^{\ast}(-x,\zeta_2)= &   1 + \beta_{21} \overline{M}_1(x,-\overline{\zeta}_1^{\ast}) + \beta_{22} \overline{M}_1(x,-\overline{\zeta}_2^{\ast})+  \beta_{23} \overline{M}_1(x,-\overline{\zeta}_3^{\ast}),  \\
                N_2^{\ast}(-x,\zeta_3)=  &  1 + \beta_{31} \overline{M}_1(x,-\overline{\zeta}_1^{\ast}) + \beta_{32} \overline{M}_1(x,-\overline{\zeta}_2^{\ast})+ \beta_{33} \overline{M}_1(x,-\overline{\zeta}_3^{\ast}),
    \end{aligned}
    \right.
\end{equation}
where
\begin{equation}
   \begin{aligned}
     \alpha_{11}=\frac{C_1^{\ast}(t)e^{2i\zeta_1^{\ast}x}}{\overline{\zeta}_1^{\ast}-\zeta_1^{\ast}},  & \qquad \alpha_{12}= \frac{C_2^{\ast}(t)e^{2i\zeta_2^{\ast}x}}{\overline{\zeta}_1^{\ast}-\zeta_2^{\ast}}, & \qquad \alpha_{13}= \frac{C_3^{\ast}(t)e^{2i\zeta_3^{\ast}x}}{\overline{\zeta}_1^{\ast}-\zeta_3^{\ast}}, \\
      \alpha_{21} =\frac{C_1^{\ast}(t)e^{2i\zeta_1^{\ast}x}}{\overline{\zeta}_2^{\ast}-\zeta_1^{\ast}},  & \qquad \alpha_{22}= \frac{C_2^{\ast}(t)e^{2i\zeta_2^{\ast}x}}{\overline{\zeta}_2^{\ast}-\zeta_2^{\ast}}, & \qquad \alpha_{23}= \frac{C_3^{\ast}(t)e^{2i\zeta_3^{\ast}x}}{\overline{\zeta}_2^{\ast}-\zeta_3^{\ast}},  \\
         \alpha_{31} =\frac{C_1^{\ast}(t)e^{2i\zeta_1^{\ast}x}}{\overline{\zeta}_3^{\ast}-\zeta_1^{\ast}},  & \qquad \alpha_{32}= \frac{C_2^{\ast}(t)e^{2i\zeta_2^{\ast}x}}{\overline{\zeta}_3^{\ast}-\zeta_2^{\ast}}, & \qquad \alpha_{33}= \frac{C_3^{\ast}(t)e^{2i\zeta_3^{\ast}x}}{\overline{\zeta}_3^{\ast}-\zeta_3^{\ast}},  \\
     \beta_{11} = \frac{\overline{C}_1^{\ast}(t)e^{-2i\overline{\zeta}_{1}^{\ast}x}}{\zeta_1^{\ast}-\overline{\zeta}_1^{\ast}}, & \qquad
     \beta_{12} =  \frac{\overline{C}_2^{\ast}(t)e^{-2i\overline{\zeta}_{2}^{\ast}x}}{\zeta_1^{\ast}-\overline{\zeta}_2^{\ast}}, & \qquad
      \beta_{13} =  \frac{\overline{C}_3^{\ast}(t)e^{-2i\overline{\zeta}_{3}^{\ast}x}}{\zeta_1^{\ast}-\overline{\zeta}_3^{\ast}},\\
      \beta_{21} = \frac{\overline{C}_1^{\ast}(t)e^{-2i\overline{\zeta}_{1}^{\ast}x}}{\zeta_2^{\ast}-\overline{\zeta}_1^{\ast}}, & \qquad
       \beta_{22} =  \frac{\overline{C}_2^{\ast}(t)e^{-2i\overline{\zeta}_{2}^{\ast}x}}{\zeta_2^{\ast}-\overline{\zeta}_2^{\ast}},  & \qquad
       \beta_{23} =  \frac{\overline{C}_3^{\ast}(t)e^{-2i\overline{\zeta}_{3}^{\ast}x}}{\zeta_2^{\ast}-\overline{\zeta}_3^{\ast}},    \\
             \beta_{31} = \frac{\overline{C}_1^{\ast}(t)e^{-2i\overline{\zeta}_{1}^{\ast}x}}{\zeta_3^{\ast}-\overline{\zeta}_1^{\ast}}, & \qquad
       \beta_{32} =  \frac{\overline{C}_2^{\ast}(t)e^{-2i\overline{\zeta}_{2}^{\ast}x}}{\zeta_3^{\ast}-\overline{\zeta}_2^{\ast}},  & \qquad
       \beta_{33} =  \frac{\overline{C}_3^{\ast}(t)e^{-2i\overline{\zeta}_{3}^{\ast}x}}{\zeta_3^{\ast}-\overline{\zeta}_3^{\ast}}.
   \end{aligned}
\end{equation}
Solving the above system, we  get
\begin{equation}
     N_2^{\ast}(-x,\zeta_j)= \frac{\det( A_j) }{ \det (A ) }, \quad j=1,2,3,
\end{equation}
where
\begin{equation}
 A=\begin{pmatrix}  \lambda_1 & \lambda_2 & \lambda_3 \\  \lambda_4 & \lambda_5 & \lambda_6 \\  \lambda_7 & \lambda_8 & \lambda_9   \end{pmatrix},
\end{equation}
and
\begin{equation}
     A_1 =  \begin{pmatrix}  1 & \lambda_2 & \lambda_3 \\  1 & \lambda_5 & \lambda_6 \\  1 & \lambda_8 & \lambda_9    \end{pmatrix},   \quad
        A_2=  \begin{pmatrix}  \lambda_1 & 1 & \lambda_3 \\  \lambda_4 & 1 & \lambda_6 \\  \lambda_7 & 1 & \lambda_9  \end{pmatrix},   \quad
        A_3=\begin{pmatrix}   \lambda_1 & \lambda_2 & 1 \\  \lambda_4 & \lambda_5 & 1 \\  \lambda_7 & \lambda_8 & 1   \end{pmatrix},
\end{equation}
with
\begin{equation}
\left\{
  \begin{aligned}
  \lambda_1 = & 1 - \alpha_{11}\beta_{11} - \alpha_{21}\beta_{12} - \alpha_{31}\beta_{13},\\
  \lambda_2 = & - \alpha_{12}\beta_{11} - \alpha_{22}\beta_{12} - \alpha_{32}\beta_{13},  \\
  \lambda_3 = & - \alpha_{13}\beta_{11} - \alpha_{23}\beta_{12} - \alpha_{33}\beta_{13}, \\
  \lambda_4 = &  - \alpha_{11}\beta_{21} - \alpha_{21}\beta_{22} - \alpha_{31}\beta_{23},    \\
  \lambda_5 = & 1 - \alpha_{12}\beta_{21} - \alpha_{22}\beta_{22} - \alpha_{32}\beta_{23},    \\
  \lambda_6 = &  - \alpha_{13}\beta_{21} - \alpha_{23}\beta_{22} - \alpha_{33}\beta_{23},    \\
  \lambda_7 = &  - \alpha_{11}\beta_{31} - \alpha_{21}\beta_{32} - \alpha_{31}\beta_{33},    \\
  \lambda_8 = &   - \alpha_{12}\beta_{31} - \alpha_{22}\beta_{32} - \alpha_{32}\beta_{33},   \\
  \lambda_9 = &  1 - \alpha_{13}\beta_{31} - \alpha_{23}\beta_{32} - \alpha_{33}\beta_{33}.   \\
  \end{aligned}
  \right.
\end{equation}
Substituting the above equations into Eq. \eqref{three}, we can obtain the formula of  three-soliton solutions. \\

\noindent
{\rotatebox{0}{\includegraphics[width=3.6cm,height=3.2cm,angle=0]{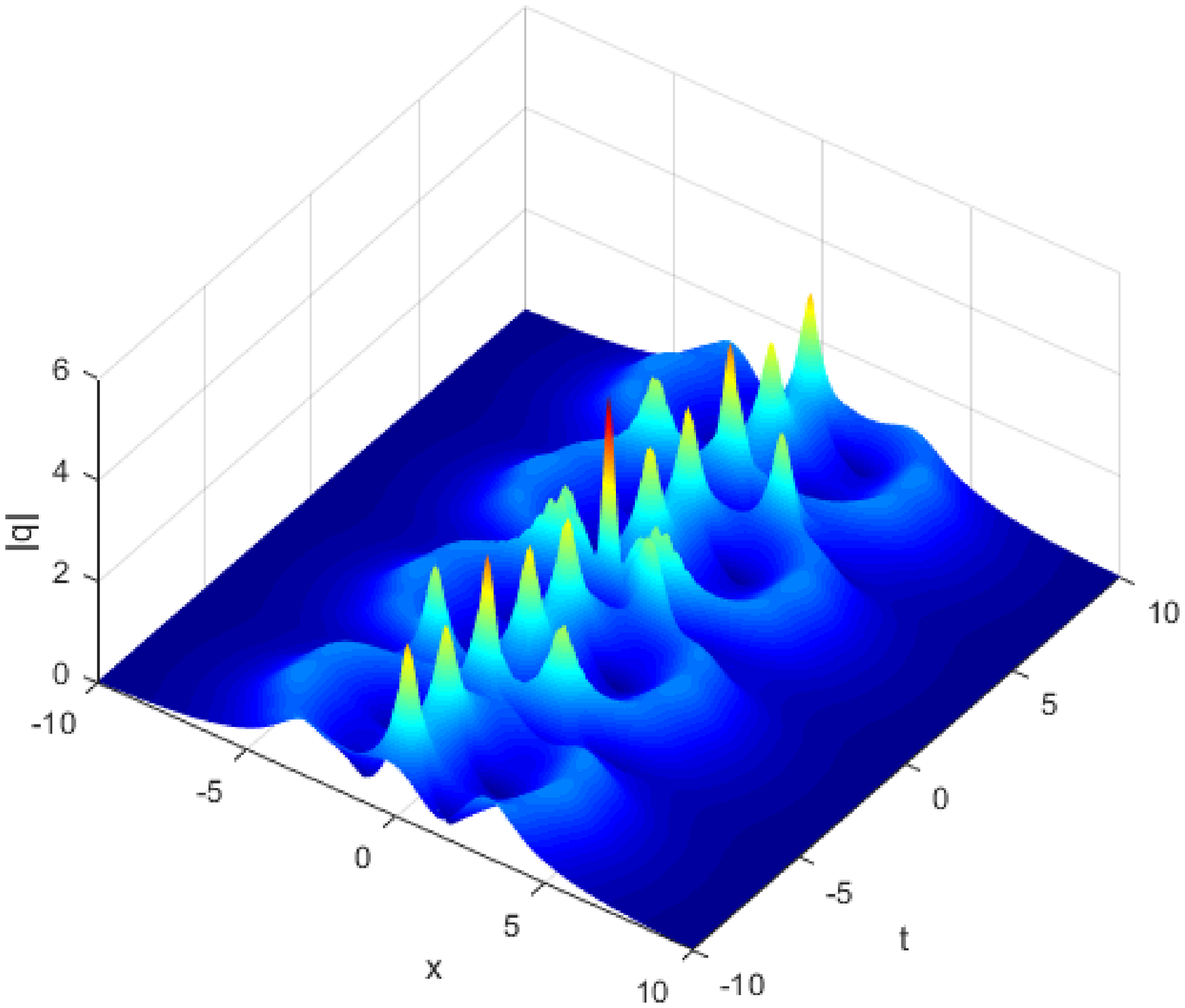}}}
~~~~
{\rotatebox{0}{\includegraphics[width=3.6cm,height=3.2cm,angle=0]{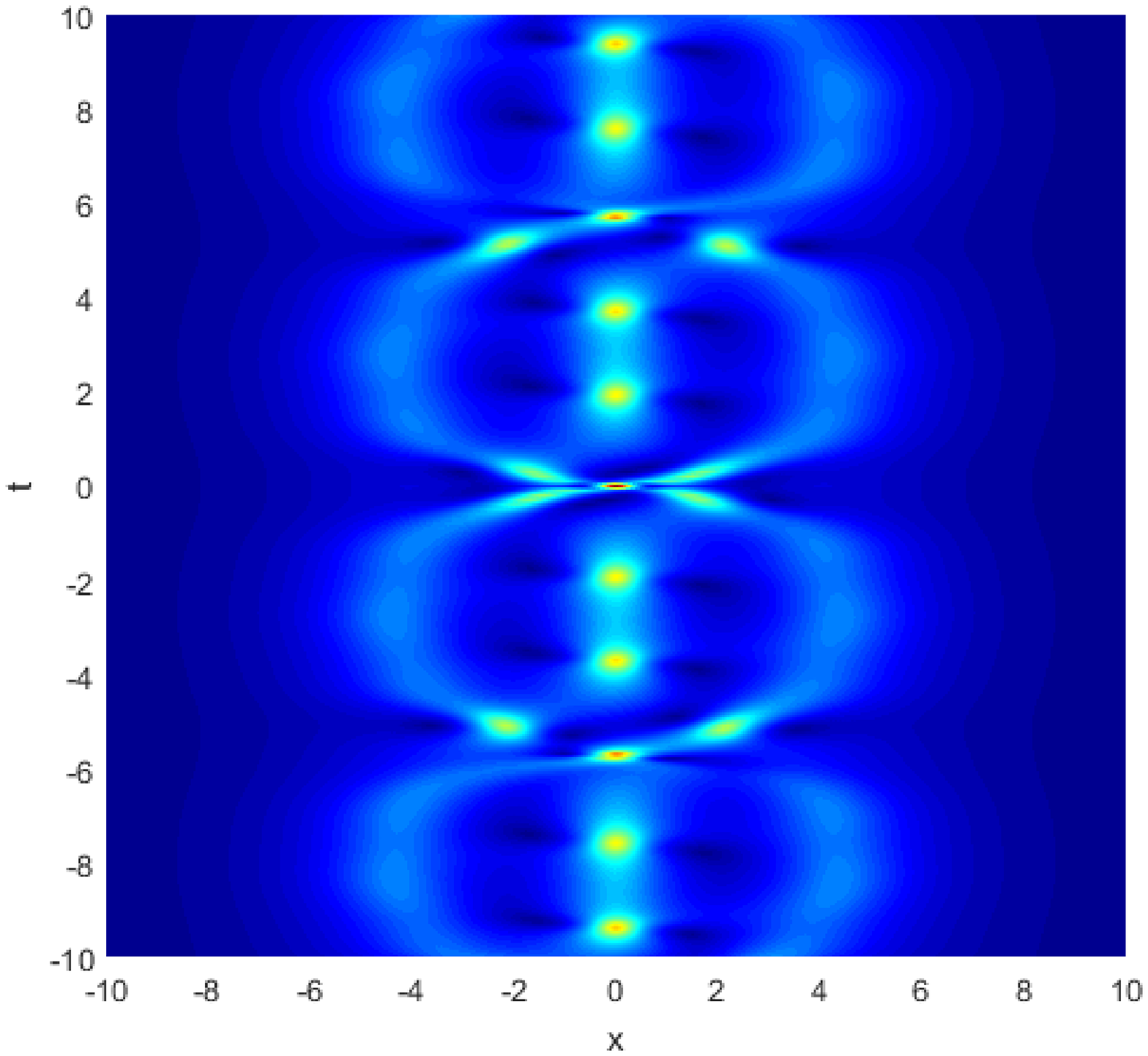}}}
~~~~
{\rotatebox{0}{\includegraphics[width=3.6cm,height=3.2cm,angle=0]{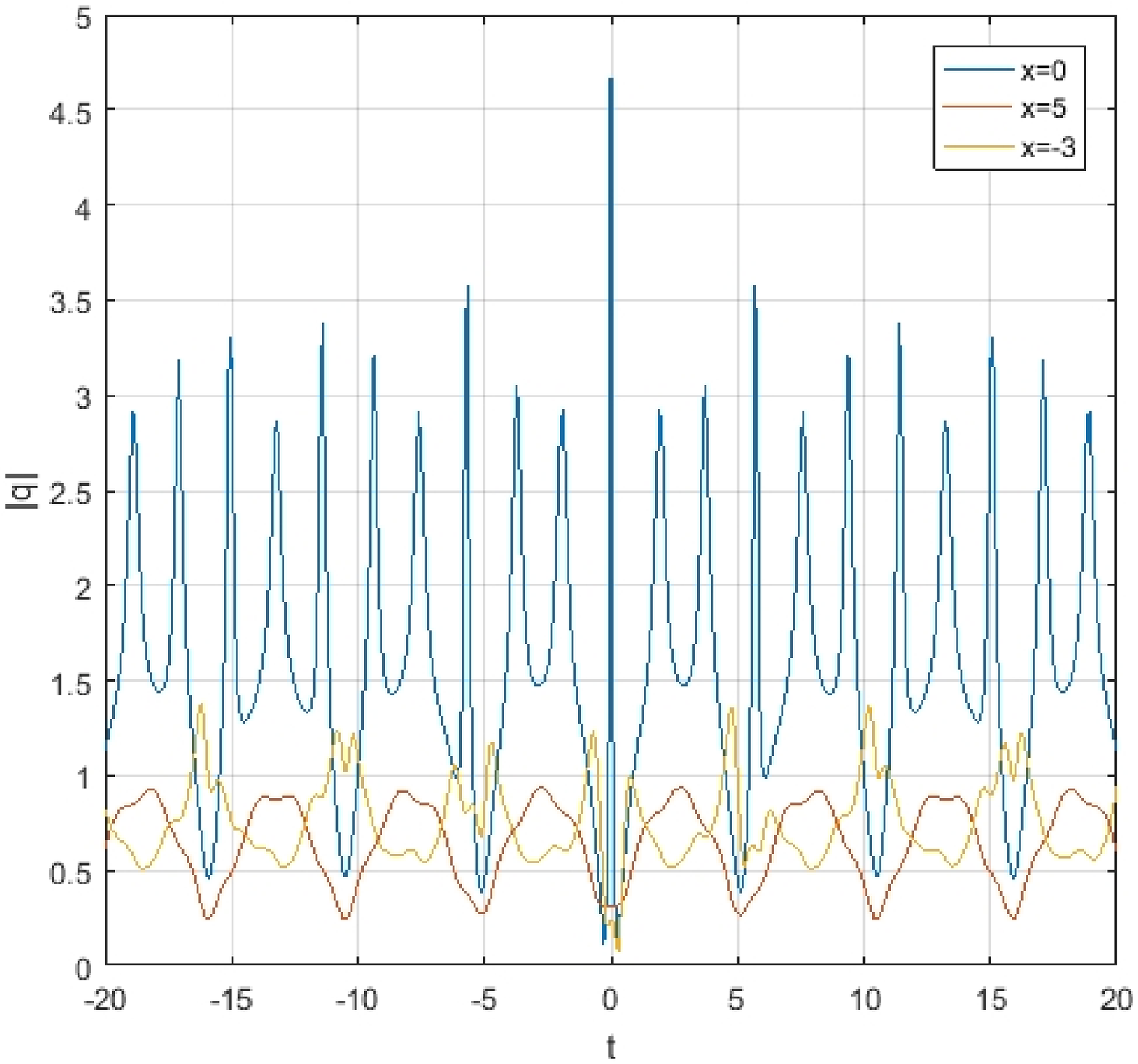}}}

$\ \qquad~~~~~~(\textbf{a})\qquad \ \qquad\qquad\qquad\qquad~(\textbf{b})
\ \qquad\qquad\qquad\qquad\qquad~(\textbf{c})$\\
\noindent
{\rotatebox{0}{\includegraphics[width=3.6cm,height=3.2cm,angle=0]{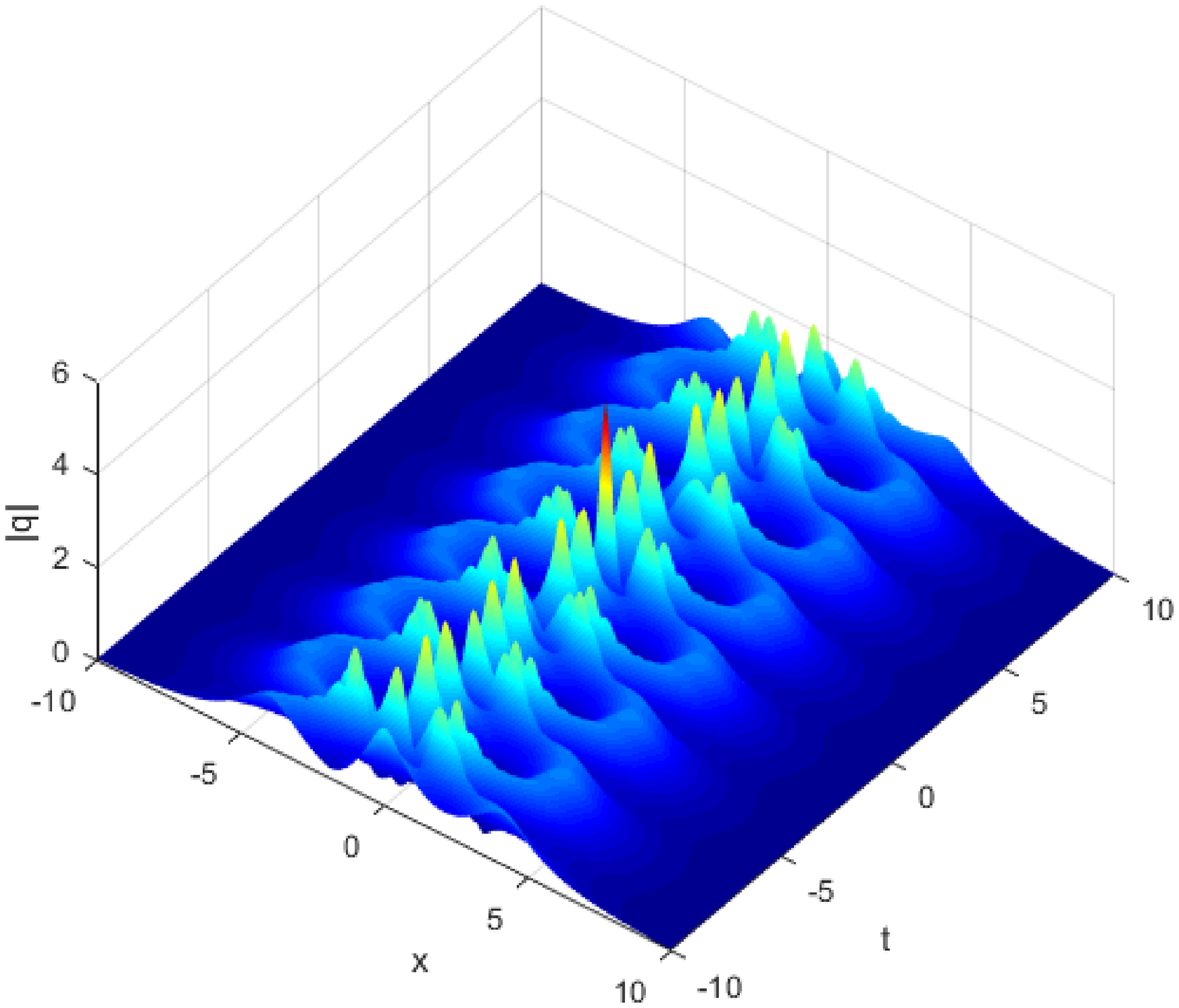}}}
~~~~
{\rotatebox{0}{\includegraphics[width=3.6cm,height=3.2cm,angle=0]{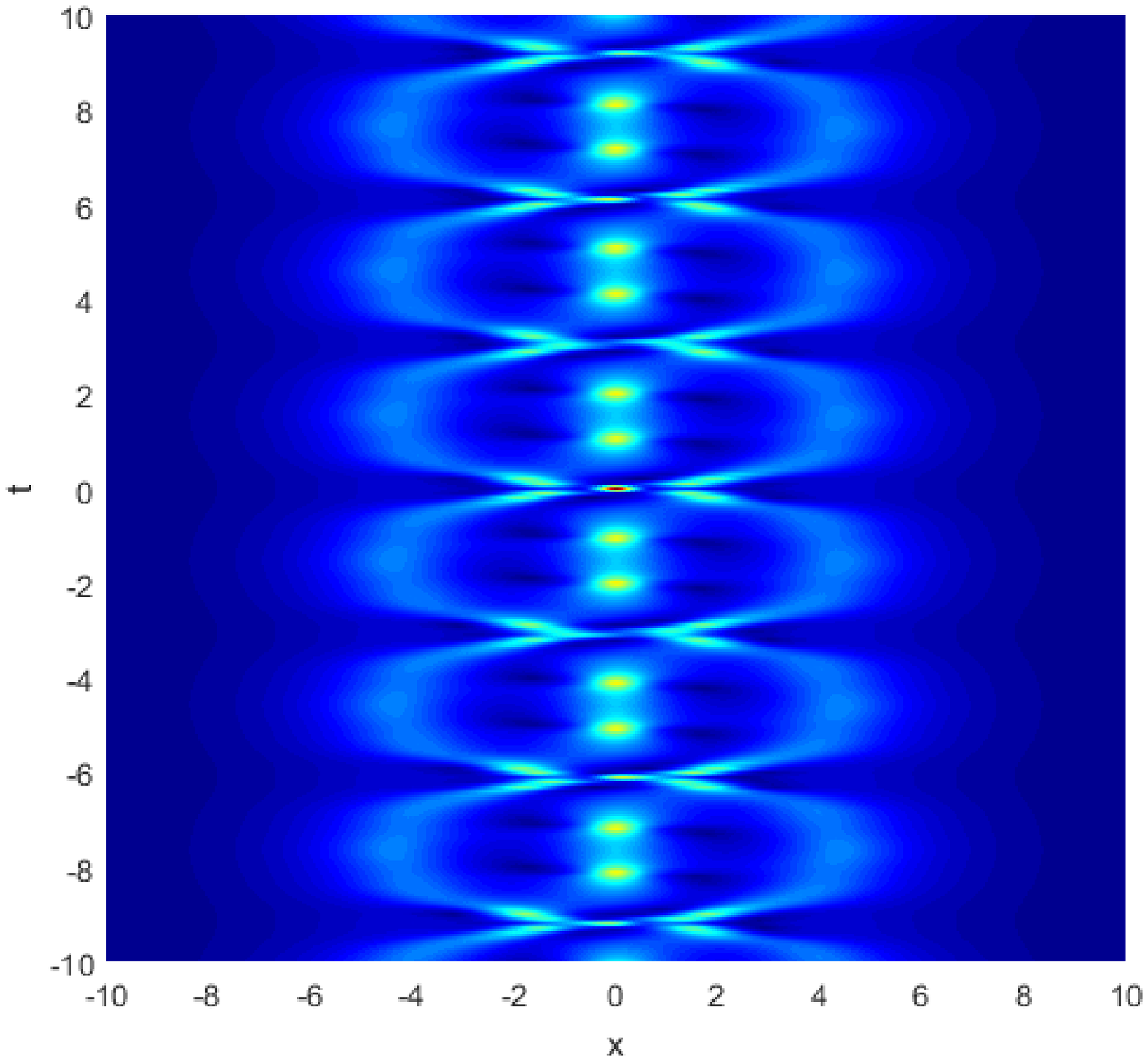}}}
~~~~
{\rotatebox{0}{\includegraphics[width=3.6cm,height=3.2cm,angle=0]{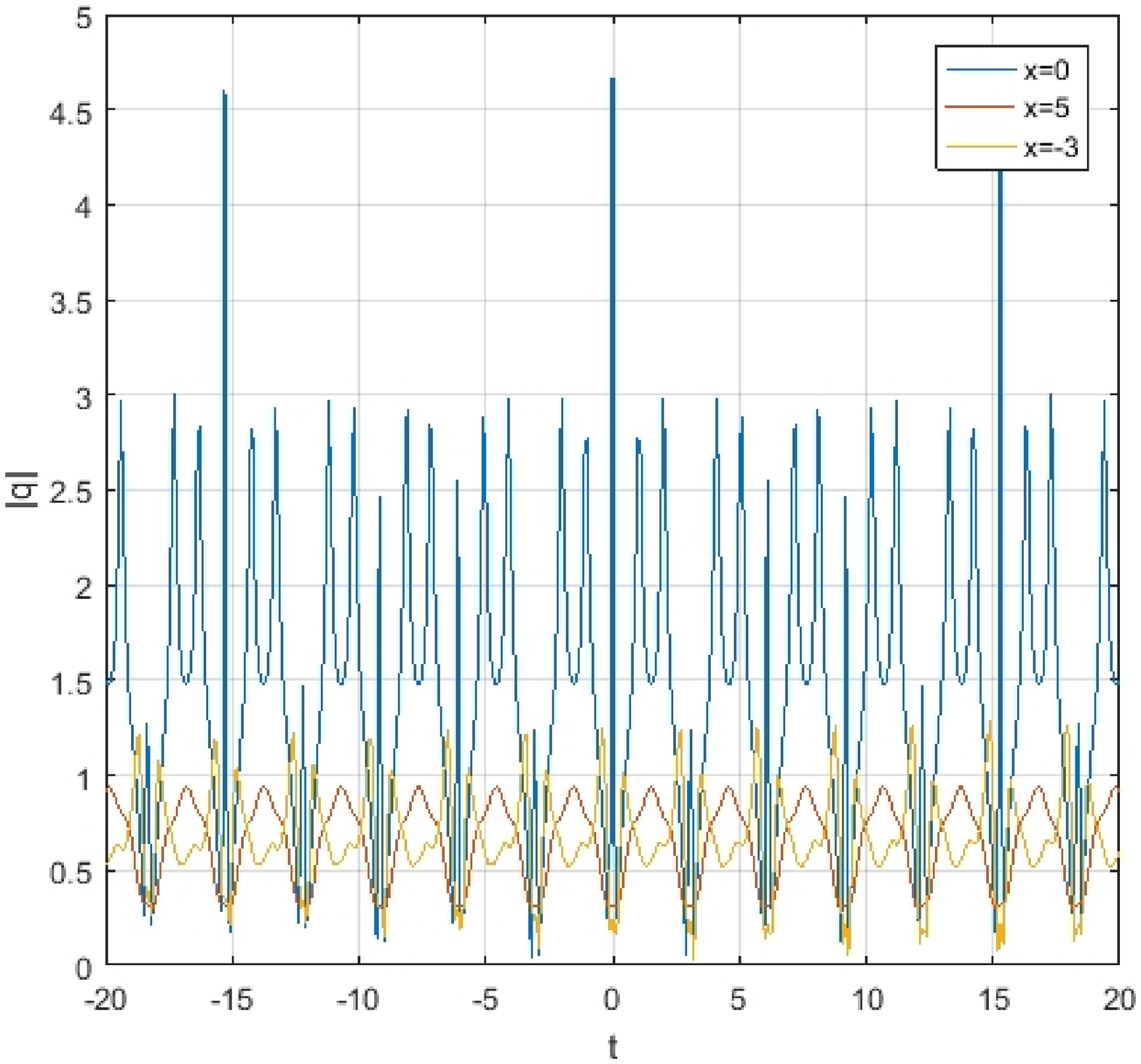}}}

$\ \qquad~~~~~~(\textbf{d})\qquad \ \qquad\qquad\qquad\qquad~(\textbf{e})
\ \qquad\qquad\qquad\qquad\qquad~(\textbf{f})$\\
\noindent
{\rotatebox{0}{\includegraphics[width=3.6cm,height=3.2cm,angle=0]{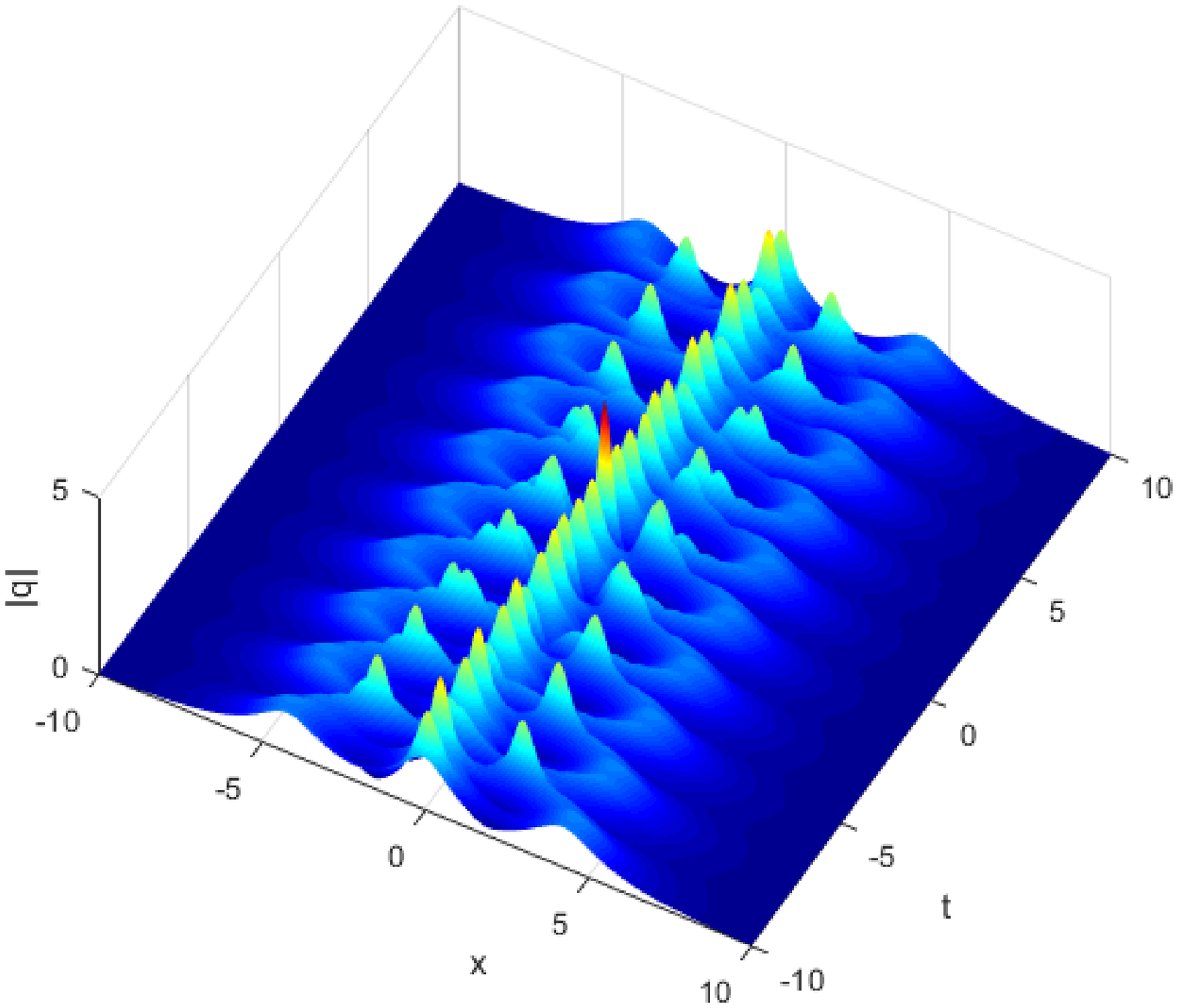}}}
~~~~
{\rotatebox{0}{\includegraphics[width=3.6cm,height=3.2cm,angle=0]{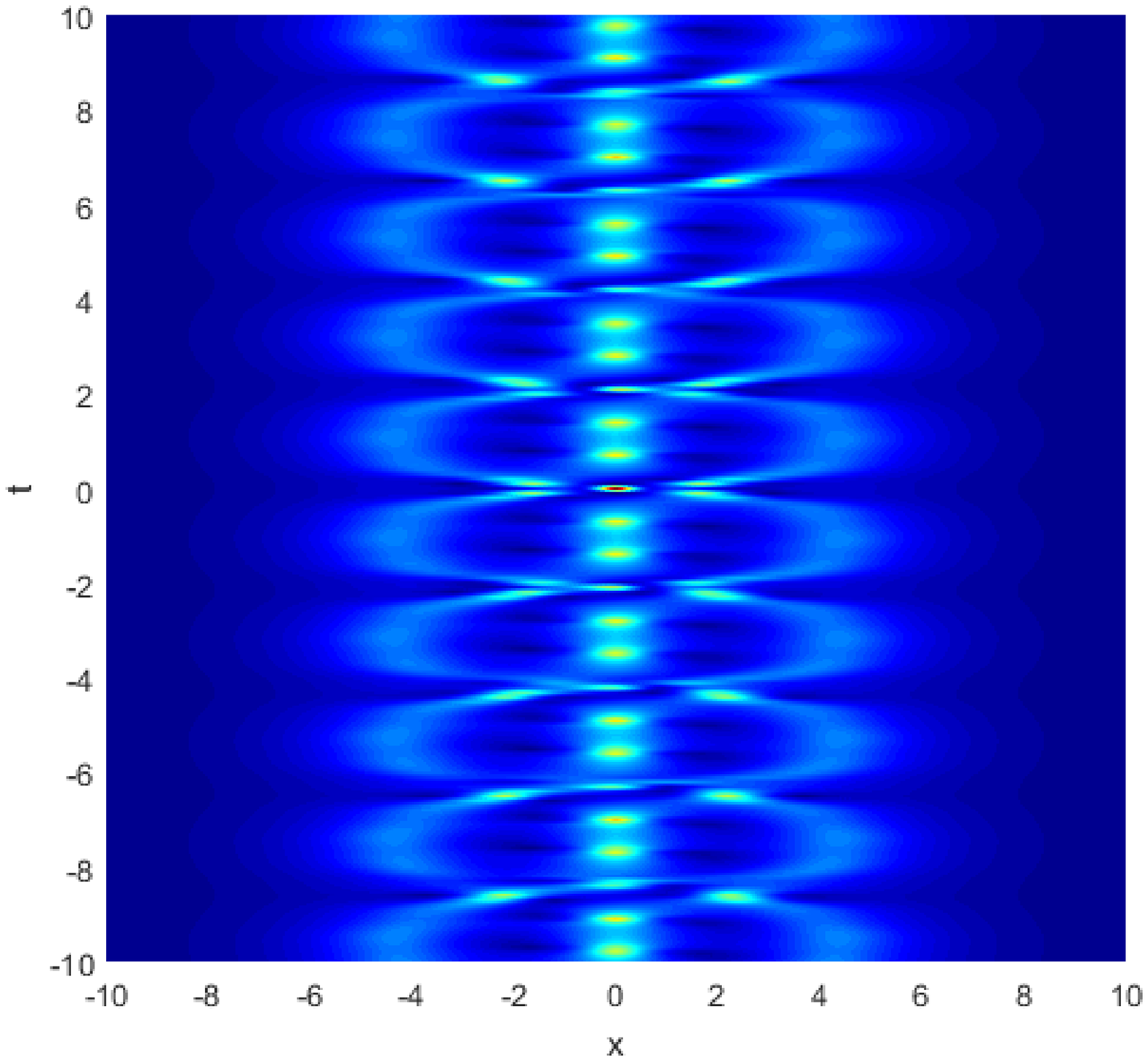}}}
~~~~
{\rotatebox{0}{\includegraphics[width=3.6cm,height=3.2cm,angle=0]{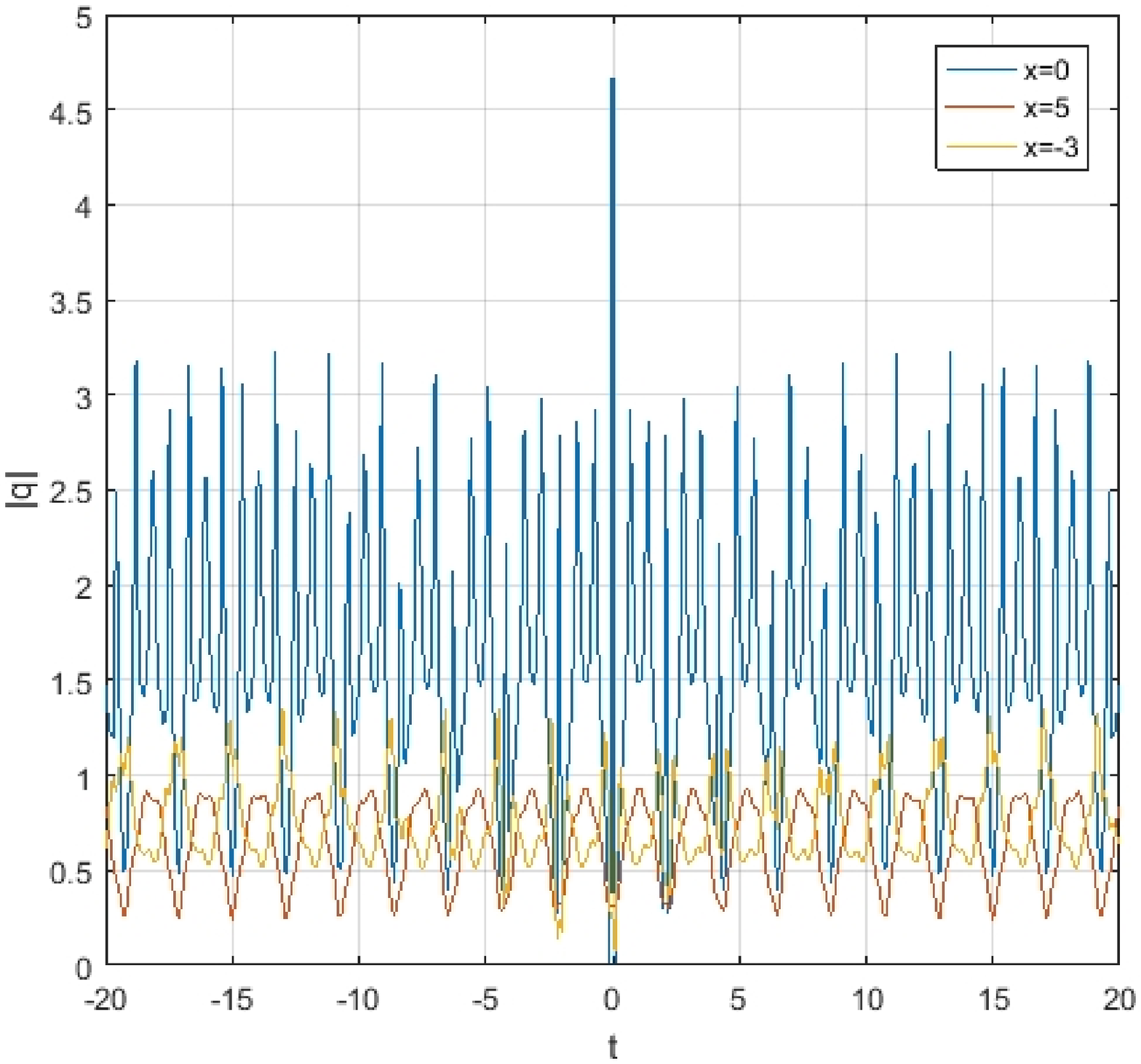}}}

$\ \qquad~~~~~~(\textbf{g})\qquad \ \qquad\qquad\qquad\qquad~(\textbf{h})
\ \qquad\qquad\qquad\qquad\qquad~(\textbf{i})$\\
\noindent { \small \textbf{Figure 4.} Three-soliton  solutions   with parameters $\theta_1=\theta_2=\theta_3=\frac{\pi}{3}$,  $\overline{\theta}_1=\overline{\theta}_2=\overline{\theta}_3=\frac{\pi}{9}$,  $\zeta_1=0.3+0.5i$,  $\zeta_2=-0.3+0.5i$, $\zeta_3=0.6i$,  $\overline{\zeta}_1=0.3-0.5i$,  $\overline{\zeta}_2=-0.3-0.5i$  and  $\overline{\zeta}_3=-0.6i$.
$\textbf{(a)(b)(c)}$: the structures and the wave propagation of the three-soliton  solutions with $\delta=0.5$,
$\textbf{(d)(e)(f)}$: the structures and the wave propagation of the three-soliton  solutions with $\delta=1$,
$\textbf{(g)(h)(i)}$: the structures and the wave propagation of the three-soliton  solutions with $\delta=2$.}  \\

In Fig. 4,  the local structure, the density and the wave propagation of three soliton solutions are shown vividly.   Different from the previous three solitons,   the three solitons here are composed of  two arc solitons on both sides  and one breathe-type soliton in the middle.  The two arc   solitons propagate forward along the left and right half of the circumference respectively, while the breathing solitons propagate forward along the diameter of the circumference, and three solitons  meet, collide elastically, and move away at the central diameter of the circumference periodically.  There is another obvious point that  by change the value of $\delta$.  The period of three soliton solutions  have changed significantly. Specifically, the period  will be shortened as the parameter $\delta$ increases which can be observed clearly form  the graphics.

 Different form Fig. 4,  the following Fig. 5  shows that the local structure and the dynamic behavior of  three ordinary solitons. The three solitons propagation along three different to the center($x=0, t=0$) and they meet at the center point. Then the three solitons collide elastically and move away along three different directions. During the whole process,  the amplitude, energy of three solitons are not changed.  It is also worth noting that the change of parameter $\delta$ has an influence on the rebound angle of two solitons on both sides. \\

\noindent
{\rotatebox{0}{\includegraphics[width=3.6cm,height=3.2cm,angle=0]{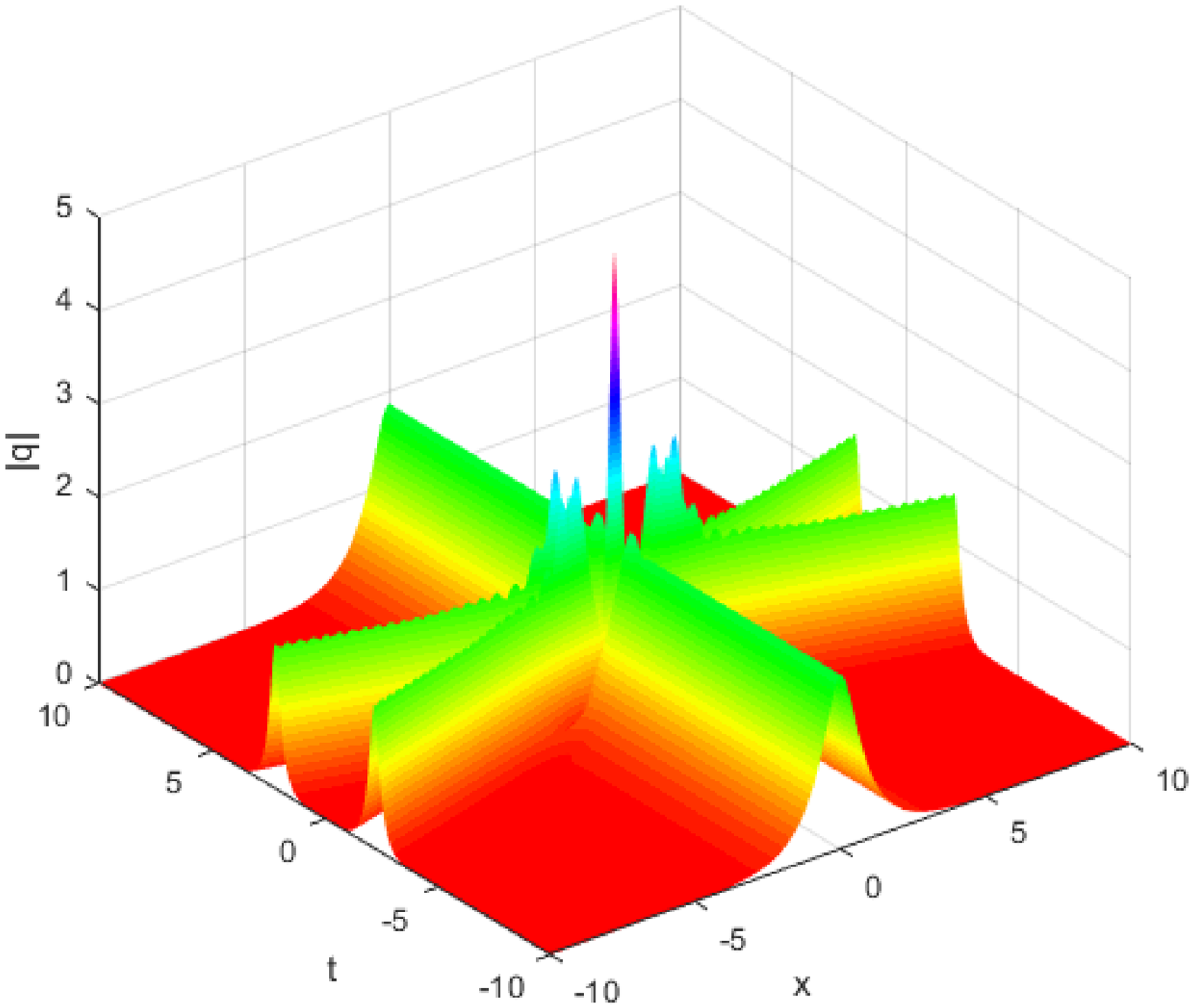}}}
~~~~
{\rotatebox{0}{\includegraphics[width=3.6cm,height=3.2cm,angle=0]{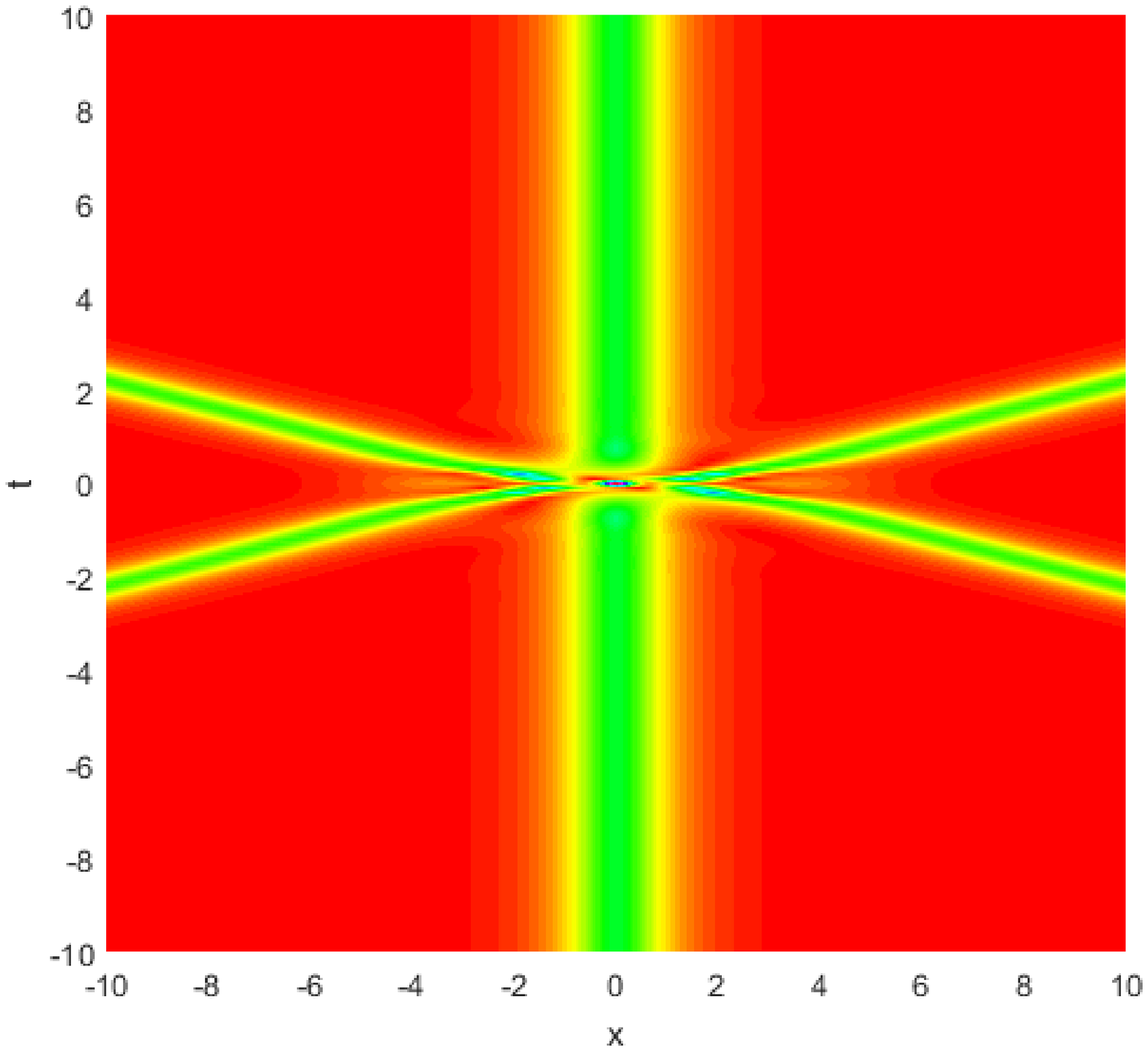}}}
~~~~
{\rotatebox{0}{\includegraphics[width=3.6cm,height=3.2cm,angle=0]{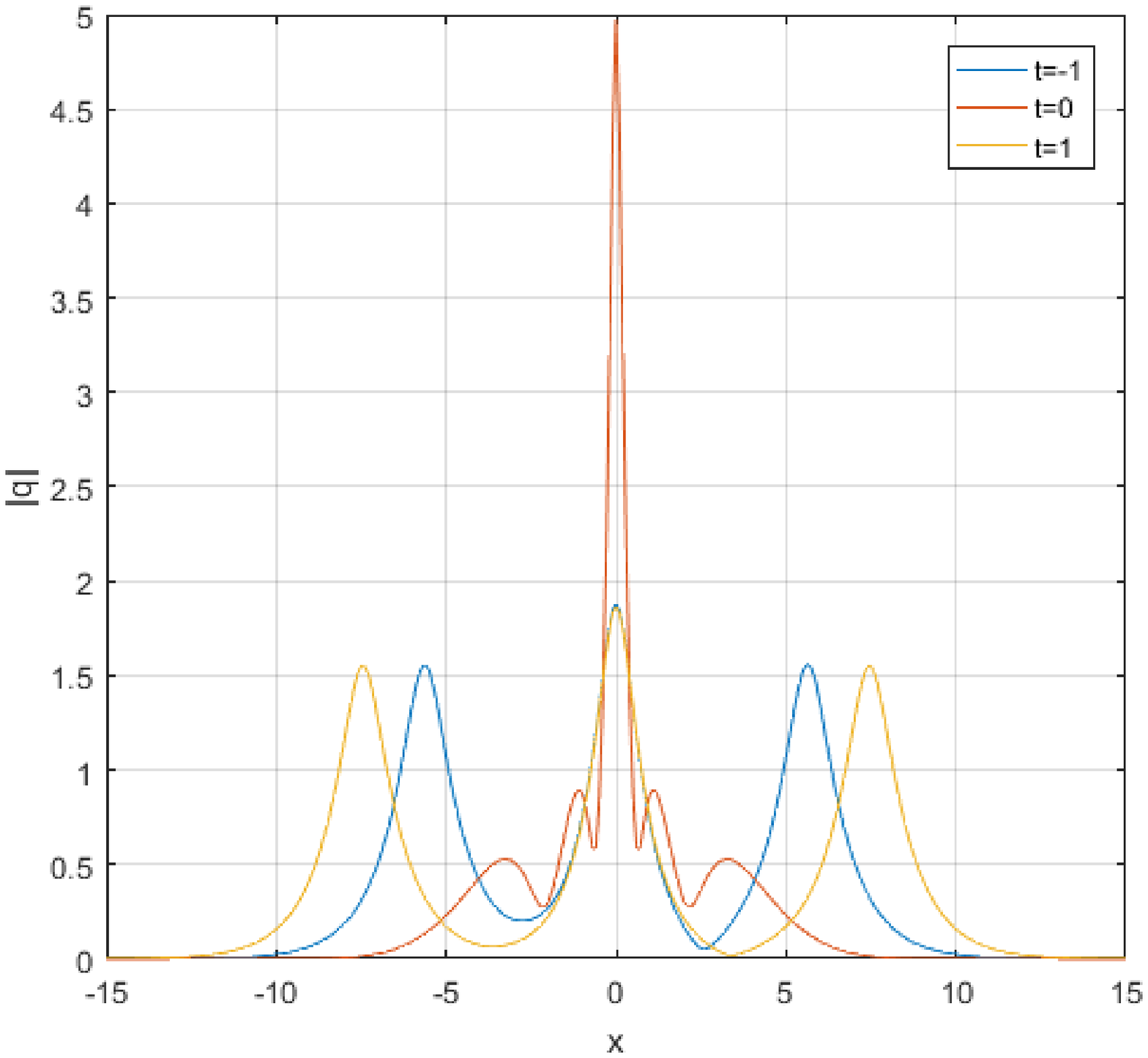}}}

$\ \qquad~~~~~~(\textbf{a})\qquad \ \qquad\qquad\qquad\qquad~(\textbf{b})
\ \qquad\qquad\qquad\qquad\qquad~(\textbf{c})$\\
\noindent
{\rotatebox{0}{\includegraphics[width=3.6cm,height=3.0cm,angle=0]{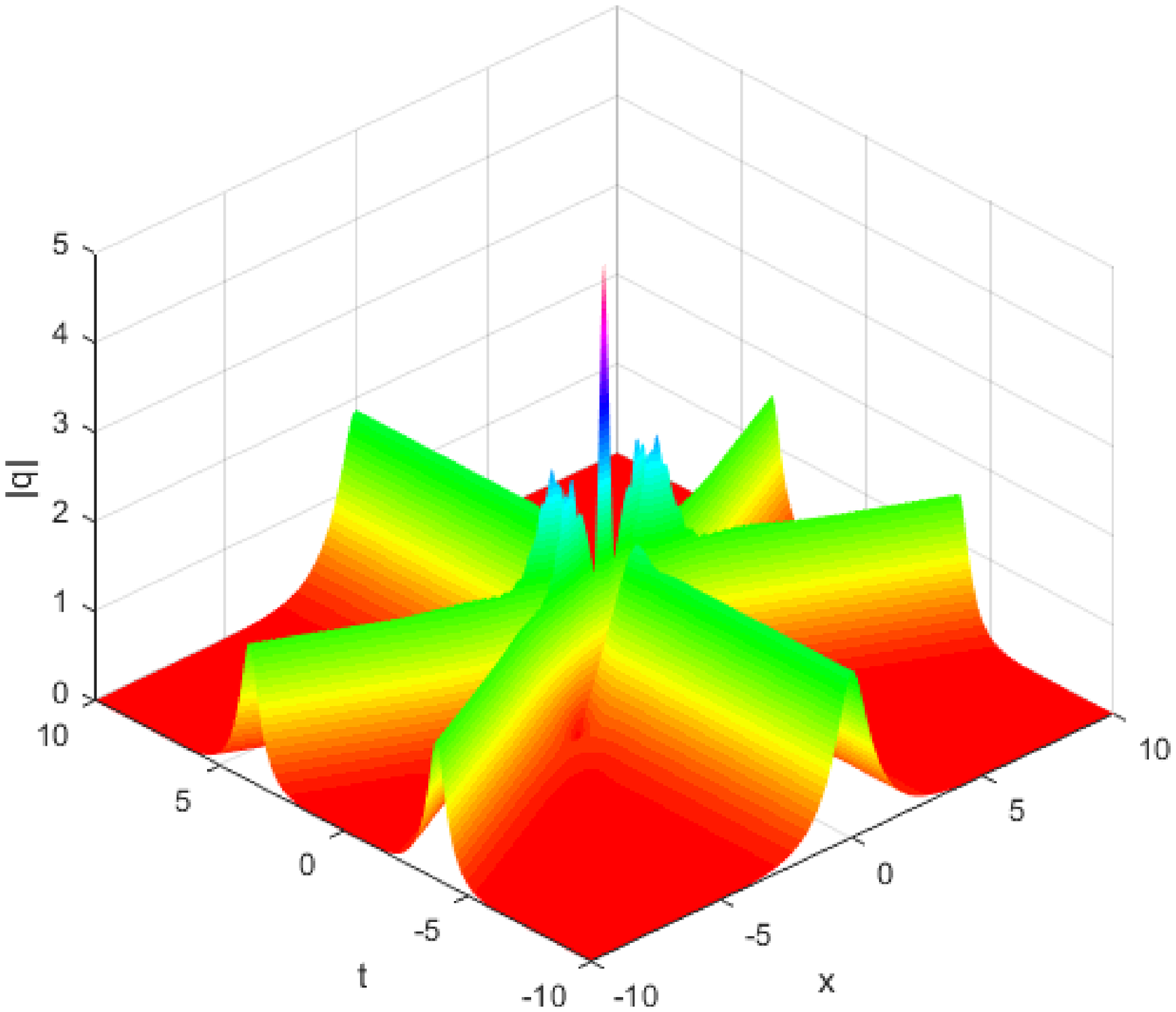}}}
~~~~
{\rotatebox{0}{\includegraphics[width=3.6cm,height=3.0cm,angle=0]{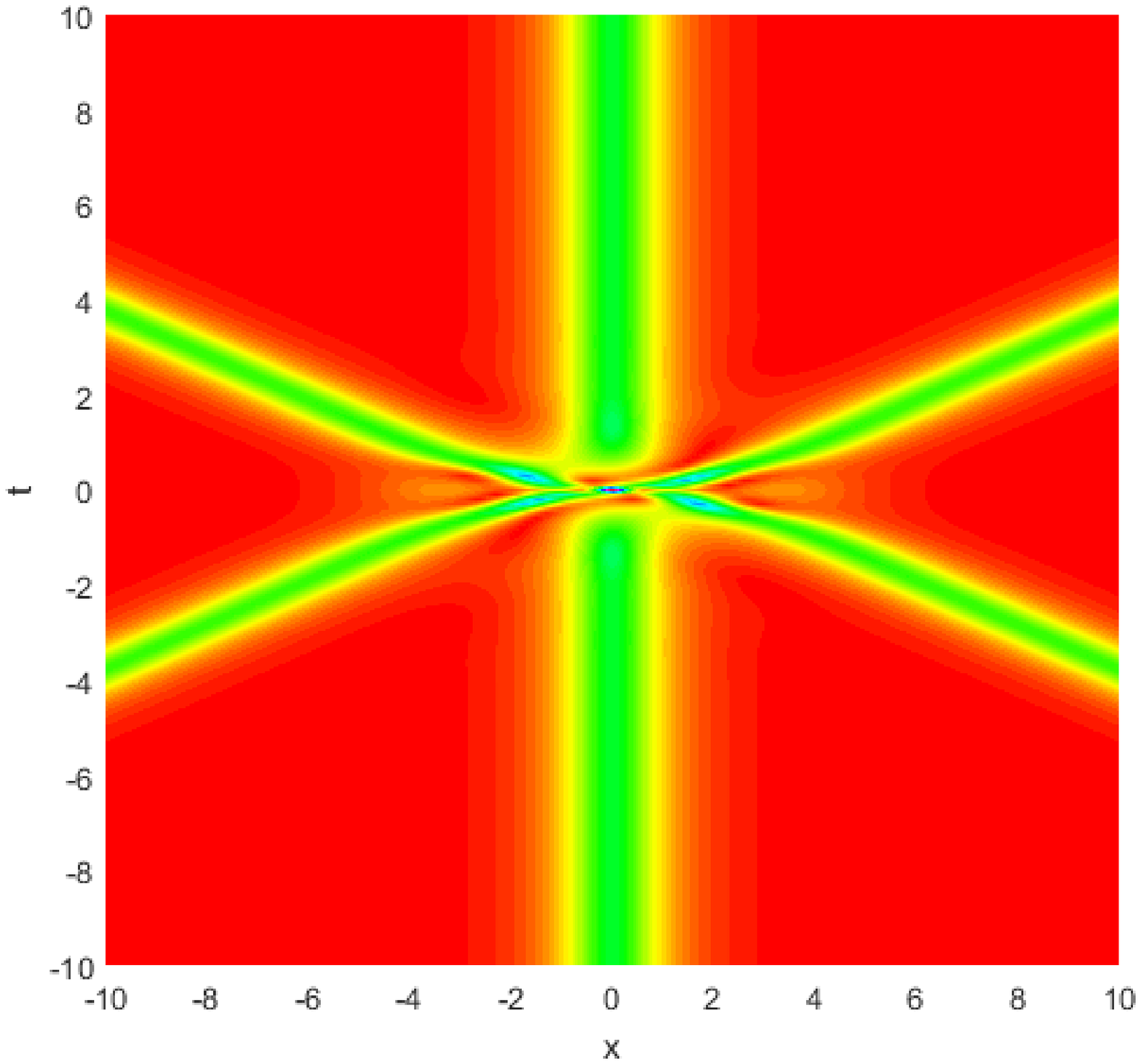}}}
~~~~
{\rotatebox{0}{\includegraphics[width=3.6cm,height=3.0cm,angle=0]{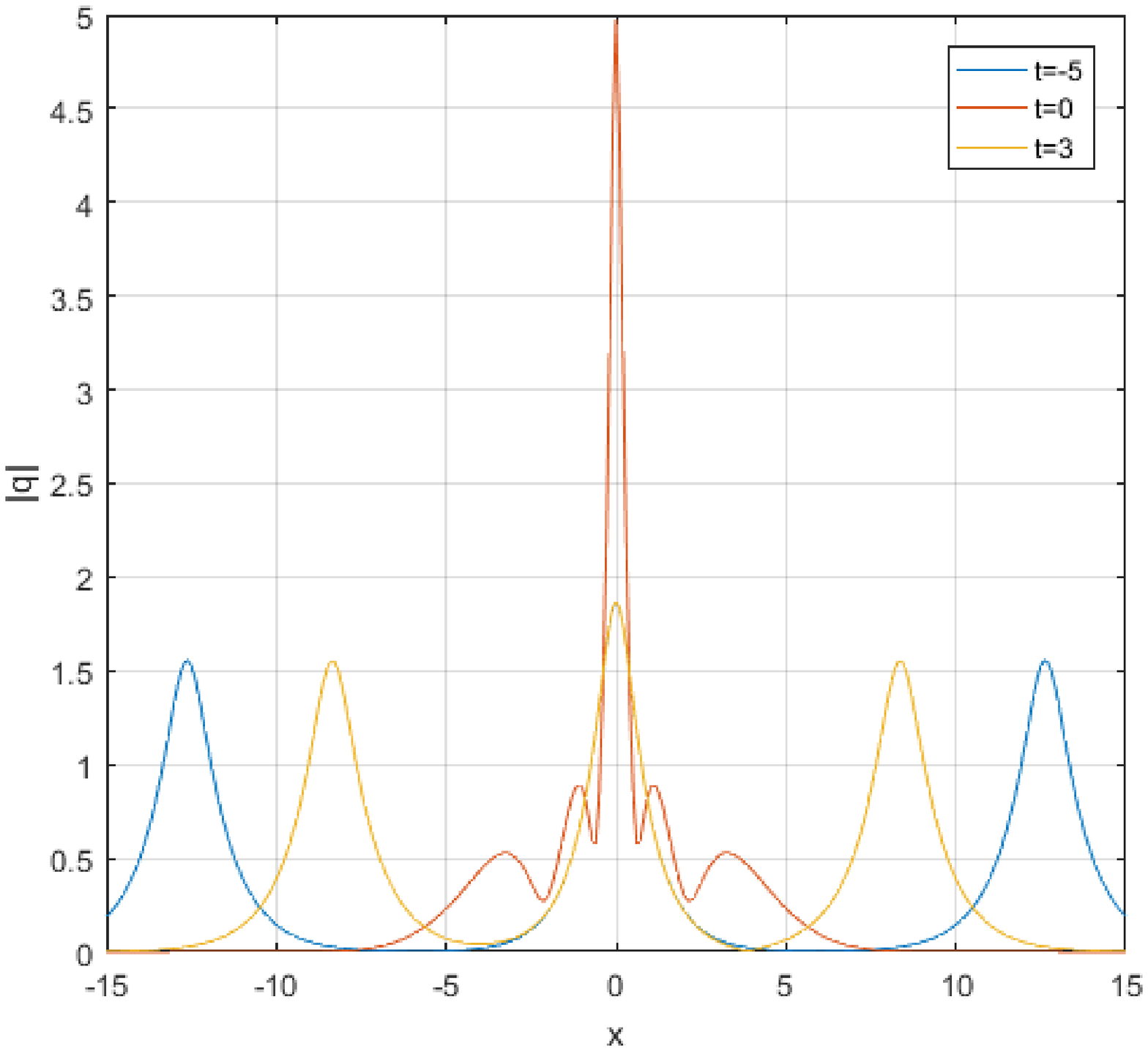}}}

$\ \qquad~~~~~~(\textbf{d})\qquad \ \qquad\qquad\qquad\qquad~(\textbf{e})
\ \qquad\qquad\qquad\qquad\qquad~(\textbf{f})$\\
\noindent
{\rotatebox{0}{\includegraphics[width=3.6cm,height=3.0cm,angle=0]{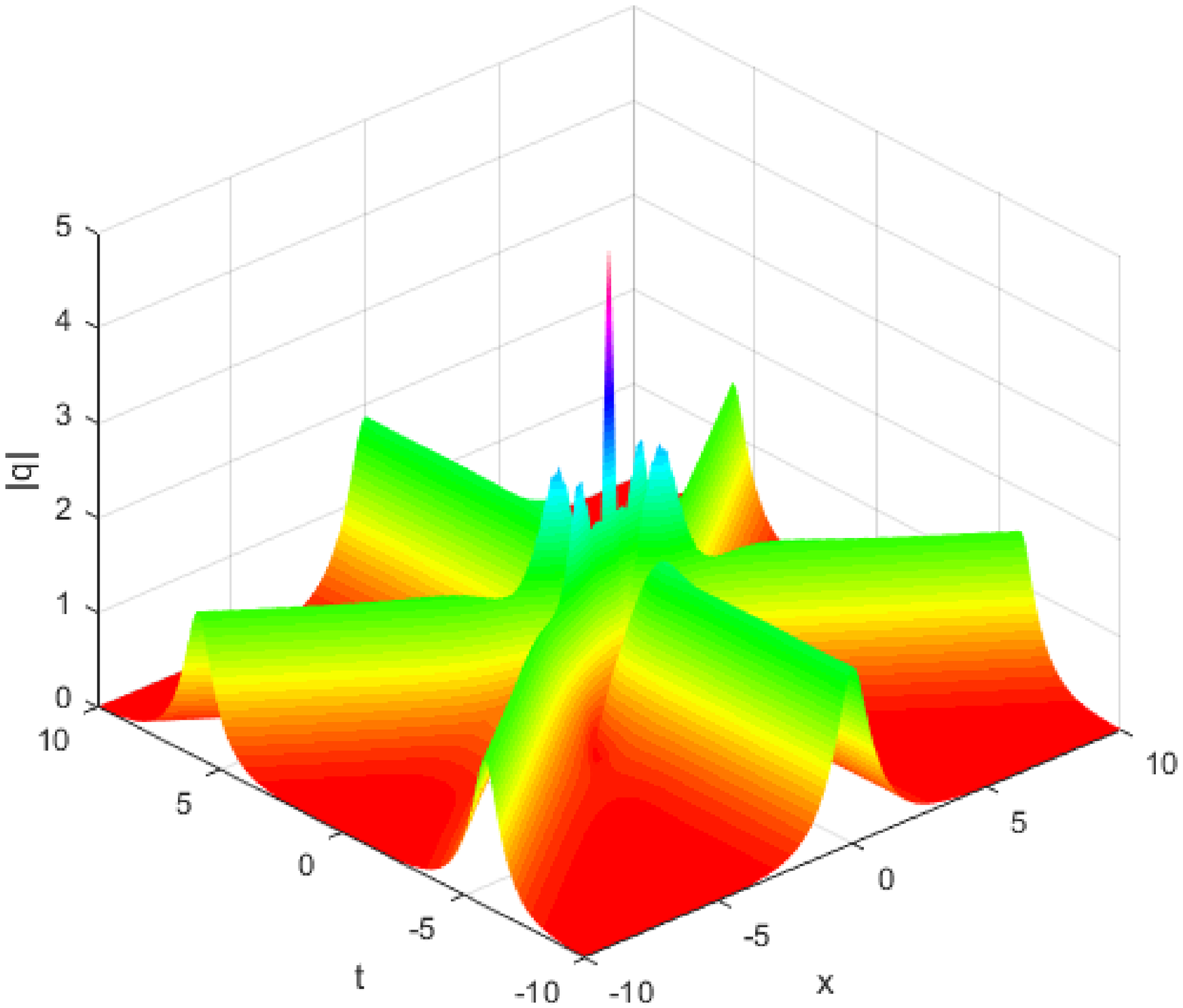}}}
~~~~
{\rotatebox{0}{\includegraphics[width=3.6cm,height=3.0cm,angle=0]{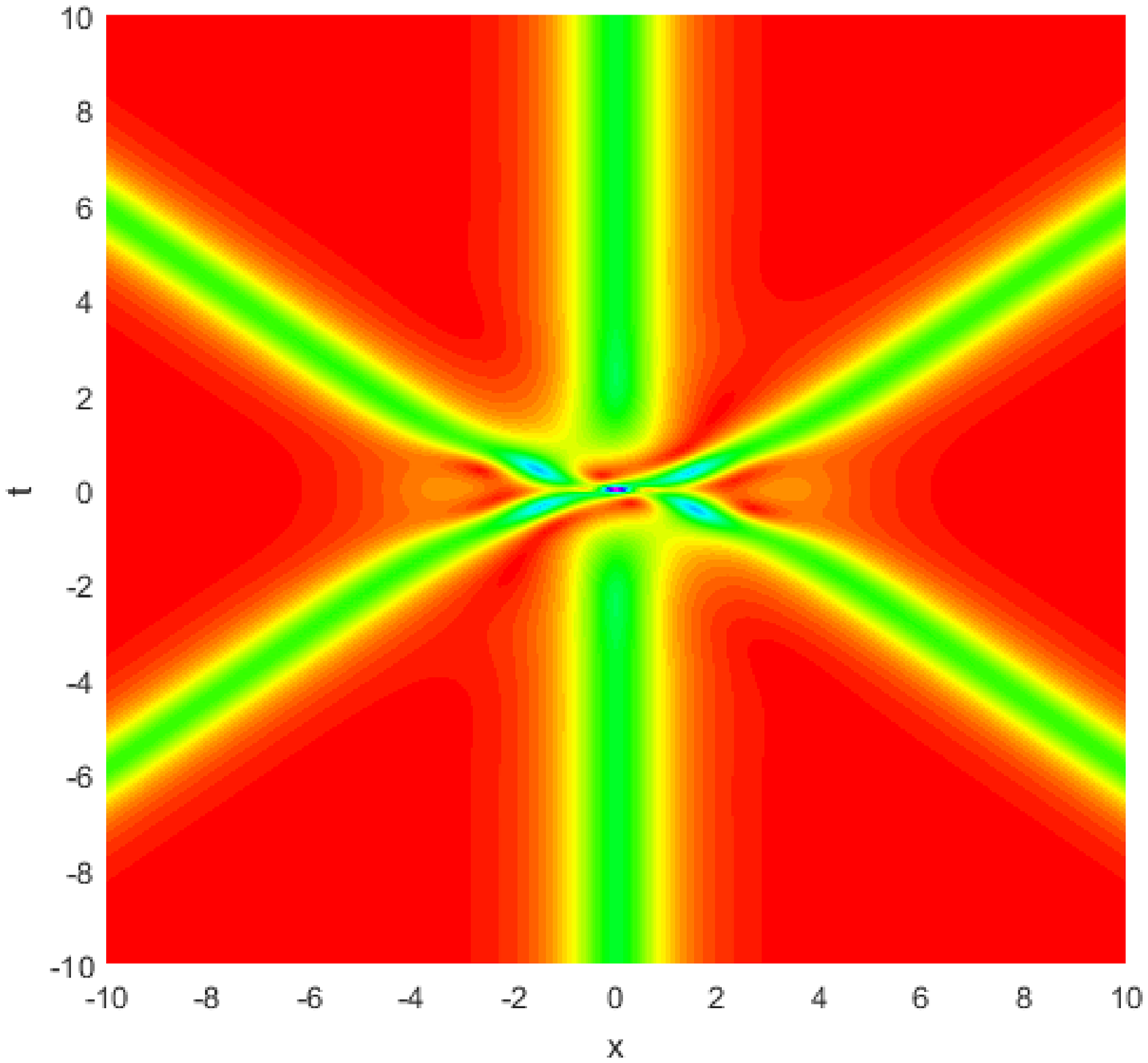}}}
~~~~
{\rotatebox{0}{\includegraphics[width=3.6cm,height=3.0cm,angle=0]{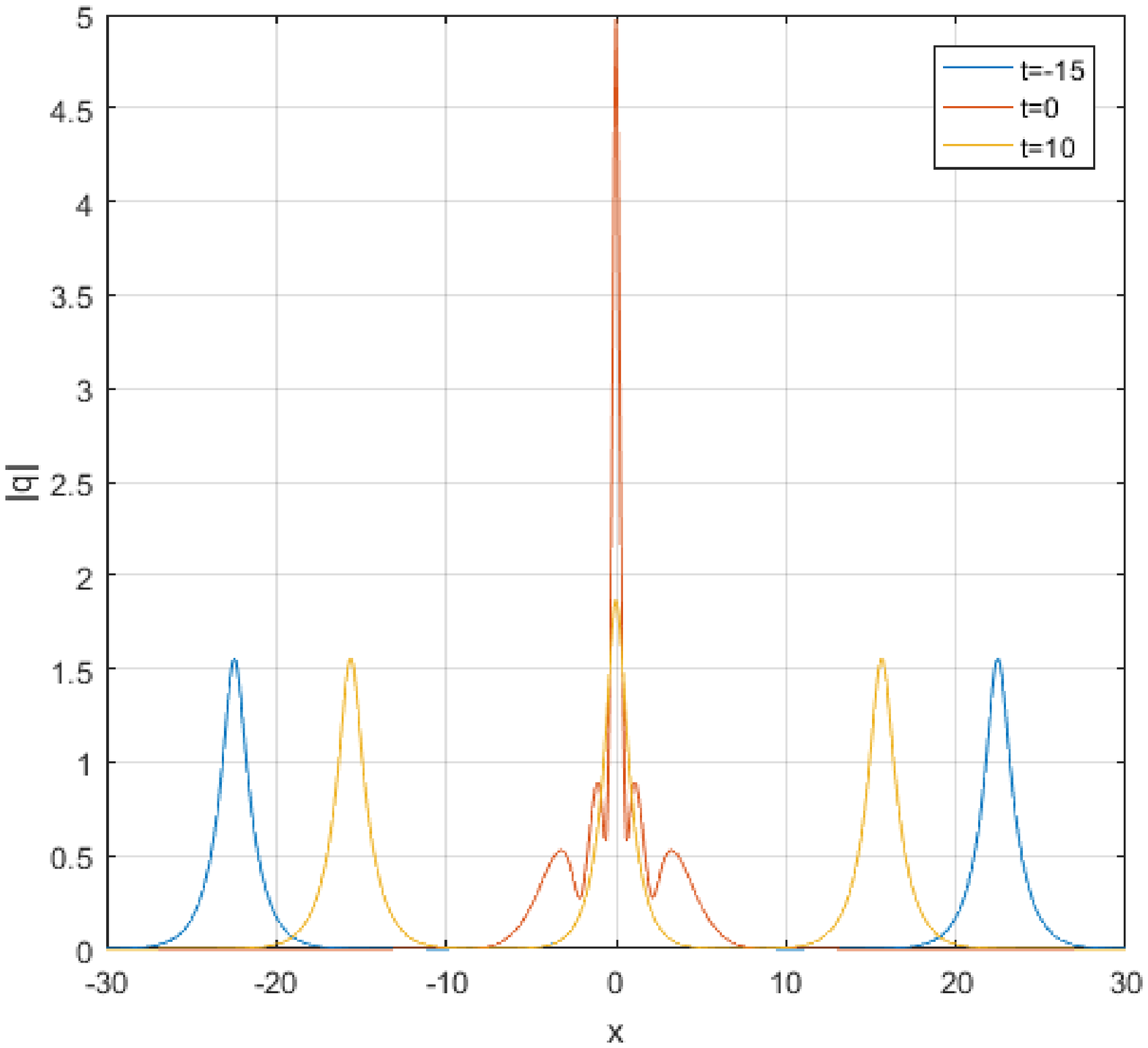}}}

$\ \qquad~~~~~~(\textbf{g})\qquad \ \qquad\qquad\qquad\qquad~(\textbf{h})
\ \qquad\qquad\qquad\qquad\qquad~(\textbf{i})$\\
\noindent { \small \textbf{Figure  5.} Three-soliton  solutions   with parameters $\theta_1=\theta_2=\theta_3=\frac{\pi}{3}$,  $\overline{\theta}_1=\overline{\theta}_2=\overline{\theta}_3=\frac{\pi}{9}$,  $\zeta_1=0.3i$,  $\zeta_2=0.5i$, $\zeta_3=0.7i$,  $\overline{\zeta}_1=-0.3i$,  $\overline{\zeta}_2=-0.5i$  and  $\overline{\zeta}_3=-0.7i$.
$\textbf{(a)(b)(c)}$: the structures and the wave propagation of the three-soliton  solutions with $\delta=3$,
$\textbf{(d)(e)(f)}$: the structures and the wave propagation of the three-soliton  solutions with $\delta=2$,
$\textbf{(g)(h)(i)}$: the structures and the wave propagation of the three-soliton  solutions with $\delta=1$.}  \\

\subsection{Four soliton solutions}
In this section,  we consider the four-soliton solutions of the nonlocal LPD equations \eqref{LPD}. Suppose the corresponding eigenvalues as follows
\begin{equation}
   \begin{aligned}
        \zeta_1 =\xi_1 + i \eta_1, &~ \zeta_2 =\xi_2+ i \eta_2, & ~ \zeta_3 =\xi_3 + i \eta_3, & ~ \zeta_4 =\xi_4 + i \eta_4, & ~ \eta_1, \eta_2, \eta_3, \eta_4 >0,  \\
        \overline{ \zeta}_1 =\overline{\xi}_1 + i \overline{\eta}_1, & ~ \overline{\zeta}_2 =\overline{\xi}_2+ i \overline{\eta}_2,  &~ \overline{\zeta}_3 =\overline{\xi}_3+ i \overline{\eta}_3,  &~ \overline{\zeta}_4 =\overline{\xi}_4+ i \overline{\eta}_4,   &~ \overline{\eta}_1, \overline{\eta}_2, \overline{\eta}_3, \overline{\eta}_4< 0. \\
   \end{aligned}
\end{equation}
Setting $J=\overline{J}=4$ into Eq. \eqref{soliton}, we find
\begin{equation}\label{four}
 \begin{aligned}
   q(x)=  & -2i C_{1}^{\ast} N_2^{\ast}(-x,\zeta_1)e^{2i\zeta_1^\ast  x} -2 i C_{2}^{\ast} N_2^{\ast}(-x,\zeta_2)e^{2i \zeta_2^\ast x}-2 i C_{3}^{\ast} N_2^{\ast}(-x,\zeta_3)e^{2i \zeta_3^\ast x} \\
       &-2 i C_{4}^{\ast} N_2^{\ast}(-x,\zeta_4)e^{2i \zeta_4^\ast x},
   \end{aligned}
\end{equation}
where $C_{j}$, $\overline{C}_{j}$, $j=1,2,3$ are the norming constants whose time evolution is given by
\begin{equation}
   \begin{aligned}
       C_{1}(t)=C_{1}(0)e^{(16i\delta \zeta_1^4 - 2i\zeta_1^2)t}, &\qquad \overline{C}_{1}(t)=\overline{C}_{1}(0)e^{(-16i\delta \overline{\zeta}_1^4 + 2i\overline{\zeta}_1^2)t}, \\
      \qquad  C_{2}(t)=C_{2}(0)e^{(16i\delta \zeta_2^4 - 2i\zeta_2^2)t},   & \qquad  \overline{C}_{2}(t)=\overline{C}_{2}(0)e^{(-16i\delta \overline{\zeta}_2^4 + 2i\overline{\zeta}_2^2)t}, \\
      C_{3}(t)=C_{3}(0)e^{(16i\delta \zeta_3^4 - 2i\zeta_3^2)t}, & \qquad \overline{C}_{3}(t)=\overline{C}_{3}(0)e^{(-16i\delta \overline{\zeta}_3^4 + 2i\overline{\zeta}_3^2)t},  \\
       C_{4}(t)=C_{4}(0)e^{(16i\delta \zeta_4^4 - 2i\zeta_4^2)t}, & \qquad \overline{C}_{4}(t)=\overline{C}_{4}(0)e^{(-16i\delta \overline{\zeta}_4^4 + 2i\overline{\zeta}_4^2)t}.
   \end{aligned}
\end{equation}
To obtain the functions $N_2^{\ast}(-x,\zeta_1)$,  $N_2^{\ast}(-x,\zeta_2)$  and $N_2^{\ast}(-x,\zeta_3)$, we need to solve the following system
\begin{equation}
  \left\{
    \begin{aligned}
        \overline{M}_1(x,-\overline{\zeta}_1^{\ast})  = &\alpha_{11} N_2^{\ast}(-x,\zeta_1) + \alpha_{12} N_2^{\ast}(-x,\zeta_2) + \alpha_{13} N_2^{\ast}(-x,\zeta_3)+ \alpha_{14} N_2^{\ast}(-x,\zeta_4),\\
        \overline{M}_1(x,-\overline{\zeta}_2^{\ast})  =& \alpha_{21} N_2^{\ast}(-x,\zeta_1) + \alpha_{22} N_2^{\ast}(-x,\zeta_2)+ \alpha_{23} N_2^{\ast}(-x,\zeta_3) + \alpha_{24} N_2^{\ast}(-x,\zeta_4),\\
       \overline{M}_1(x,-\overline{\zeta}_3^{\ast})   =& \alpha_{31} N_2^{\ast}(-x,\zeta_1) + \alpha_{32} N_2^{\ast}(-x,\zeta_2)+ \alpha_{33} N_2^{\ast}(-x,\zeta_3)+ \alpha_{34} N_2^{\ast}(-x,\zeta_4),\\
       \overline{M}_1(x,-\overline{\zeta}_4^{\ast})   =& \alpha_{41} N_2^{\ast}(-x,\zeta_1) + \alpha_{42} N_2^{\ast}(-x,\zeta_2)+ \alpha_{43} N_2^{\ast}(-x,\zeta_3)+ \alpha_{44} N_2^{\ast}(-x,\zeta_4),\\
        N_2^{\ast}(-x,\zeta_1)=  &  1 + \beta_{11} \overline{M}_1(x,-\overline{\zeta}_1^{\ast}) +  \beta_{12} \overline{M}_1(x,-\overline{\zeta}_2^{\ast})+  \beta_{13} \overline{M}_1(x,-\overline{\zeta}_3^{\ast})+  \beta_{14} \overline{M}_1(x,-\overline{\zeta}_4^{\ast}),  \\
        N_2^{\ast}(-x,\zeta_2)=  &  1 + \beta_{21} \overline{M}_1(x,-\overline{\zeta}_1^{\ast}) + \beta_{22} \overline{M}_1(x,-\overline{\zeta}_2^{\ast})+  \beta_{23} \overline{M}_1(x,-\overline{\zeta}_3^{\ast}) +  \beta_{24} \overline{M}_1(x,-\overline{\zeta}_4^{\ast}),  \\
                N_2^{\ast}(-x,\zeta_3)= &   1 + \beta_{31} \overline{M}_1(x,-\overline{\zeta}_1^{\ast}) + \beta_{32} \overline{M}_1(x,-\overline{\zeta}_2^{\ast})+ \beta_{33} \overline{M}_1(x,-\overline{\zeta}_3^{\ast})+ \beta_{34} \overline{M}_1(x,-\overline{\zeta}_4^{\ast}),  \\
                N_2^{\ast}(-x,\zeta_4)=  &  1 + \beta_{41} \overline{M}_1(x,-\overline{\zeta}_1^{\ast}) + \beta_{42} \overline{M}_1(x,-\overline{\zeta}_2^{\ast})+ \beta_{43} \overline{M}_1(x,-\overline{\zeta}_3^{\ast})+ \beta_{44} \overline{M}_1(x,-\overline{\zeta}_4^{\ast}),
    \end{aligned}
    \right.
\end{equation}
where
\begin{equation}
   \alpha_{ij} =\frac{C_j^{\ast}(t)e^{2i\zeta_j^{\ast}x}}{\overline{\zeta}_i^{\ast}-\zeta_j^{\ast}},   \quad  \beta_{ij} = \frac{\overline{C}_j^{\ast}(t)e^{-2i\overline{\zeta}_{j}^{\ast}x}}{\zeta_i^{\ast}-\overline{\zeta}_j^{\ast}},  \quad  1\leq  i,j \leq 4.
\end{equation}
Solving the above system, we  get

\begin{equation}
     N_2^{\ast}(-x,\zeta_j)= \frac{\det( A_j) }{ \det (A ) }, \quad j=1,2,3,4,
\end{equation}
where
\begin{equation}
 A=\begin{pmatrix} \lambda_1 & \lambda_2 & \lambda_3 & \lambda_4 \\  \lambda_5 & \lambda_6 & \lambda_7 & \lambda_8 \\  \lambda_9 & \lambda_{10} & \lambda_{11} & \lambda_{12} \\  \lambda_{13} & \lambda_{14} & \lambda_{15} & \lambda_{16}      \end{pmatrix},
\end{equation}
and
\begin{equation}
\begin{aligned}
     A_1 =  \begin{pmatrix} 1 & \lambda_2 & \lambda_3 & \lambda_4 \\ 1 & \lambda_6 & \lambda_7 & \lambda_8 \\  1 & \lambda_{10} & \lambda_{11} & \lambda_{12} \\  1 & \lambda_{14} & \lambda_{15} & \lambda_{16}      \end{pmatrix},   \quad
        A_2=  \begin{pmatrix} \lambda_1 & 1 & \lambda_3 & \lambda_4 \\  \lambda_5 & 1 & \lambda_7 & \lambda_8 \\  \lambda_9 & 1 & \lambda_{11} & \lambda_{12} \\  \lambda_{13} & 1 & \lambda_{15} & \lambda_{16}      \end{pmatrix},  \\
\end{aligned}
\end{equation}

\begin{equation}
\begin{aligned}
        A_3=\begin{pmatrix} \lambda_1 & \lambda_2 & 1 & \lambda_4 \\  \lambda_5 & \lambda_6 & 1 & \lambda_8 \\  \lambda_9 & \lambda_{10} & 1 & \lambda_{12} \\  \lambda_{13} & \lambda_{14} & 1 & \lambda_{16}      \end{pmatrix},  \quad
         A_4= \begin{pmatrix} \lambda_1 & \lambda_2 & \lambda_3 & 1 \\  \lambda_5 & \lambda_6 & \lambda_7 & 1 \\  \lambda_9 & \lambda_{10} & \lambda_{11} & 1 \\  \lambda_{13} & \lambda_{14} & \lambda_{15} &  1      \end{pmatrix},
\end{aligned}
\end{equation}
with
\begin{equation}
\left\{
  \begin{aligned}
  \lambda_1 = & 1 - \alpha_{11}\beta_{11} - \alpha_{21}\beta_{12} - \alpha_{31}\beta_{13}- \alpha_{41}\beta_{14},~~
  \lambda_2 =  - \alpha_{12}\beta_{11} - \alpha_{22}\beta_{12} - \alpha_{32}\beta_{13}- \alpha_{42}\beta_{14},  \\
  \lambda_3 = & - \alpha_{13}\beta_{11} - \alpha_{23}\beta_{12} - \alpha_{33}\beta_{13}- \alpha_{43}\beta_{14}, ~~
  \lambda_4 =   - \alpha_{14}\beta_{11} - \alpha_{24}\beta_{12} - \alpha_{34}\beta_{13}- \alpha_{44}\beta_{14}, \\
  \lambda_5 = &  - \alpha_{11}\beta_{21} - \alpha_{21}\beta_{22} - \alpha_{31}\beta_{23}- \alpha_{41}\beta_{24},  ~~  
  \lambda_6 =  1 - \alpha_{12}\beta_{21} - \alpha_{22}\beta_{22} - \alpha_{32}\beta_{23}- \alpha_{42}\beta_{24},    \\
  \lambda_7 = &  - \alpha_{13}\beta_{21} - \alpha_{23}\beta_{22} - \alpha_{33}\beta_{23}- \alpha_{43}\beta_{24}, ~~   
  \lambda_8 =   - \alpha_{14}\beta_{21} - \alpha_{24}\beta_{22} - \alpha_{34}\beta_{23}- \alpha_{44}\beta_{24}, \\
  \lambda_9 = &  - \alpha_{11}\beta_{31} - \alpha_{21}\beta_{32} - \alpha_{31}\beta_{33}-\alpha_{41}\beta_{34},  ~~  
  \lambda_{10} =   - \alpha_{12}\beta_{31} - \alpha_{22}\beta_{32} - \alpha_{32}\beta_{33}-\alpha_{42}\beta_{34},   \\
  \lambda_{11} = &  1 - \alpha_{13}\beta_{31} - \alpha_{23}\beta_{32} - \alpha_{33}\beta_{33}-\alpha_{43}\beta_{34}, ~~  
  \lambda_{12} =   - \alpha_{14}\beta_{31} - \alpha_{24}\beta_{32} - \alpha_{34}\beta_{33}- \alpha_{44}\beta_{34},  \\
  \lambda_{13} = &  - \alpha_{11}\beta_{41} - \alpha_{21}\beta_{42} - \alpha_{31}\beta_{43}-\alpha_{41}\beta_{44}, ~~   
  \lambda_{14} =    - \alpha_{12}\beta_{41} - \alpha_{22}\beta_{42} - \alpha_{32}\beta_{43}-\alpha_{42}\beta_{44},   \\
  \lambda_{15} = &   - \alpha_{13}\beta_{41} - \alpha_{23}\beta_{42} - \alpha_{33}\beta_{43}-\alpha_{43}\beta_{44}, ~~  
  \lambda_{16} =  1 - \alpha_{14}\beta_{41} - \alpha_{24}\beta_{42} - \alpha_{34}\beta_{43}- \alpha_{44}\beta_{44}.  
  \end{aligned}
  \right.
\end{equation}
Substituting the above equations into Eq. \eqref{four}, we can obtain the formula of  four-soliton solutions.

\noindent
{\rotatebox{0}{\includegraphics[width=3.6cm,height=3.5cm,angle=0]{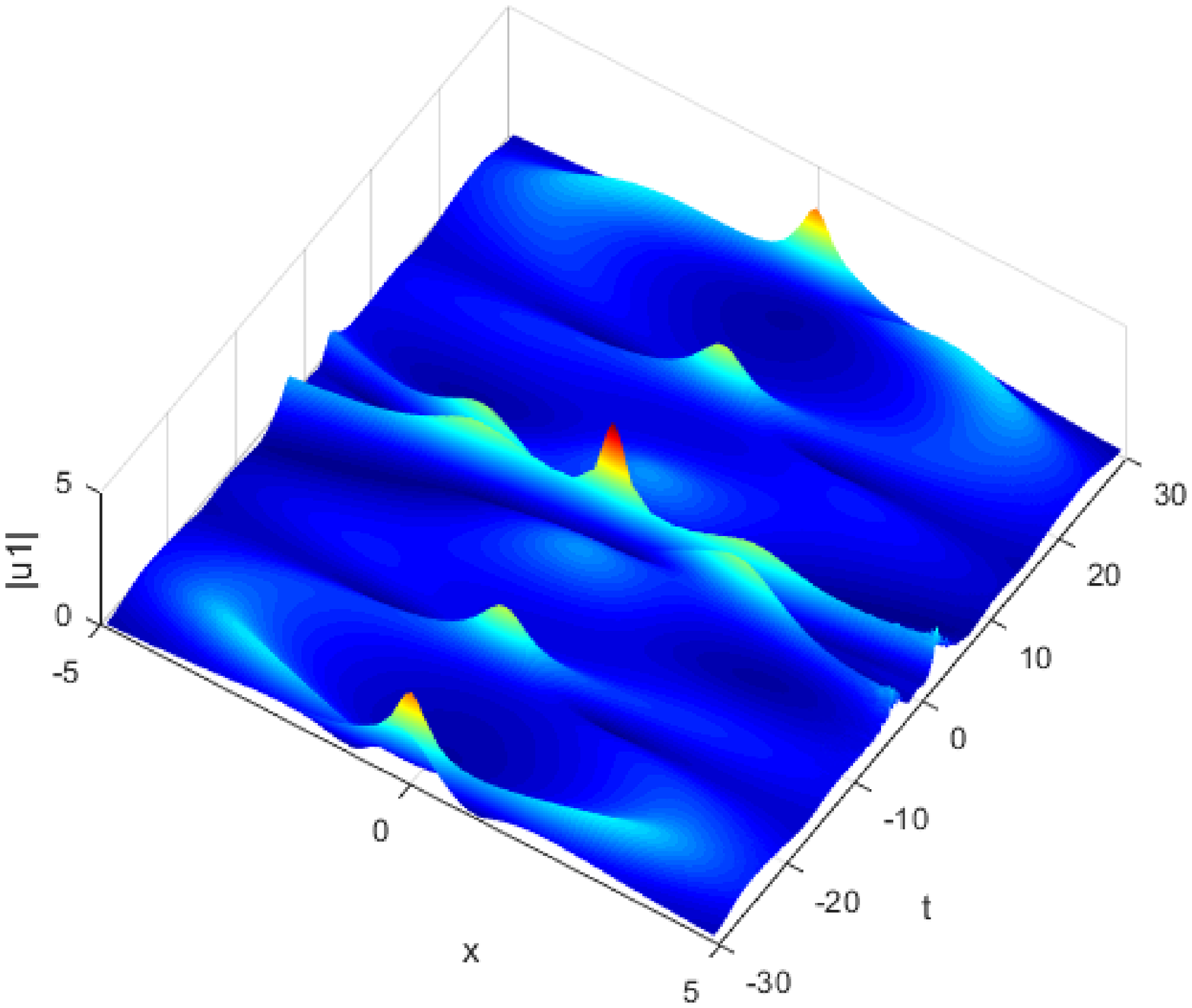}}}
~~~~
{\rotatebox{0}{\includegraphics[width=3.6cm,height=3.5cm,angle=0]{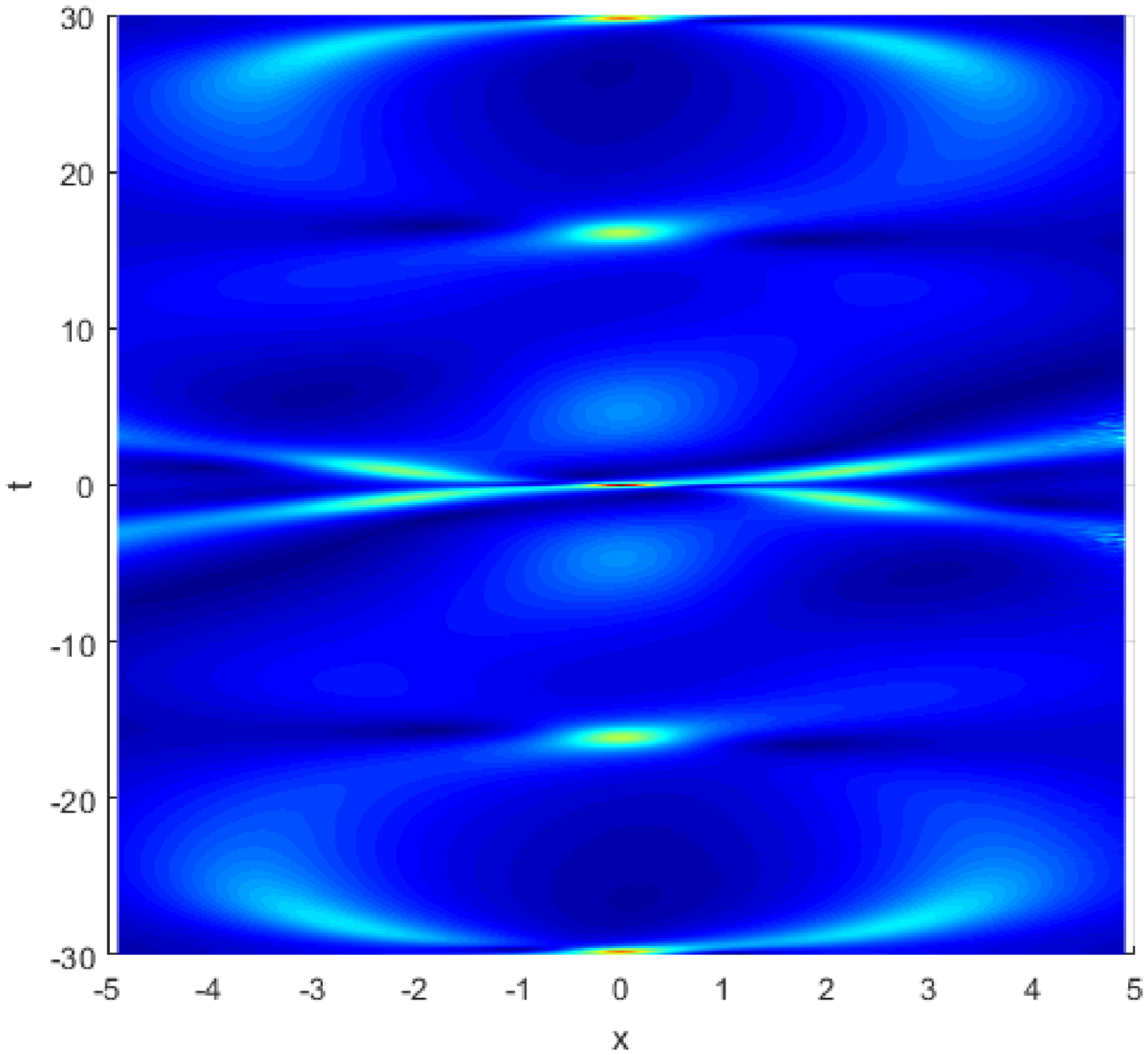}}}
~~~~
{\rotatebox{0}{\includegraphics[width=3.6cm,height=3.5cm,angle=0]{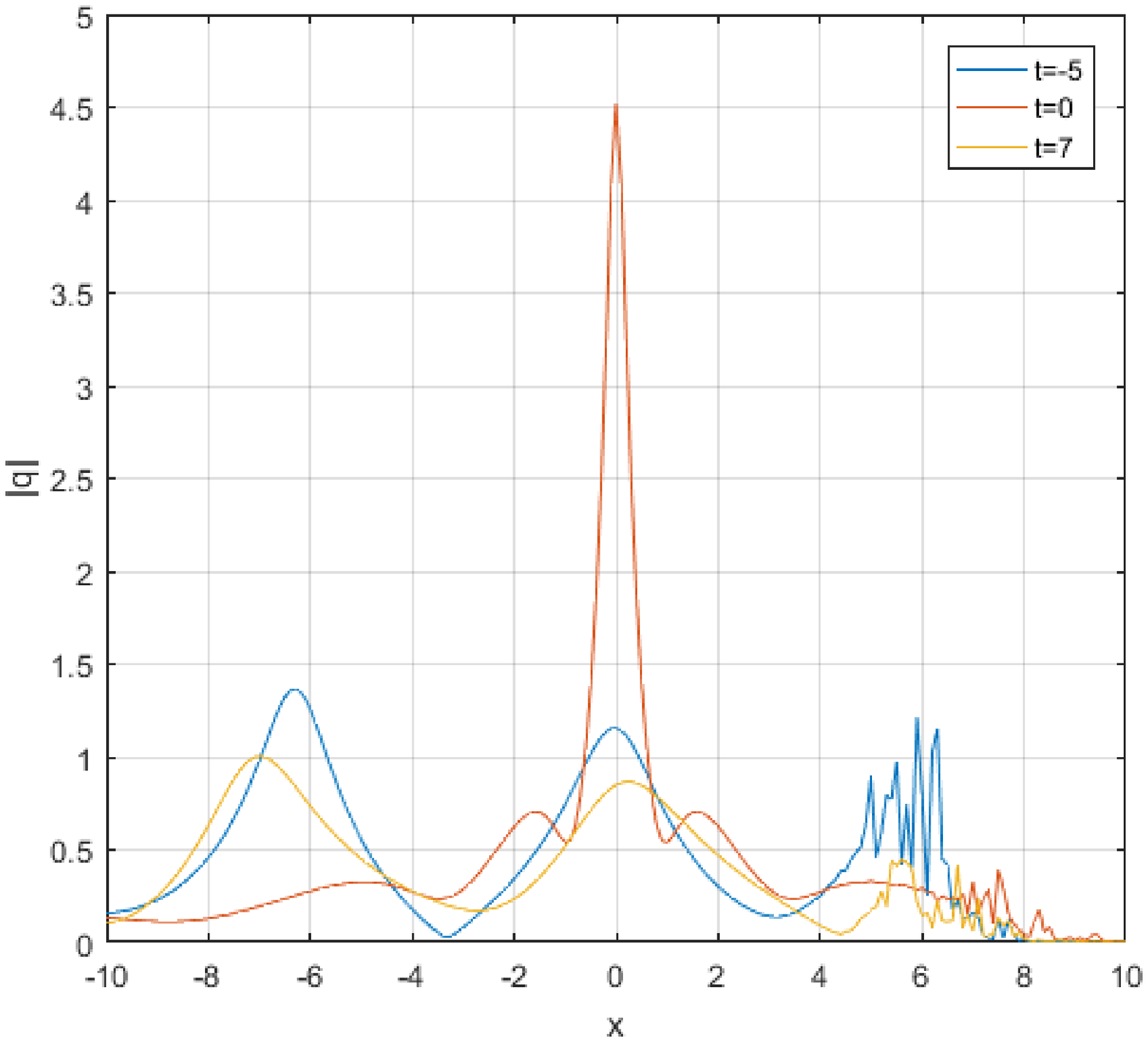}}}

$\ \qquad~~~~~~(\textbf{a})\qquad \ \qquad\qquad\qquad\qquad~(\textbf{b})
\ \qquad\qquad\qquad\qquad\qquad~(\textbf{c})$\\
\noindent { \small \textbf{Figure 6.} Four-soliton  solution   with parameters  $\delta=1$, $\theta_1=\theta_2=\theta_3=\theta_4=\frac{\pi}{6}$,  $\overline{\theta}_1=\overline{\theta}_2=\overline{\theta}_3=\overline{\theta}_4=\frac{\pi}{8}$, $\zeta_1=0.1i$, $\zeta_2=0.2i$, $\zeta_3=0.3i$, $\zeta_4=0.4i$, $\overline{\zeta}_1=-0.1i$,  $\overline{\zeta}_2=-0.2i$,  $\overline{\zeta}_3=-0.3i$  and  $\overline{\zeta}_4=-0.4i$.
$\textbf{(a)}$: the structures of the four-soliton  solution,
$\textbf{(b)}$: the density plot,
$\textbf{(c)}$: the wave propagation of the four-soliton  solution.}  \\

\noindent
{\rotatebox{0}{\includegraphics[width=3.6cm,height=4.0cm,angle=0]{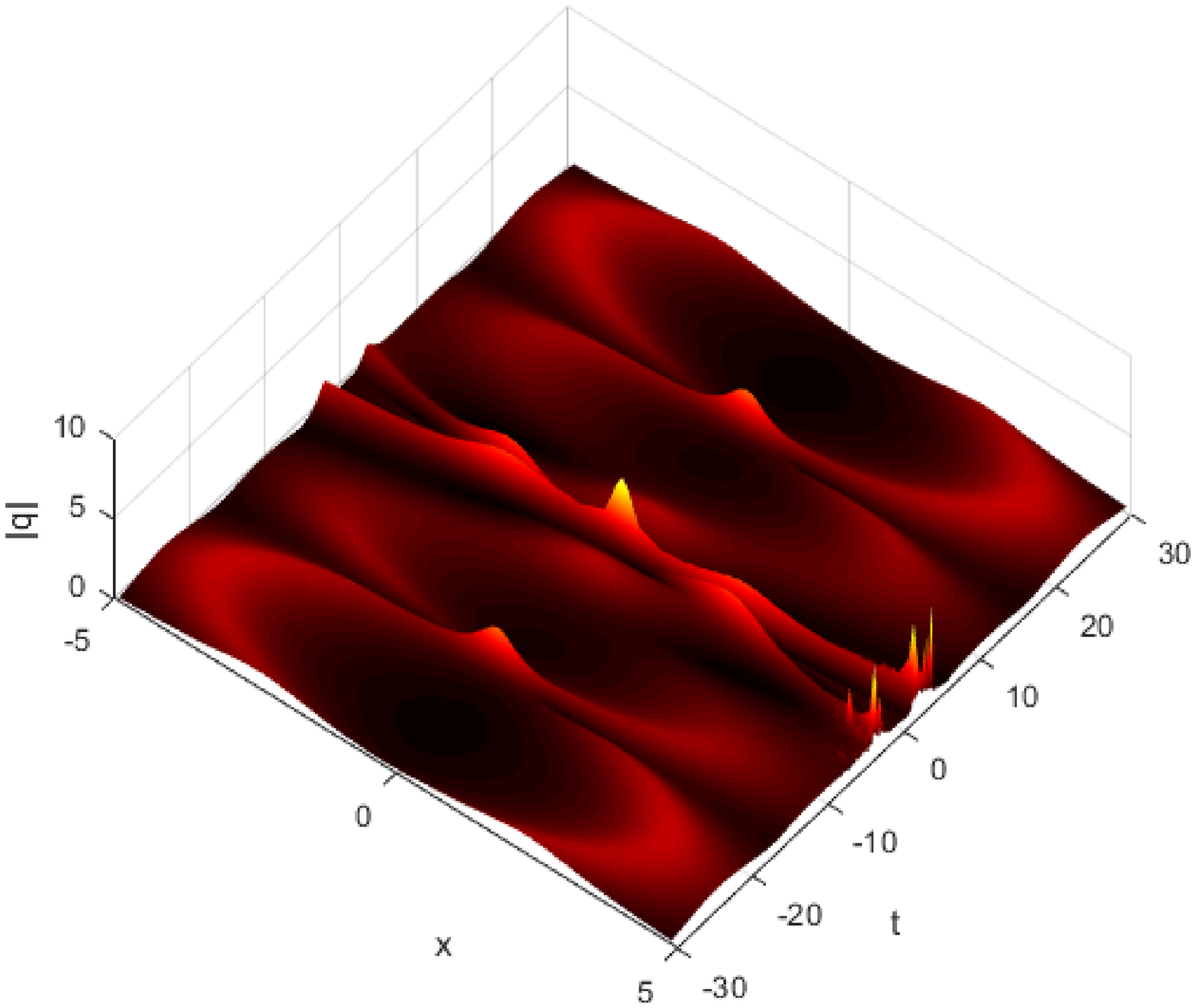}}}
~~~~
{\rotatebox{0}{\includegraphics[width=3.6cm,height=3.6cm,angle=0]{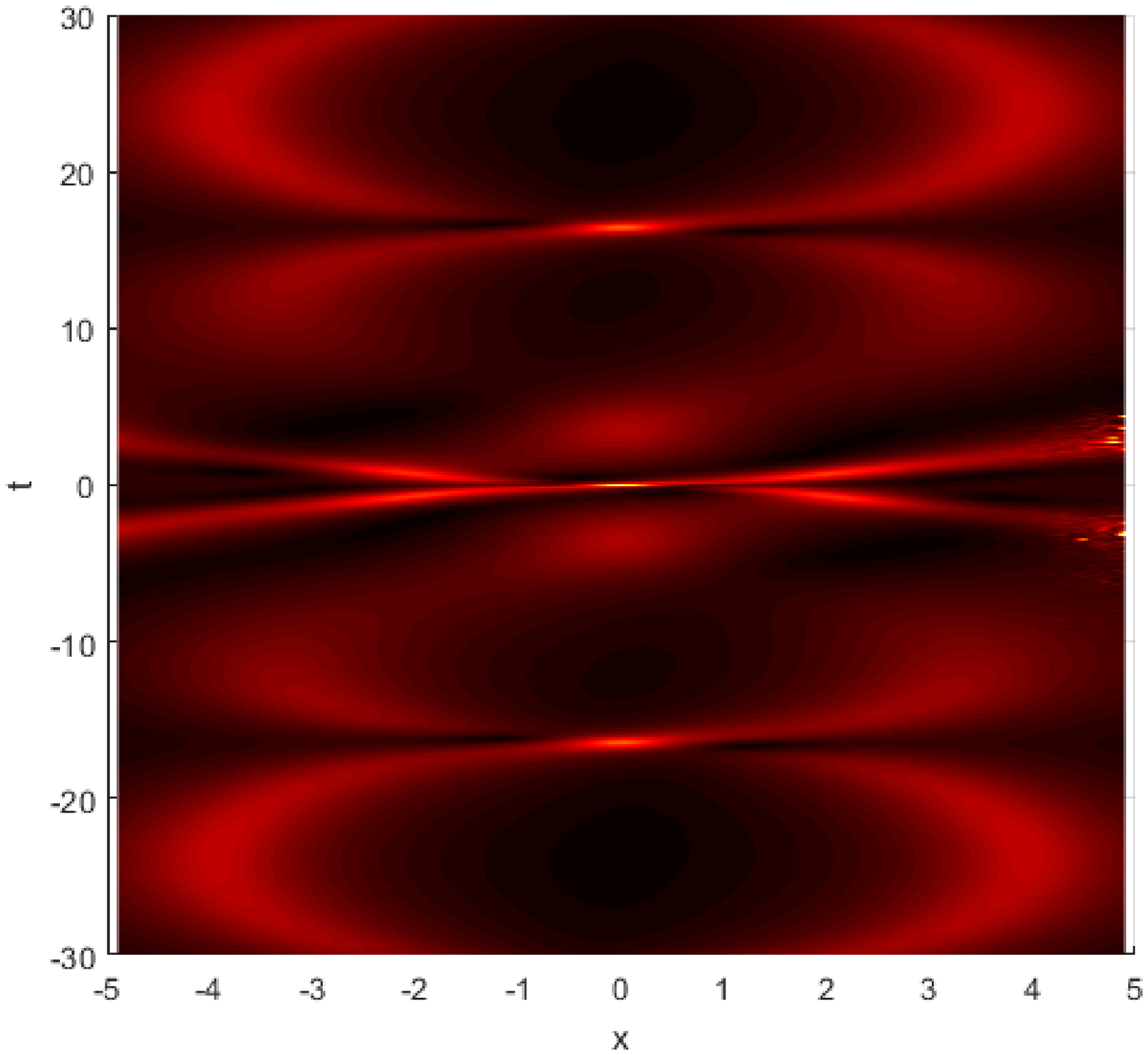}}}
~~~~
{\rotatebox{0}{\includegraphics[width=3.6cm,height=3.6cm,angle=0]{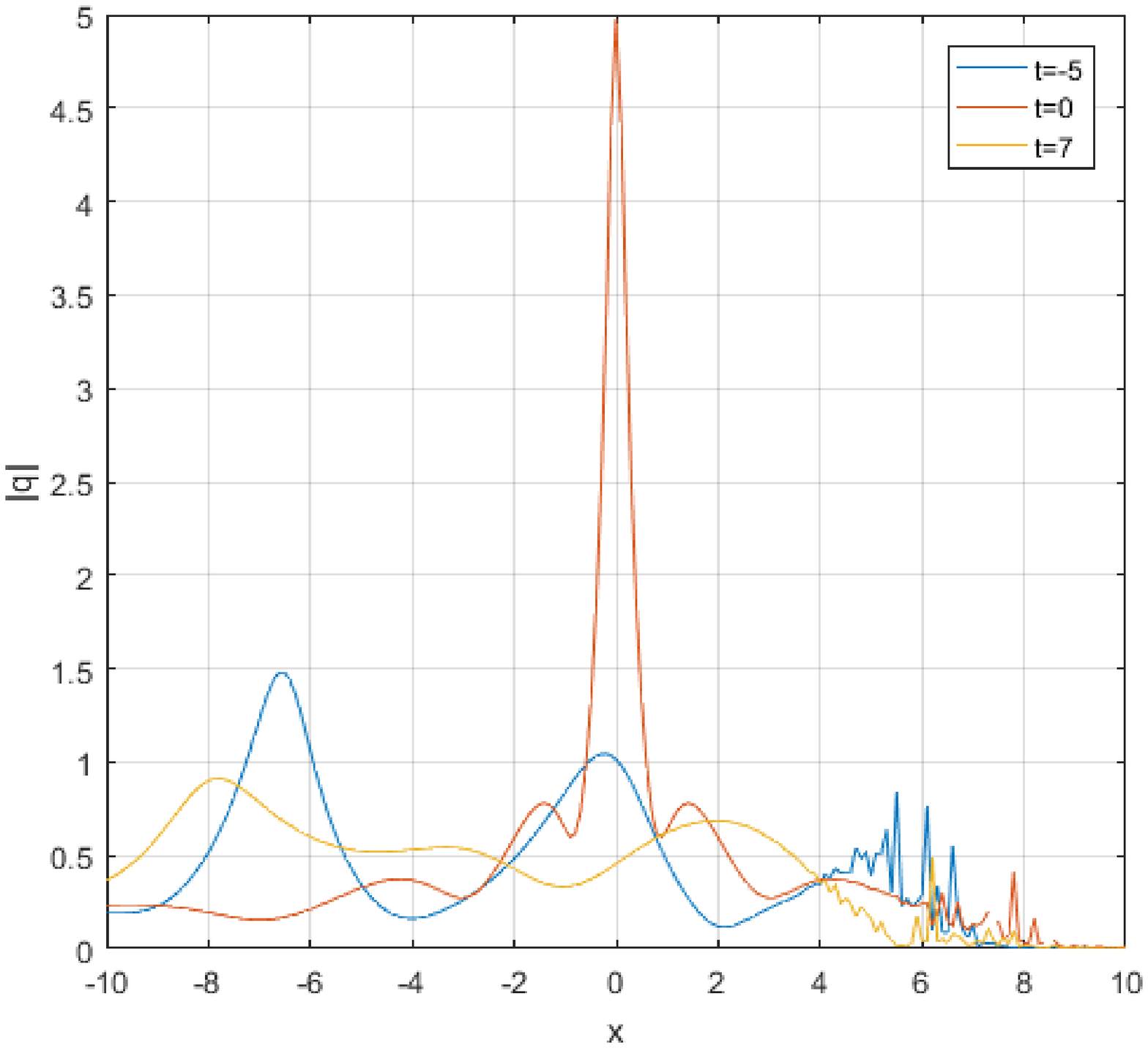}}}

$\ \qquad~~~~~~(\textbf{a})\qquad \ \qquad\qquad\qquad\qquad~(\textbf{b})
\ \qquad\qquad\qquad\qquad\qquad~(\textbf{c})$\\
\noindent { \small \textbf{Figure 7.} Four-soliton  solution   with parameters  $\delta=1$, $\theta_1=\theta_2=\theta_3=\theta_4=\frac{\pi}{6}$,  $\overline{\theta}_1=\overline{\theta}_2=\overline{\theta}_3=\overline{\theta}_4=\frac{\pi}{8}$, $\zeta_1=0.1+0.2i$, $\zeta_2=-0.1+0.2i$, $\zeta_3=0.3i$, $\zeta_4=0.4i$, $\overline{\zeta}_1=0.1-0.2i$,  $\overline{\zeta}_2=-0.1-0.2i$,  $\overline{\zeta}_3=-0.3i$  and  $\overline{\zeta}_4=-0.4i$.
$\textbf{(a)}$: the structures of the four-soliton  solution,
$\textbf{(b)}$: the density plot,
$\textbf{(c)}$: the wave propagation of the four-soliton  solution.}  \\

Figs. 6 and 7 present the local structure and the dynamic behavior of four soliton solutions. The four solitons include two arc-shaped solitons  and two ordinary solitons. Four solitons meet, collide   and move away at the center point($x=0,t=0$).
Moreover, before and after the collision, the properties of four  solitons have no changed.  In the following Fig. 8,  by select special parameters, we obtain  another form of four solitons which is center symmetry about the center point. It is worth nothing that the energy of four solitons is exchanged according to symmetry relation.  For example,  the energy exchange between the leftmost soliton and the rightmost soliton  and  the energy exchange between  the two solitons in the middle.

\noindent
{\rotatebox{0}{\includegraphics[width=3.6cm,height=4.0cm,angle=0]{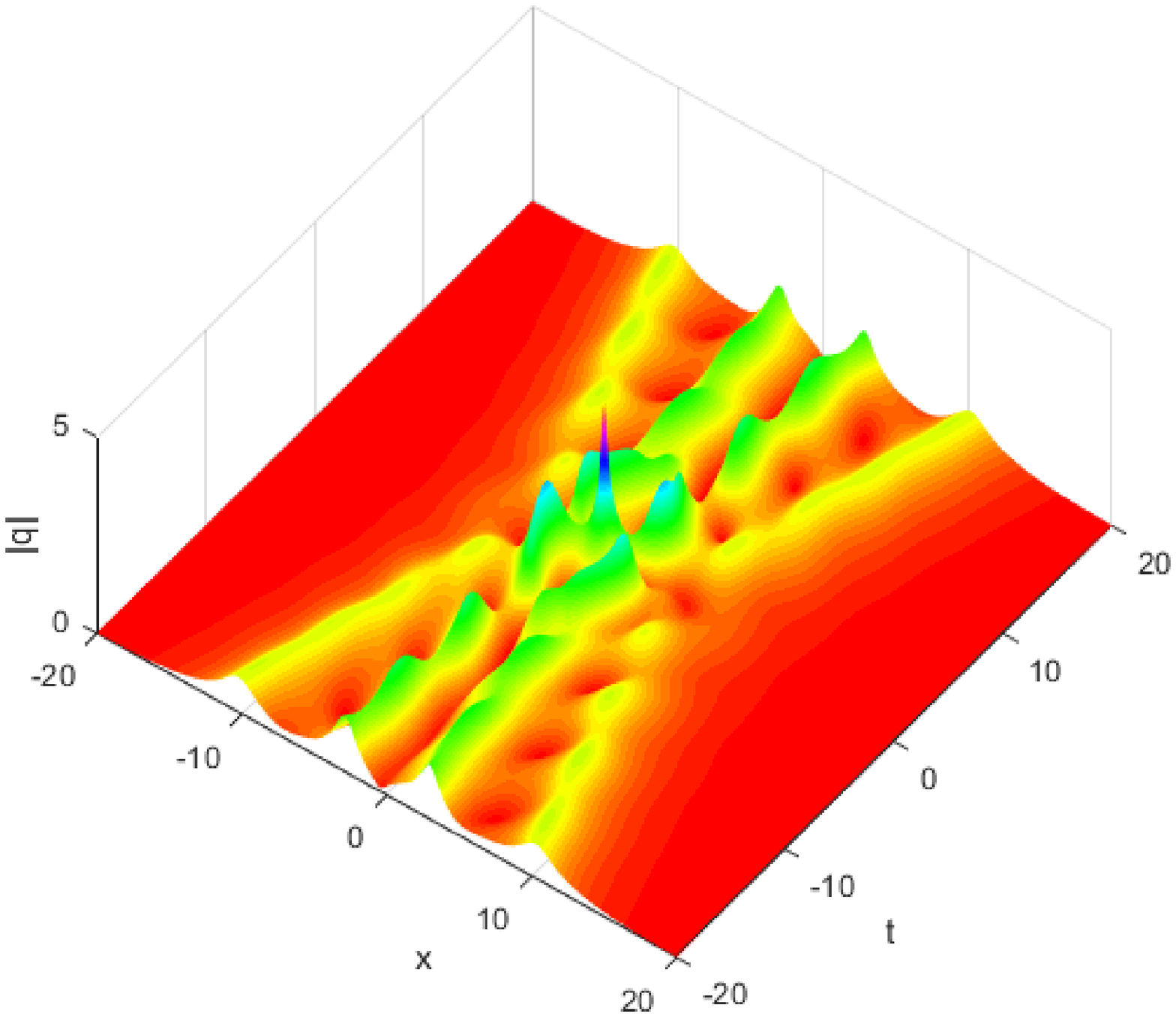}}}
~~~~
{\rotatebox{0}{\includegraphics[width=3.6cm,height=3.6cm,angle=0]{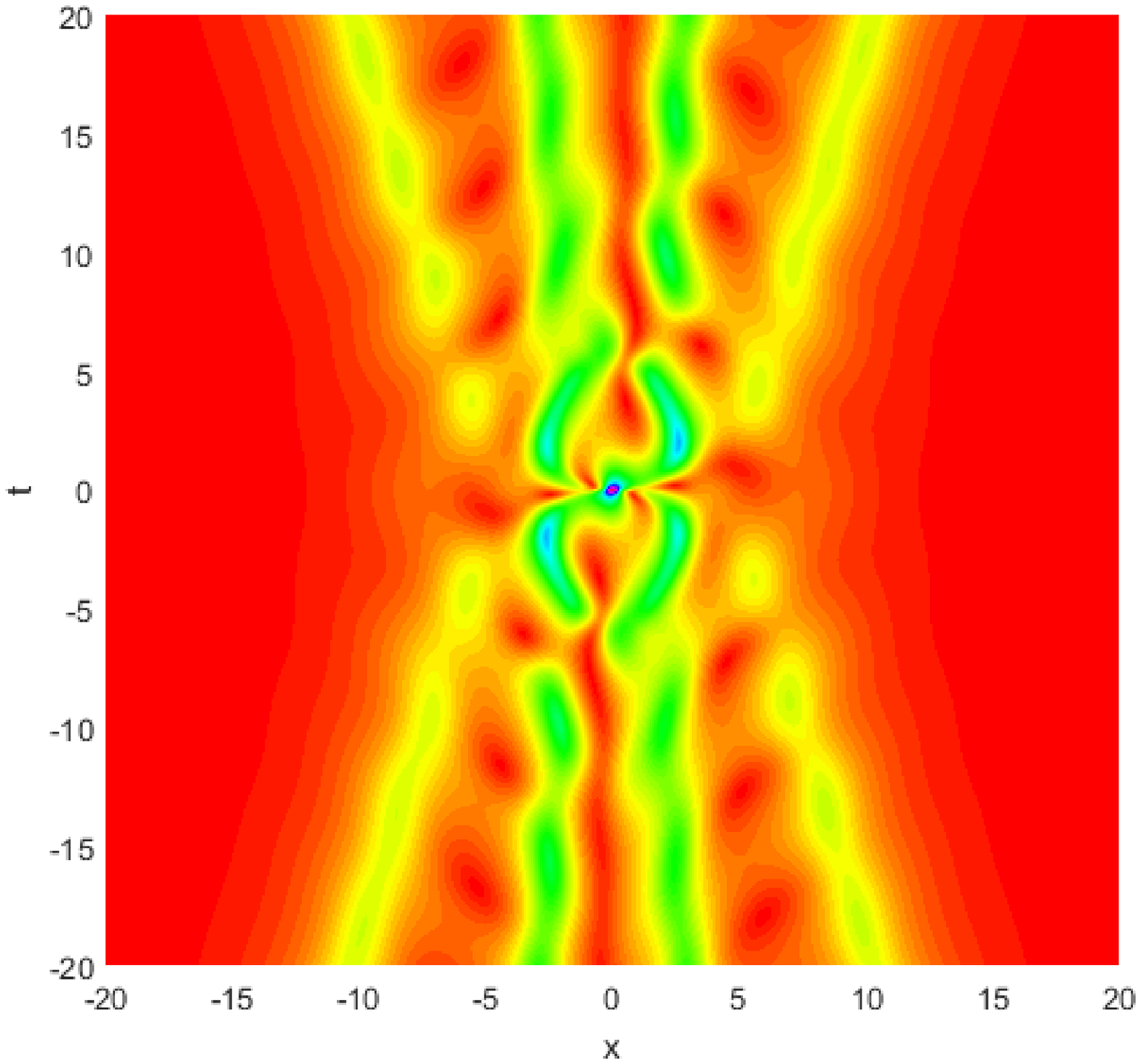}}}
~~~~
{\rotatebox{0}{\includegraphics[width=3.6cm,height=3.6cm,angle=0]{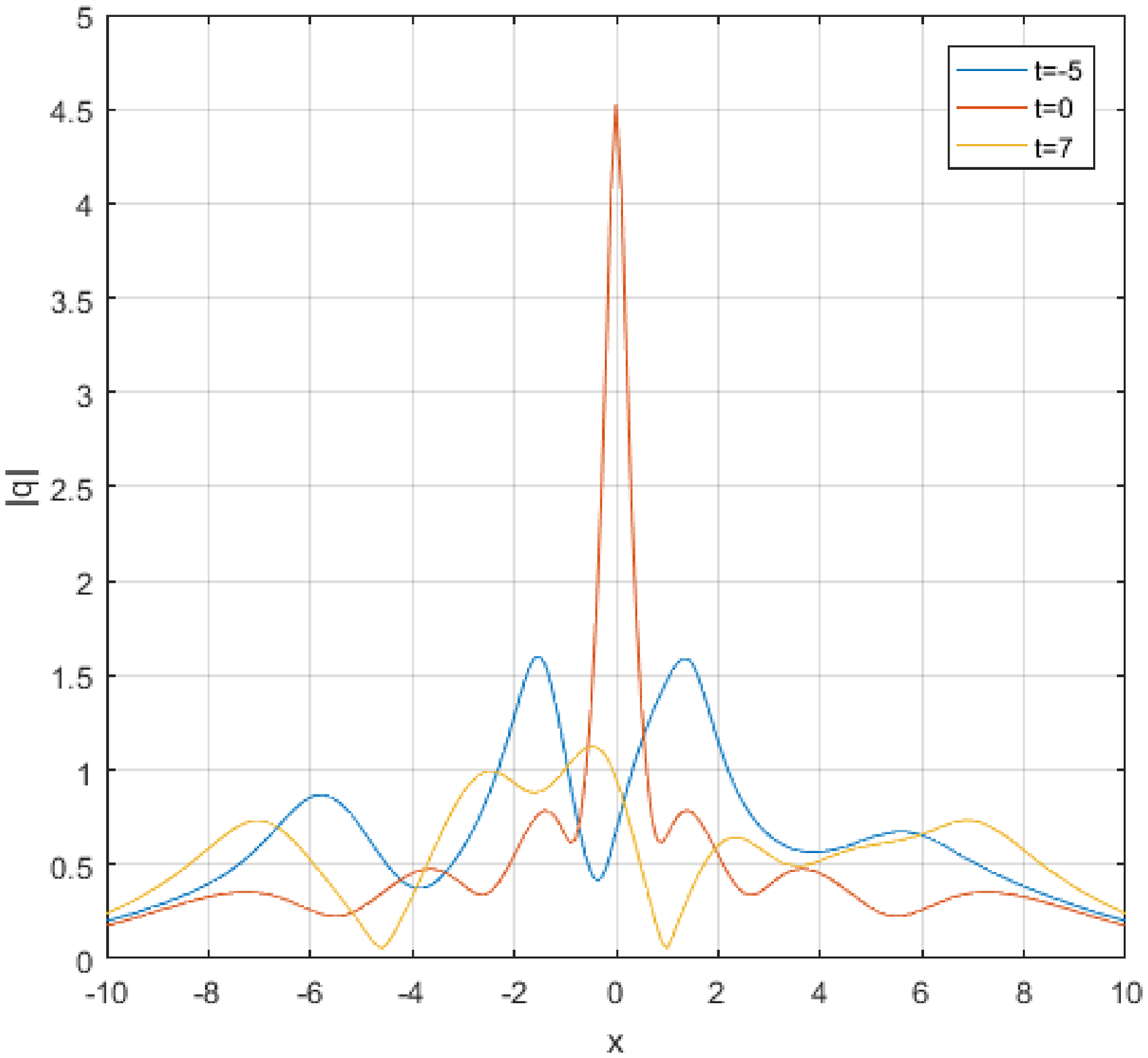}}}

$\ \qquad~~~~~~(\textbf{a})\qquad \ \qquad\qquad\qquad\qquad~(\textbf{b})
\ \qquad\qquad\qquad\qquad\qquad~(\textbf{c})$\\
\noindent { \small \textbf{Figure 8.} Four-soliton  solution   with parameters $\delta=1$, $\theta_1=\theta_2=\theta_3=\theta_4=\frac{\pi}{6}$,  $\overline{\theta}_1=\overline{\theta}_2=\overline{\theta}_3=\overline{\theta}_4=\frac{\pi}{8}$, $\zeta_1=0.1+0.2i$, $\zeta_2=-0.1+0.2i$, $\zeta_3=0.4+0.3i$, $\zeta_4=-0.4+0.3i$, $\overline{\zeta}_1=0.1-0.2i$,  $\overline{\zeta}_2=-0.1-0.2i$,  $\overline{\zeta}_3=0.4-0.3i$  and  $\overline{\zeta}_4=-0.4-0.3i$.
$\textbf{(a)}$: the structures of the four-soliton  solution,
$\textbf{(b)}$: the density plot,
$\textbf{(c)}$: the wave propagation of the four-soliton  solution.}

\section{Eigenvlaues  and  conserved quantities under some special initial conditions}
  Before this section,  we consider  pure soliton solutions  of the objective equation \eqref{LPD} under the condition $\rho(\xi)=\overline{\rho}(\xi)=0$.  However, as for more general initial condition $q(x,0)$ and $r(x,0)$,  the reflectionless case  may not hold,  which makes the objective equation unsolvable by using IST method.   Next, we will  analysis the eigenvalues the conserved quantities under some special initial conditions.

  \subsection{Rectangular wave}
   In what follows, we  study the following   rectangular   initial condition
\begin{equation}
    q(x,0)= \left\{
    \begin{aligned}
     0,  & \quad  x \in (-\infty,0), \\
     h,  & \quad  x \in (0,L), \\
     0,    &  \quad x \in (L,\infty), \\
    \end{aligned}
    \right.
\end{equation}
where $h$ and $L$ are real and positive constants. Under the symmetry relation $r(x,0)=-q^{\ast}(-x,0)$, we obtain the initial data of $r(x,t)$
\begin{equation}
    r(x,0)= \left\{
    \begin{aligned}
     0,  & \quad  x \in (-\infty,-L), \\
     -h,  & \quad  x \in (-L,0), \\
     0,    &  \quad x \in (0,\infty). \\
    \end{aligned}
    \right.
\end{equation}
According  to the $t$-independent scattering problem,   we have
\begin{equation}\label{x-sca}
\left\{
   \begin{aligned}
   \phi_{1,x} & = - i \zeta \phi_1 + q(x,t) \phi_2, \\
   \phi_{2,x} & = i \zeta \phi_2  + r(x,t) \phi_1.
   \end{aligned}
\right.
\end{equation}
Instituting the above initial condition into  Eq. \eqref{x-sca} and solving the ordinary differential equations, we have
\begin{equation}
\left\{
\begin{aligned}
   0<x<L, & \quad \begin{pmatrix} \phi_{1}(x,\zeta) \\  \phi_2(x,\zeta)   \end{pmatrix}= \begin{pmatrix}  \frac{h}{2i\zeta}c_1 e^{i\zeta x} + c_2 e^{ - i \zeta x } \\ c_1 e^{i\zeta x}  \end{pmatrix},  \\
   -L <x< 0, & \quad \begin{pmatrix} \phi_{1}(x,\zeta) \\  \phi_2(x,\zeta)   \end{pmatrix}= \begin{pmatrix}  \tilde{c}_1 e^{-i\zeta x}  \\  \tilde{c}_2 e^{  i \zeta x } + \frac{h}{2i \zeta} \tilde{c}_1 e^{-i\zeta x}  \end{pmatrix}.
\end{aligned}
\right.
\end{equation}
In order to match the values of eigenfunctions at the critical point $x=0$, $x=-L$, we obtain
\begin{equation}
    \begin{aligned}
 \tilde{c}_1=1,   & \qquad   c_1=\frac{h}{2i \zeta} \left(1-e^{2i\zeta L} \right),   \\
 \tilde{c}_2= - \frac{h}{2i \zeta} e^{2i\zeta L},         & \qquad  c_2=1+ \left(\frac{h}{2i\zeta} \right)^2 (e^{2i\zeta L}-1).
    \end{aligned}
\end{equation}
At the same time,  according to Eqs. \eqref{boun}  and \eqref{left}, when $x> L$, we have
\begin{equation}
   \phi(x,t) = \begin{pmatrix}  a(\zeta)e^{-i\zeta x} \\ b(\zeta) e^{i \zeta x}     \end{pmatrix}.
\end{equation}
In the process of matching the value of eigenfunction at $x=L$, we find
\begin{equation}\label{a1}
   \begin{aligned}
      a(\zeta)= &1+ \left(\frac{h}{2i\zeta} \right)^2  (e^{2i\zeta L}-1  ) - \left(\frac{h}{2i \zeta} \right)^2   (e^{4i\zeta L} - e^{2i\zeta L}  ),  \\
      b(\zeta)= &-h e^{i\zeta L} \frac{\sin(\zeta L)}{\zeta},
   \end{aligned}
\end{equation}
then the eigenvalues, i.e. the zeros of $a(\zeta)$, can be given  implicitly by
\begin{equation}
   e^{2i \zeta L} - 1  \pm \frac{2i \zeta}{h} = 0.
\end{equation}
Besides,  the asymptotic  behavior of $a(\zeta)$ for  large  and small $\zeta$  can be derived  from Eq. \eqref{a1}
\begin{equation}
       \begin{aligned}
         & a(\zeta) \sim  1-  \frac{h^2}{(2i\zeta)^2},   & \qquad  \zeta \to \infty,  \\
        &  a(\zeta)  \sim  1-h^2 L^2, & \qquad \zeta \to 0.
       \end{aligned}
\end{equation}
With the aid of the lagre $\zeta$ asymptotic behavior of $a(\zeta)$ and Eq. \eqref{conser3}, we find  that  the conserved quantities satisfy
\begin{equation}
   \mathcal{C}_{2n}=0,  \qquad  \mathcal{C}_{2n+1}=- \frac{h^{2n+2}}{n+1}, \qquad n=0,1,2,\dots.
\end{equation}

\subsection{Arcuated  wave}
 In the second example, we consider the  following  arcuated   initial condition
\begin{equation}
    q(x,0)= \left\{
    \begin{aligned}
     0,  & \quad  x \in (-\infty,0), \\
     -x^2 +L x,  & \quad  x \in (0,L), \\
     0,    &  \quad x \in (L,\infty), \\
    \end{aligned}
    \right.
\end{equation}
where $L$ is real and positive constant. Under the symmetry relation $r(x,0)=-q^{\ast}(-x,0)$, we obtain the initial data of $r(x,t)$
\begin{equation}
    r(x,0)= \left\{
    \begin{aligned}
     0,  & \quad  x \in (-\infty,-L), \\
     x^2+Lx,  & \quad  x \in (-L,0), \\
     0,    &  \quad x \in (0,\infty). \\
    \end{aligned}
    \right.
\end{equation}
Instituting the above initial condition into the scattering problem \eqref{x-sca} and solving the ordinary differential equations, we have
\begin{equation}
\left\{
\begin{aligned}
   0<x<L, & \quad \begin{pmatrix} \phi_{1}(x,\zeta) \\  \phi_2(x,\zeta)   \end{pmatrix}= \begin{pmatrix}  c_1 e^{-i \zeta x} + c_2 e^{i \zeta x} \left[ \frac{1}{2i \zeta} (-x^2 +L x) +\frac{1}{(2i \zeta)^2}(2x-L)- \frac{2}{(2i \zeta)^3}  \right]  \\ c_2 e^{i \zeta x}  \end{pmatrix},  \\
   -L <x< 0, & \quad \begin{pmatrix} \phi_{1}(x,\zeta) \\  \phi_2(x,\zeta)   \end{pmatrix}= \begin{pmatrix}  \tilde{c}_1 e^{-i\zeta x} \\  \tilde{c}_1 e^{-i\zeta x} \left[ -\frac{1}{2i \zeta} (x^2+Lx) -\frac{1}{(2i \zeta)^2} (2x +L) -\frac{2}{(2i\zeta)^3}  \right] +\tilde{c}_2 e^{i \zeta x} \end{pmatrix}.
\end{aligned}
\right.
\end{equation}
Matching the value of the eigenfunction at $x=0$ and $-L$, we find
\begin{equation}
\begin{aligned}
   \tilde{c}_1 & =1,\\
  \tilde{c}_2 & =e^{2i\zeta L} (\frac{2}{(2i\zeta)^3}- \frac{L}{(2i\zeta)^2}), \\
   c_1  & = 1 + e^{2i\zeta L} \left( \frac{4}{(2i\zeta)^6} - \frac{L^2}{(2i\zeta)^4}  \right) -\left( \frac{2}{(2i\zeta)^3} +\frac{L}{(2i\zeta)^2}  \right)^2,  \\
   c_2 & = e^{2i \zeta L} \left(  \frac{2}{(2i\zeta)^3} - \frac{L}{(2i\zeta)^2}  \right) - \left( \frac{2}{(2i \zeta)^3} + \frac{L}{(2i\zeta)^2}   \right).   \\
\end{aligned}
\end{equation}
Simliar to case 1,   when $x> L$, we have
\begin{equation}
   \phi(x,t) = \begin{pmatrix}  a(\zeta)e^{-i\zeta x} \\ b(\zeta) e^{i \zeta x}     \end{pmatrix}.
\end{equation}
In the process of matching the value of eigenfunction at $x=L$, we find
\begin{equation}
   \begin{aligned}
      a(\zeta)= & 1 - \left[  e^{2i \zeta L}  \left( \frac{2}{(2i\zeta)^3} - \frac{L}{(2i\zeta)^2}  \right)    -     \left( \frac{2}{(2i\zeta)^3} + \frac{L}{(2i\zeta)^2}  \right)            \right]^2, \\
      b(\zeta)= & e^{i \zeta L}  \left( \frac{2}{(2i\zeta)^3} - \frac{L}{(2i\zeta)^2}  \right) - e^{-i\zeta L}  \left( \frac{2}{(2i\zeta)^3} + \frac{L}{(2i\zeta)^2}  \right)   ,
   \end{aligned}
\end{equation}
then the eigenvalues, i.e. the zeros of $a(\zeta)$, can be given  implicitly by
\begin{equation}
     e^{2i \zeta L}  \left( \frac{2}{(2i\zeta)^3} - \frac{L}{(2i\zeta)^2}  \right)    -     \left( \frac{2}{(2i\zeta)^3} + \frac{L}{(2i\zeta)^2}  \right) \pm 1 = 0.
\end{equation}
Besides,  the asymptotic  behavior of $a(\zeta)$ for  large  and small $\zeta$  can be derived  from Eq. \eqref{a1}
\begin{equation}
       \begin{aligned}
         & a(\zeta) \sim  1-\frac{L^2}{(2i\zeta)^4},   & \qquad  \zeta \to \infty,  \\
        &  a(\zeta)  \sim  1- \frac{L^6}{36}, & \qquad \zeta \to 0.
       \end{aligned}
\end{equation}
With the aid of the lagre $\zeta$ asymptotic behavior of $a(\zeta)$ and Eq. \eqref{conser3}, we find  that  the conserved quantities satisfy
\begin{equation}
  \mathcal{C}_{n}= \left\{
  \begin{aligned}
        -\frac{L^{2(m+1)}}{m+1}, & \quad n=4m+3, m=0,1,2,\dots,  \\
        0,   &   \quad  else.
  \end{aligned}
  \right.
\end{equation}

\subsection{Trianglular  wave}
 In the second example, we consider the  following  triangular  initial condition
\begin{equation}
    q(x,0)= \left\{
    \begin{aligned}
     0,  & \quad  x \in (-\infty,0), \\
     L-|x-L|,  & \quad  x \in (0,2L), \\
     0,    &  \quad x \in (2L,\infty), \\
    \end{aligned}
    \right.
\end{equation}
where $L$ is real and positive constant.  Under the symmetry relation $r(x,0)=-q^{\ast}(-x,0)$, we obtain the initial data of $r(x,t)$
\begin{equation}
    r(x,0)= \left\{
    \begin{aligned}
     0,  & \quad  x \in (-\infty,-2L), \\
     -L+|x+L|,  & \quad  x \in (-2L,0), \\
     0,    &  \quad x \in (0,\infty). \\
    \end{aligned}
    \right.
\end{equation}
 Instituting the above initial condition into the scattering problem \eqref{x-sca} and solving the ordinary differential equations, we have
\begin{equation}
\left\{
\begin{aligned}
   -2L<x<-L, & \quad \begin{pmatrix} \phi_{1}(x,\zeta) \\  \phi_2(x,\zeta)   \end{pmatrix}= \begin{pmatrix}  c_1 e^{-i \zeta x}   \\ c_2 e^{i \zeta x} +c_1 e^{-i \zeta x}\left(\frac{x+2L}{2i\zeta}+\frac{1}{(2i\zeta)^2}   \right)   \end{pmatrix},  \\
   -L <x< 0, & \quad \begin{pmatrix} \phi_{1}(x,\zeta) \\  \phi_2(x,\zeta)   \end{pmatrix}= \begin{pmatrix}  c_3 e^{-i\zeta x} \\  c_3 e^{-i\zeta x} \left( -\frac{x}{2i \zeta}   -\frac{1}{(2i \zeta)^2} \right) +  c_4 e^{i \zeta x} \end{pmatrix},  \\
       0<x< L, & \quad \begin{pmatrix} \phi_{1}(x,\zeta) \\  \phi_2(x,\zeta)   \end{pmatrix}= \begin{pmatrix}  c_5 e^{i\zeta x} \left( \frac{x}{2i\zeta} -\frac{1}{(2i\zeta)^2} \right) +c_6 e^{-i\zeta x}  \\  c_5 e^{i\zeta x}    \end{pmatrix},  \\
         L <x< 2L, & \quad \begin{pmatrix} \phi_{1}(x,\zeta) \\  \phi_2(x,\zeta)   \end{pmatrix}= \begin{pmatrix}
         c_8 e^{-i \zeta x}  + c_7 e^{i \zeta x} \left( \frac{2L-x}{2i\zeta} + \frac{1}{(2i\zeta)^2}  \right)  \\  c_7 e^{i\zeta x}
          \end{pmatrix}.  \\
\end{aligned}
\right.
\end{equation}
Matching the value of the eigenfunction at $x=-2L,-L,0,L$,  we find
\begin{equation}
\begin{aligned}
   &c_1 =1, \quad c_2=-\frac{1}{(2i\zeta)^2} e^{4i \zeta L}, \quad c_3 =1,  \quad c_4=\frac{1}{(2i\zeta)^2}(2e^{2i\zeta L}- e^{4i\zeta L}),  \\
  & c_5=-\frac{1}{(2i\zeta)^2} (e^{2i\zeta L} -1)^2, \quad c_6=1-\frac{1}{(2i\zeta)^4} (e^{2i\zeta L} -1)^2,  \\
  & c_7=-\frac{1}{(2i\zeta)^2} (e^{2i\zeta L} -1)^2, \quad c_8=1-\frac{1}{(2i\zeta)^4} (e^{2i\zeta L} -1)^2 + \frac{2e^{2i\zeta L}}{(2i\zeta )^4} (e^{2i\zeta L}-1)^2.
\end{aligned}
\end{equation}
Simliarly,   when $x> 2 L$, we have
\begin{equation}
   \phi(x,t) = \begin{pmatrix}  a(\zeta)e^{-i\zeta x} \\ b(\zeta) e^{i \zeta x}     \end{pmatrix}.
\end{equation}
In the process of matching the value of eigenfunction at $x=2L$, we find
\begin{equation}
   \begin{aligned}
      a(\zeta)= & 1 - \frac{1}{(2i\zeta)^4} (e^{2i\zeta L}-1)^4, \\
      b(\zeta)= & - e^{2i\zeta L} \frac{(\sin(\zeta L))^2}{\zeta^2} ,
   \end{aligned}
\end{equation}
then the eigenvalues, i.e. the zeros of $a(\zeta)$, can be given  implicitly by
\begin{equation}
     e^{2i\zeta L} -1 \pm 2i\zeta =0, \qquad  or \qquad e^{2i\zeta L} -1 \pm 2 \zeta =0.
\end{equation}
However, since $a(\zeta)=a^{\ast}(-\zeta^{\ast})$, the eigenvalues are determined uniquely  by
\begin{equation}
     e^{2i\zeta L} -1 \pm 2i\zeta =0.
\end{equation}
Besides,  the asymptotic  behavior of $a(\zeta)$ for  large  and small $\zeta$  can be derived  from Eq. \eqref{a1}
\begin{equation}
       \begin{aligned}
         & a(\zeta) \sim  1-  \frac{1}{(2i\zeta)^4},   & \qquad  \zeta \to \infty,  \\
        &  a(\zeta)  \sim  1- L^4, & \qquad \zeta \to 0.
       \end{aligned}
\end{equation}
With the aid of the large $\zeta$ asymptotic behavior of $a(\zeta)$ and Eq. \eqref{conser3}, we find  that  the conserved quantities satisfy
\begin{equation}
  \mathcal{C}_{n}= \left\{
  \begin{aligned}
        -\frac{1}{m+1}, & \quad n=4m+3, m=0,1,2,\dots,  \\
        0,   &   \quad  else.
  \end{aligned}
  \right.
\end{equation}

\section{Conclusions}
In this work,  a detailed  study of the inverse scattering transform for a new  nonlocal LPD equation is carried out. Firstly, by an ingenious method, the  local and global  conservation laws  of nonlocal LPD equation is obtained, which establish the integrability   as an infinite dimensional Hamilton dynamic system. The direct scattering problem is constructed and some critical symmetries are obtained.  Afterwards, with the aid of the novel Left-Right RH approach, the inverse scattering problem is established. Furthermore, the potential function is recovered successfully.  By introducing the reflectionless case,  the soliton solutions of the nonlocal LPD equation are given.  In order to understand the dynamic  behavior of soliton solutions more intuitively,  we take $J=\overline{J}=1,2,3,4$ and select some special parameters as examples to present some interesting  phenomenon, such as breather-type solitons,  arc  solitons, three solitons,  four solitons, etc.  Meanwhile, we also discuss the influence of parameter $\delta$ on soliton solutions.  Besides,  under some special cases of initial condition such as  rectangular  wave,  arc wave and  triangular wave, we consider the zeros of the scattering data $a(\zeta)$ and  the conserved quantities.

\section*{Acknowledgements}
This work was supported by  the Natural Science Foundation of Jiangsu Province under Grant No. BK20181351, the National Natural Science Foundation of China under Grant No. 11975306, the Six Talent Peaks Project in Jiangsu Province under Grant No. JY-059, the Qinglan Project of Jiangsu Province of China,  and the Fundamental Research Fund for the Central Universities under the Grant Nos. 2019ZDPY07 and 2019QNA35.

\end{document}